\documentclass[aps,prb,superscriptaddress,amsfonts,amsmath,amssymb,twocolumn,showpacs,floatfix]{revtex4}

\usepackage{graphicx}
\usepackage{amsmath,amssymb}

\usepackage[utf8]{inputenc}
\usepackage{xcolor}

\def\jr{\ensuremath{J_{\rm{rung}}} }
\def\tr{\ensuremath{t_{\rm{rung}}} }
\def\jl{\ensuremath{J_{\rm{leg}}} }
\def\tl{\ensuremath{t_{\rm{leg}}} }
\def\vr{\ensuremath{V_{\rm{rep}}} }
\def\t{\ensuremath{\tau} }

\begin{document}
\title{Effective models of doped quantum ladders of non-Abelian anyons}
\author{Medha Soni}
%\email[]{medha.soni@irsamc.ups-tlse.fr}
\affiliation{Laboratoire de Physique Th\'eorique UMR-5152, CNRS and
Universit\'e de Toulouse, F-31062 France}
\author{Matthias Troyer}
\affiliation{Institut f\"ur Theoretische Physik, ETH-H\"onggerberg, 8093 Z\"urich, Switzerland}
\author{Didier Poilblanc}
%\email[]{didier.poilblanc@irsamc.ups-tlse.fr}
\affiliation{Laboratoire de Physique Th\'eorique UMR-5152, CNRS and
Universit\'e de Toulouse, F-31062 France}
\date{\today}

\begin{abstract}
Quantum spin models have been studied extensively in one and higher dimensions. Furthermore, these systems have been doped with holes to study $t$--$J$ models of $SU(2)$ spin-1/2. Their anyonic counterparts can be built from non-Abelian anyons, such as Fibonacci anyons described by $SU(2)_3$ theories, which are quantum deformations of the $SU(2)$ algebra. Inspired by the physics of $SU(2)$ spins, several works have explored ladders of Fibonacci anyons and also one-dimensional (1D) $t$--$J$ models. 
Here we aim to explore the combined effects of extended dimensionality and doping by 
 studying ladders composed of coupled chains of interacting itinerant Fibonacci anyons.
  We show analytically that in the limit of strong rung couplings these models can be mapped onto effective 1D models. These effective models can either be gapped models of hole pairs, or gapless models described by $t$--$J$ (or modified $t$--$J$--$V$) chains of Fibonacci anyons, whose spectrum exhibits a fractionalization into charge and anyon degrees of freedom. The charge degrees of freedom are described by the hardcore boson spectra while the anyon sector is given by a chain of localized interacting anyons. By using exact diagonalizations for two-leg and three-leg ladders, we 
show that indeed the doped ladders show exactly the same behavior as that of $t$--$J$ chains. 
In the strong ferromagnetic rung limit, we can obtain a new model that hosts two different kinds of Fibonacci particles - which we denote as the \emph{heavy} $\tau$'s and \emph{light} $\tau$'s. These two particle types carry the same (non-Abelian) topological charge but different (Abelian) electric charges. Once again, we map the two-dimensional ladder onto an effective chain carrying these heavy and light $\tau$'s. We perform a finite size scaling analysis to show the appearance of gapless modes for certain anyon densities whereas a topological gapped phase is suggested for another density regime. 
\end{abstract}

\maketitle

\section{Introduction}\label{introduction}
Quantum statistics is an important aspect of quantum mechanics and it lays down the rules for identifying different classes of particles.
In three dimensions, the exchange of two particles in a quantum system can result in a phase change of either 0 or $\pi$ for the wave function, leading to bosons and fermions. The scenario is very different in two-dimensions (2D) and it is well known that such quantum systems give rise to  anyonic quasiparticle excitations.\cite{Leinaas77,Wilczek82,Canright90} These anyons are observed as excitations, localized disturbances of the ground state in systems with topological order. Rich behavior emerges especially for non-Abelian anyons, for which the exchange of two anyons is described by a unitary matrix.\cite{Fred89, Frohlich90} 

Models of interacting non-Abelian anyons draw motivation from quantum spin models, and in particular the Heisenberg model, which has been studied for a wide range of lattices including one-dimension (1D) chains and ladders of $SU(2)$ spins. 
Non-Abelian anyons can be described as so-called quantum deformations of the usual $SU(2)$ spins. Formally, they are described by $SU(2)_k$ Chern-Simons theories. These $SU(2)_k$ theories are obtained by `deforming' the $SU(2)$ algebra so as to retain only the first $k+1$ representations. The most widely studied models are the $k=2$ and $k=3$ theories that describe `Ising' and `Fibonacci' anyons respectively. 

Non-Abelian anyons are expected to be present in topologically ordered systems including certain fractional quantum Hall states,\cite{clarke13, MooreRead, read-rezayi, clarke14} $p$-wave superconductors,\cite{Alicea11} spin models,\cite{levin05,bonesteel12,kapit13,palumbo14} solid state heterostructures\cite{ReadGreen, Kitaev01, Fu08, Sao10, Lutchyn10,mong14} and rotating Bose-Einstein condensates.\cite{Cooper01} In particular, Fibonacci anyons occur as quasi holes in the $\mathbb{Z}_3$ Read-Rezayi state which might be able to describe the $\nu=12/5$ fractional quantum Hall state.\cite{slingerland01}  While several experiments have shown evidence for emergent Majorana modes,\cite{Mourik12,das12,rokhinson12, finck13, churcill13} experimental evidence of Fibonacci anyons is yet to come.

In the absence of interactions, non-Abelian anyons  have an exponentially degenerate ground state manifold. On this ground state manifold exchange (braiding) of non-Abelian anyons acts in a non-commutative way, by bringing about a non-trivial transformation in the degenerate manifold of the many-quasiparticle Hilbert space. This  has generated interest since  it can be used  for topological fault-tolerant quantum computation.\cite{Preskill97,Stern13, Nayak08, Freedman03, Kitaev03, Kitaev06} 

Interactions between non-Abelian anyons can be modeled by generalizations of the spin-1/2 Heisenberg model. These models  have been studied for chains of Fibonacci anyons\cite{intro} including nearest neighbour couplings\cite{golden} and also longer range couplings\cite{maj-ghosh} have been studied. Chains of higher spin quasiparticles have also been explored,\cite{gils09,gils13, PEF13,PEF14,PEF142} critical phases have been identified and the corresponding conformal field theory (CFT) has been obtained from numerical simulations.\cite{RNCP12} The effect of disorder for chains of Fibonacci anyons has been investigated.\cite{Bone07, Fidk08}

As a step towards understanding the collective behavior of itinerant non-Abelian anyons, these chains can be doped with mobile holes, inspired by the electronic $t$--$J$ model.\cite{zhang88,Hubbard63,Gutz63,Kan63} Electrons  confined in 1D can exhibit the phenomenon of ``spin-charge separation'', where the spectrum can be interpreted in terms of two independent pieces, one arising from electric charge without spin, and the other from a spinon without any charge.\cite{And67}
Analogously low-energy effective $t$--$J$ models have been analyzed for the case of doped chains of Fibonacci (and Ising) anyons. These models exhibit a fractionalization of the spectrum into charge and anyonic degrees of freedom,\cite{frac,itinerant_long} an extension of a phenomenon that exists in Luttinger liquids.\cite{Tom50,Lutt63,Haldane81}

As a step towards two dimensions,  anyon models have been investigated on  chains coupled to form so-called quantum \emph{ladders} of non-Abelian anyons, which provide anyonic generalizations of the 2D quantum magnets.\cite{ladder,liquid}

In this paper, we combine the models mentioned above, by studying {\em doped} quantum ladders  of Fibonacci anyons consisting of two or three chains.
The structure of the rest of the paper is as follows: Section~\ref{model}  begins with a brief introduction of non-Abelian anyons and introduces our model. 
In section ~\ref{isolated_rungs}, we discuss the strong rung coupling limit drawing analogy with the spin ladder models, listing out all the possible quantum states and the phase diagram for the model under study. 
In section \ref{coupled_rungs}, we list all the phases that arise for weakly coupled rungs.
The subsequent sections are dedicated to the discussion of all these phases. 
Section \ref{sec:effpaired} discusses the hard-core boson models and section \ref{sec:effgolden} describes the golden chain phases. 
In section \ref{sec:efftj}, we analyse the effective $t-J$ models for two and three-leg ladders, presenting the numerical results for the same. We extend the phenomenon of spin-charge separation to (doped) ladders of Fibonacci anyons. 
In section~\ref{fm_rungs}, we introduce a new model of heavy and light Fibonacci anyons that is obtained for certain density regimes when the rung couplings are FM. 
We conclude by summarising our results in section~\ref{conclude}. 

%%%%%%%%%%%%%%%%%%%%%%%%%%%%%%%%%%%%%%%%%%%%%
%%		the model
%%%%%%%%%%%%%%%%%%%%%%%%%%%%%%%%%%%%%%%%%%%%%

\section{The model}\label{model}

\subsection{Introduction to non-Abelian anyons}\label{intro_anyons}

In this paper we  focus on Fibonacci anyons that are described by the $SU(2)_3$ theories.  $SU(2)_k$ Chern-Simons theories\cite{witten,intro} are so-called quantum deformations of the $SU(2)$ algebra. Their degrees of freedom are encoded by  $\lq$topological charges' $j$, which are generalized angular momenta. In contrast to $SU(2)$, in $SU(2)_k$ theories the total `spin' $j$ is limited to be  $j=0,\frac{1}{2},\cdots,\frac{k}{2}$. 

Akin to the tensor product of spins,  non-Abelian anyons can be $\lq$fused' according to  fusion rules given by
\begin{equation}
 j_1 \times j_2 = \sum_{j_3 = |{j_1-j_2}| }^{min\{j_1+j_2,k-j_1-j_2\}} j_3   .
\end{equation}

For example, for the fusion of two anyons with  $j_{1,2} = \frac{1}{2}$, these rules would mean $\frac{1}{2}\otimes \frac{1}{2} = 0 \oplus 1$ (for $k \ge2$). Similarly, for the case of Fibonacci anyons $(k=3)$, when $j_{1,2}=1$ the fusion rule reads as $1 \otimes 1 = 0 \oplus 1$. Note that this is different from what one would obtain under a  tensor product of $SU(2)$ spin-1 particles.
In the limit $k \rightarrow \infty$, however, we recover the $SU(2)$ algebra and the rule simply describes the tensor product of two ordinary $SU(2)$ spins. 

\begin{figure}[t]
\begin{center}
\includegraphics[width=\columnwidth]{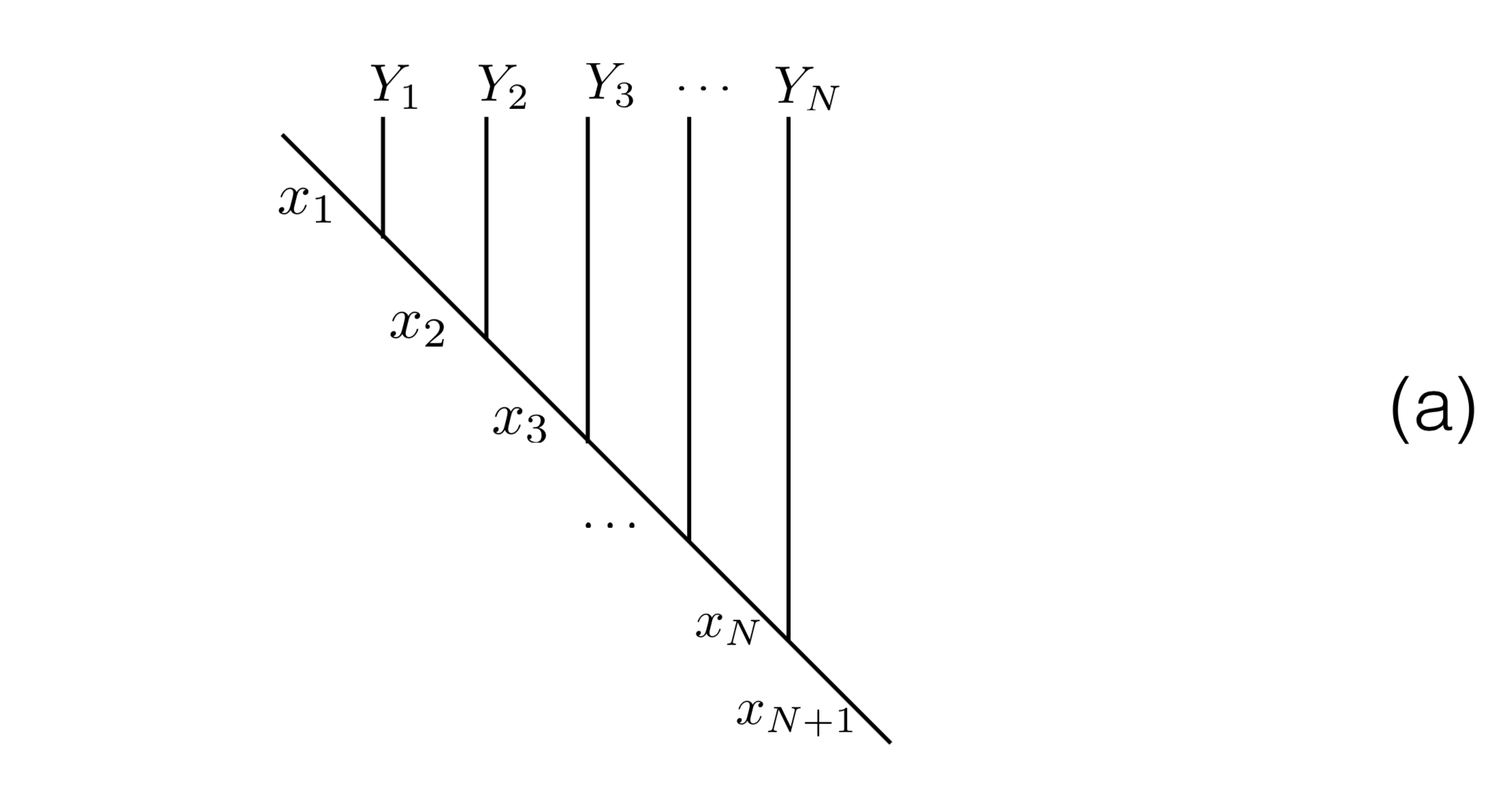}
\vskip 0.4truecm
\includegraphics[width=\columnwidth]{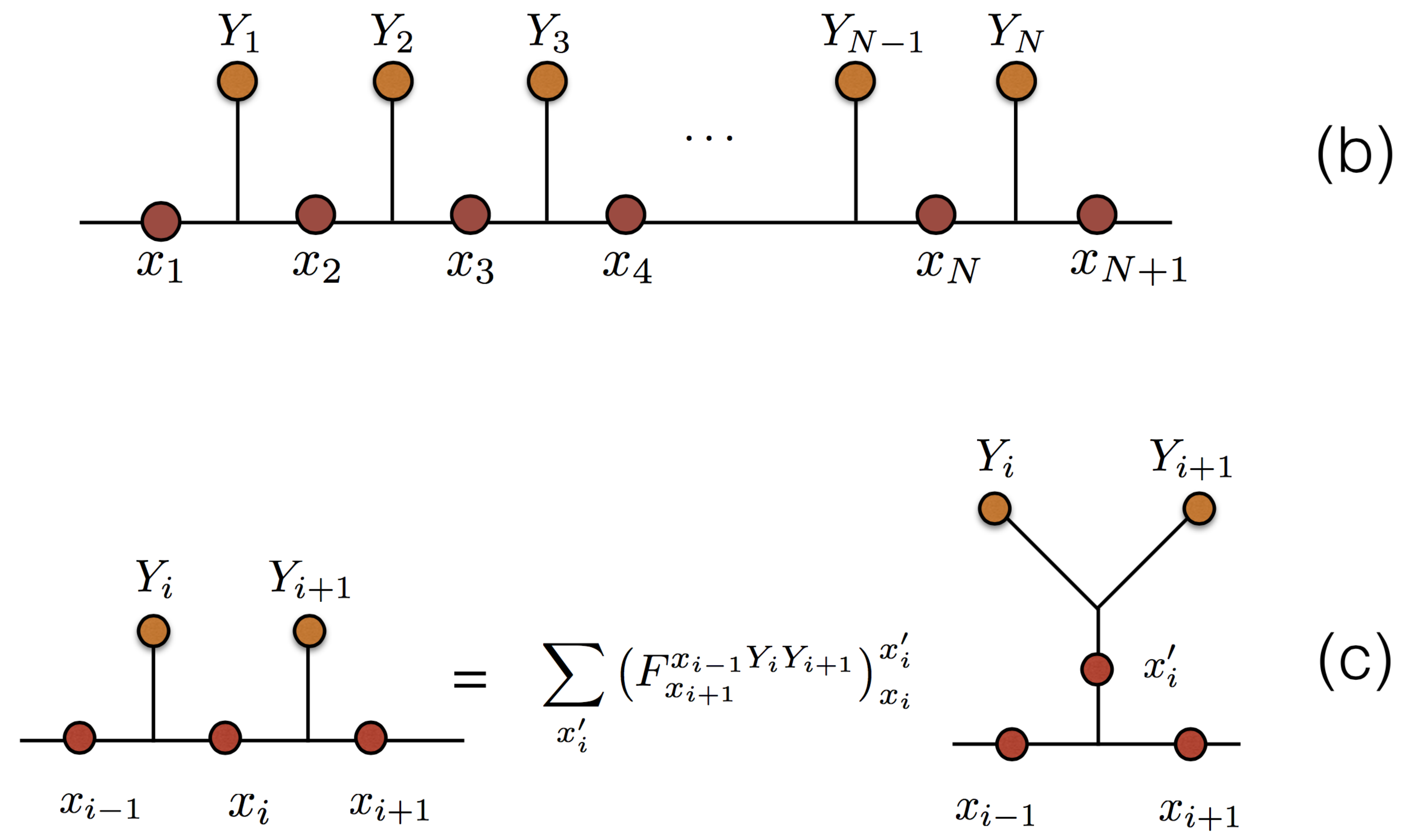}
\caption{(a) Illustration of the standard fusion tree with site labels $Y_i$ (that can be either \t or $\mathbf{1}$) and bond labels $x_i$. (b) The flat version of the fusion tree. (c) A basis change to a different fusion tree using an $F$-move. }
\label{fig:3figs}
\end{center}
\end{figure}

In the rest of this   manuscript, we focus only to the Fibonacci theory with $k=3$, unless otherwise stated. There the allowed values for the topological charges are $j=0,\frac{1}{2},1,\frac{3}{2}$. But if we look closely at the fusion rules, we can make the identification $ 0 \leftrightarrow \frac{3}{2}$ and $ 1 \leftrightarrow \frac{1}{2}$. 
Thus, the Fibonacci theory has two distinct types of \emph{particles} 
which we denote as $\mathbf{1}$ for the trivial particle with $j=0$ and $\tau$ for the Fibonacci anyon with $j=1$ respectively. Using these, the fusion rules read
\begin{equation}
\begin{aligned}
\mathbf{1}\otimes \mathbf{1}&= \mathbf{1}\\
\tau \otimes \mathbf{1}&= \mathbf{1}\otimes  \tau = \tau \\
\tau \otimes \tau &= \mathbf{1}\oplus \tau   .\\
\end{aligned}
\end{equation}

\begin{figure}[bt]
\includegraphics[width= \columnwidth]{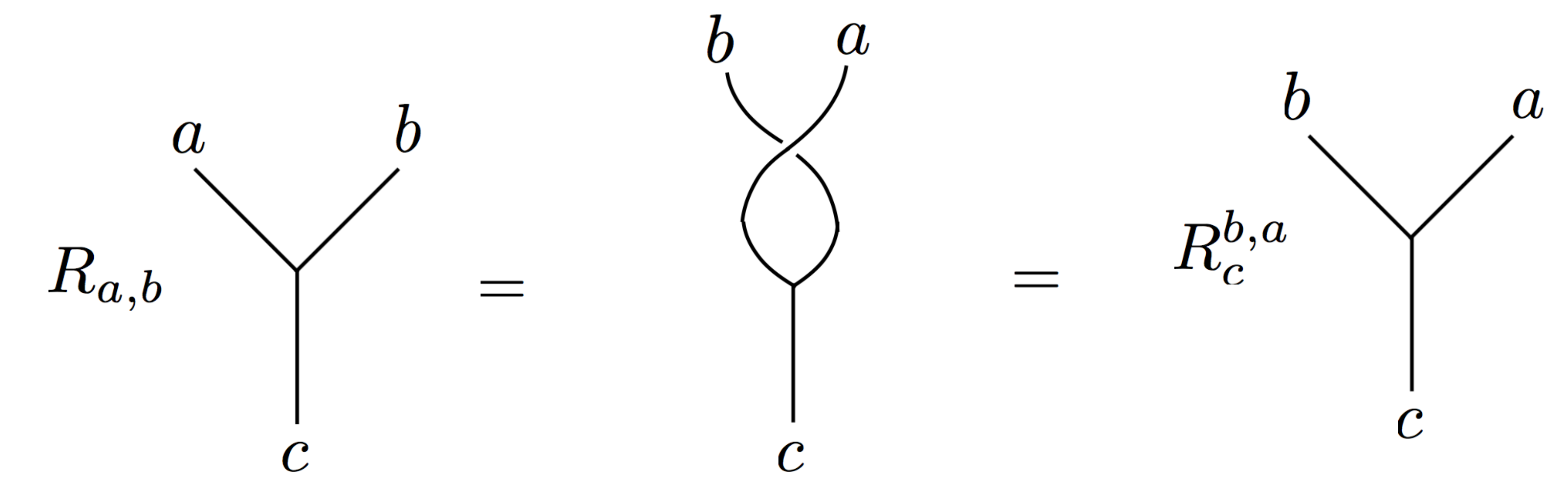}
\caption{Convention for the right handed braid to exchange two particles. }
\label{fig:R}
\end{figure}

We represent a system of $N$ anyons by means of a \emph{fusion tree} as shown in Fig.~\ref{fig:3figs}(a), where the anyonic charges of the individual anyons are labelled by $Y_i$. The fusion outcome of successive fusion of the anyons are encoded by the `bond' labels (links) in the fusion tree, labelled by $x_i$ in Fig.~\ref{fig:3figs}(a).
The constraints on the bond labels due to fusion rules which must be satisfied at each  vertex significantly reduce  the size of the internal Hilbert space. 
For $N$ Fibonacci anyons ($\tau$) the Hilbert space  grows asymptotically as $\phi^N$ where $\phi=(\sqrt5+1)/2$ is the golden ratio.  From now on we draw a flat version of the fusion tree, as shown in Fig.~\ref{fig:3figs}(b).

\begin{figure*}
\includegraphics[width= 1.7\columnwidth]{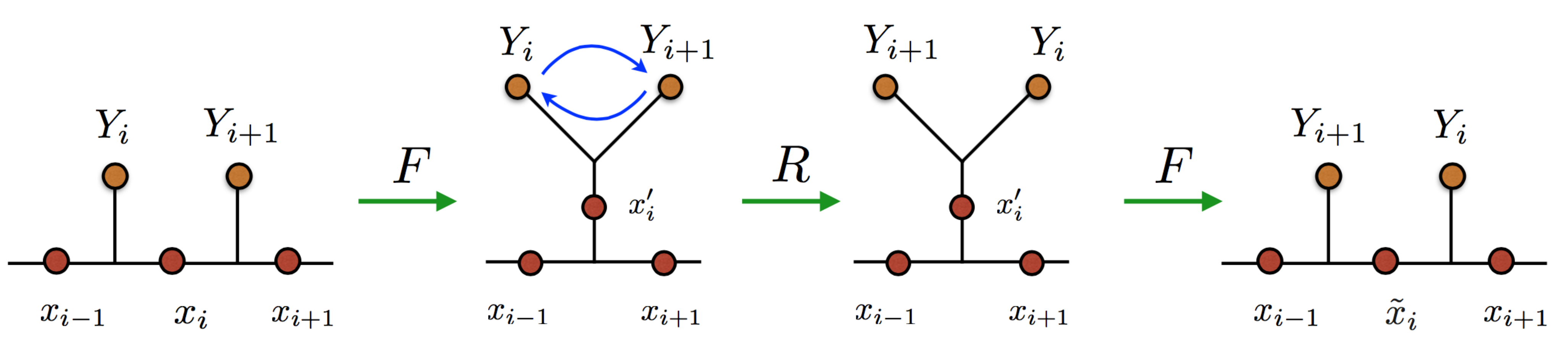}
\caption{Schematic representation of a braid on the fusion tree.}
\label{fig:FRF}
\end{figure*}

To perform an operation on nearest neighbor anyons, it is advantageous  to change to a different basis in which  the two-particle fusion outcome is explicit. This is done via the $F$-move shown schematically in Fig.~\ref{fig:3figs}(c). 
The $F$-move depends on the two site labels $Y_i$ and $Y_{i+1}$ and the bond labels $x_{i-1}$ and $x_{i+1}$. If at least one of these four labels is $\mathbf{1}$, then there is only one choice of bond labels that satisfies the fusion algebra and the $F$-move is trivial. A non-trivial matrix is obtained only when all the four labels are \t anyons.
Specialising to the case where $Y_i=Y_{i+1} =\t$, the labels for the three bonds $|x_{i-1},x_i,x_{i+1}\rangle$  allowed by the fusion rules are
\begin{equation}\label{eq:basis}
\{|{\bf 1},\tau,{\bf 1}\rangle, |{\bf 1},\tau,\tau\rangle, |\tau,\tau,{\bf 1}\rangle, |\tau,{\bf 1},\tau\rangle, |\tau,\tau,\tau\rangle\} 
\end{equation}
which transforms to a new basis $|x_{i-1},x'_i,x_{i+1}\rangle$ after the $F$-move:
\begin{equation}\nonumber
 \{|{\bf 1},{\bf 1},{\bf 1}\rangle,|{\bf 1},\tau,\tau\rangle,|\tau,\tau,{\bf 1}\rangle, |\tau,{\bf 1},\tau\rangle, |\tau,\tau,\tau\rangle\}  .
 \end{equation} 
 Using these bases the $F$-matrix is represented as,
 \begin{equation}\label{eq:f}
 F=\begin{bmatrix}
1&&&&\\
&1&&&\\
&&1&&\\
&&&\phi^{-1}&\phi^{-1/2}\\
&&&\phi^{-1/2}&-\phi^{-1}
\end{bmatrix}  ,
\end{equation}
where as mentioned above we have a non-trivial $2\times 2$ submatrix only when also  $x_{i-1}=x_{i+1}=\t$.

Another operation that we need to perform on nearest neighbor anyons is that of exchanging (or braiding) them. In Fig.~\ref{fig:R}, we show our convention for a right-handed braid. The left-hand braiding is the inverse of the process shown here. Under a right-handed braid anyons $a$ and $b$ pick up a phase $R^{b,a}_c$ depending on the anyon types, $a$ and $b$, that are undergoing an exchange and their fusion outcome $c$. Note that whenever $a$ or $b$ are 
$ \mathbf{1}$, the phase is trivial. Non-trivial phases  are only obtained are for $a=b=\t$:
\begin{equation}\label{eq:r}
R^{\tau\tau}_{\mathbf{1}} = e^{+4\pi i/5},       R^{\tau\tau}_{\tau} =  e^{-3\pi i/5}  .
\end{equation}

In order to implement a braid on the standard fusion tree, we first have to change basis using an $F$-move to make the fusion outcome of the two anyons explicit, and then braid. This process is shown schematically in Fig.~\ref{fig:FRF}.
This is represented by a Braid matrix $B$ acting on the bond labels $|x_{i-1},x_i,x_{i+1}\rangle$. 
The only non-trivial Braid matrix is obtained when both the sites are occupied by \t anyons. In the basis of Eq.~(\ref{eq:basis}) we obtain:
\begin{widetext}
\begin{equation}\label{eq:braid}
 B= F R F = \begin{bmatrix}
e^{4i\pi/5}&0&0&0&0\\
0&e^{-3i\pi/5}&0&0&0\\
0&0&e^{-3i\pi/5}&0&0\\
0&0&0&\frac{1}{\phi^2}e^{4i\pi/5}+\frac{1}{\phi}e^{-3i\pi/5}&\frac{1}{\phi^{3/2}}(e^{4i\pi/5}-e^{-3i\pi/5})\\
0&0&0&\frac{1}{\phi^{3/2}}(e^{4i\pi/5}-e^{-3i\pi/5})&\frac{1}{\phi^2}e^{-3i\pi/5}+\frac{1}{\phi}e^{4i\pi/5} \\
\end{bmatrix}  .
\end{equation}
\end{widetext}

Note that when the two site labels are a ${\mathbf{1}}$ and a $\tau$, the $F$-moves and the exchange phases are all trivial.  The Braid matrix is effectively the hopping of the anyon to the adjacent site. When both the site labels are ${\mathbf{1}}$, the Braid matrix is simply given by the identity matrix. 

\subsection{Golden chain model}\label{goldenchain}
\begin{figure}[b]
\begin{center}
\includegraphics[width=\columnwidth]{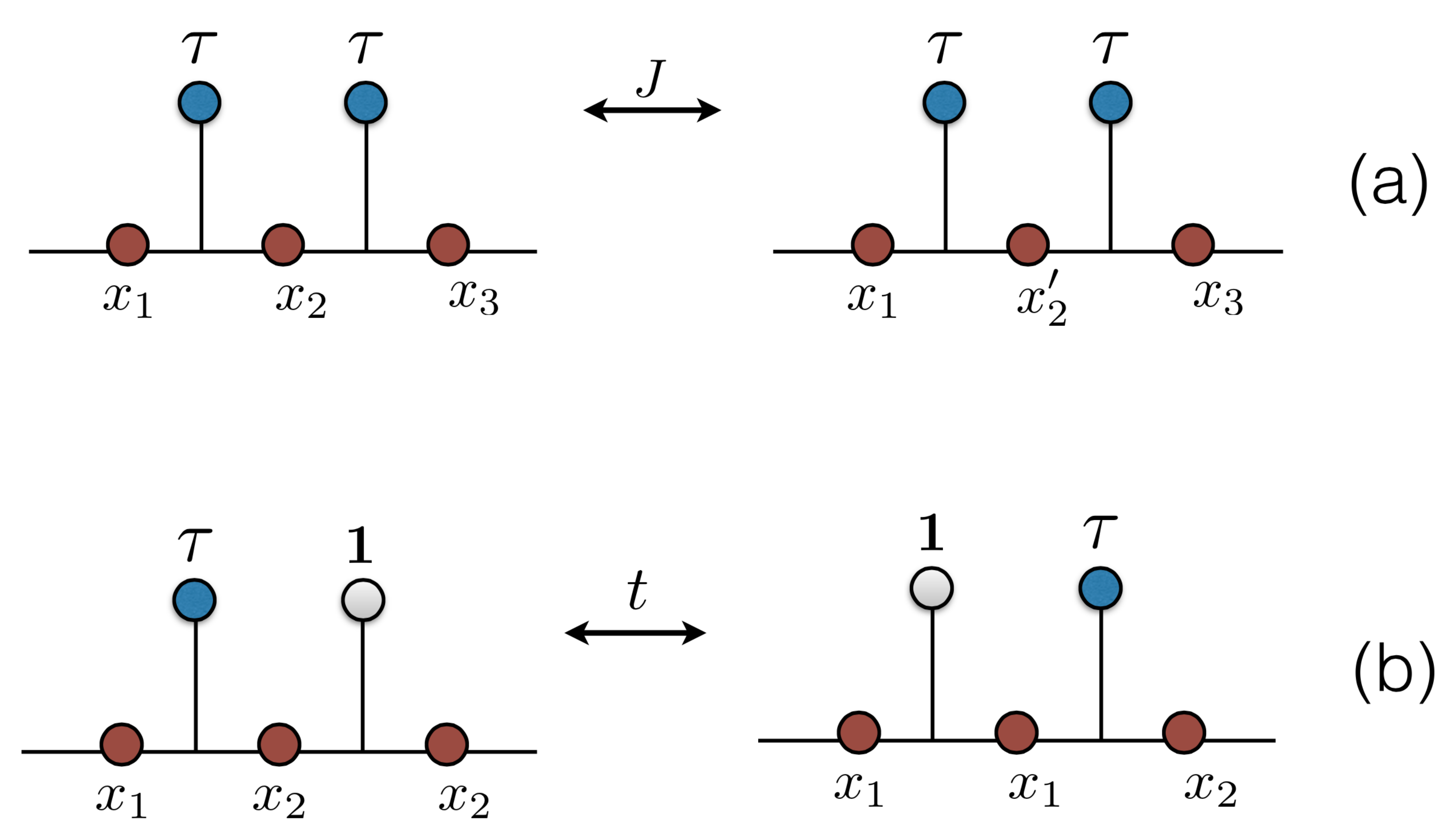}
\caption{(a) Nearest neighbor magnetic interaction of amplitude $J$. (b) The kinetic hopping (of amplitude $t$) of an anyon to its nearest neighbor vacant site. The blue circles represent \t anyons while the white circles denote vacant sites or holes.}
\label{fig:tJ}
\end{center}
\end{figure}

In this section we review the so-called `golden chains', consisting of 1D arrays of localized Fibonacci anyons with pairwise interactions between nearest neighbors.\cite{golden}
In this model the Hamiltonian for the magnetic interactions between anyons is defined in analogy to the Heisenberg exchange interaction. We assign an energy $-J$ if the fusion outcome of two interacting anyons is trivial. For AFM couplings $(J>0)$, this favours the fusion outcome of two neighbouring anyons to be  trivial, while for FM couplings $(J<0)$, the fusion of two anyons is preferred to be  $\tau$. This interaction between nearest neighbor anyons is depicted schematically in Fig.~\ref{fig:tJ}(a) and is implemented
  by projecting on the identity fusion channel \begin{equation}H_{\rm{mag}}=J h_{\rm mag}=-J(F P^{\mathbf{1}} F^{-1})  .\end{equation} where $F$ is the operator corresponding to the $F$-move (see Eq.~(\ref{eq:f})) and $P^\mathbf{1}$ is an operator that projects onto the $\mathbf{1}$ state. In the basis of Eq.~(\ref{eq:basis}), the matrix representation for the (dimensionless) magnetic interaction can be written explicitly as:
\begin{equation}\label{eq:hmag}
 {h_{\rm{mag}}}=-\begin{bmatrix}
1&&&&\\
&0&&&\\
&&0&&\\
&&&\phi^{-2}&\phi^{-3/2}\\
&&&\phi^{-3/2}&\phi^{-1}
\end{bmatrix}  .
\end{equation}

Golden chains with  AFM couplings are described by the $k=3$ restricted solid on solid (RSOS) model which is a CFT with central charge $c=7/10$. For ferromagnetic couplings the corresponding CFT is that of the critical 3-state Potts model with $c=4/5$.\cite{golden,PDF97}

\subsection{Itinerant Fibonacci anyons in 1D}\label{tJ}
To model itinerant anyons we introduce holes, {\it i.e.} sites with a trivial anyon $\bf 1$  on some of the sites.  The holes and $\tau$ anyons are labelled by different $U(1)$ (electric) charges %, counting the number of $\t$ particles 
and anyonic (non-Abelian) charges. The $\tau$ anyons (referred to  simply as `anyons' or `\t particles' hereafter) can move on the chain which results in an additional kinetic energy contribution. This kinetic process is schematically shown in Fig.~\ref{fig:tJ}(b). It involves the hopping of a particle, along with its electric and anyonic charge  to a neighbouring site. 

\begin{figure}[b]
\begin{center}
\includegraphics[width= 0.9 \columnwidth]{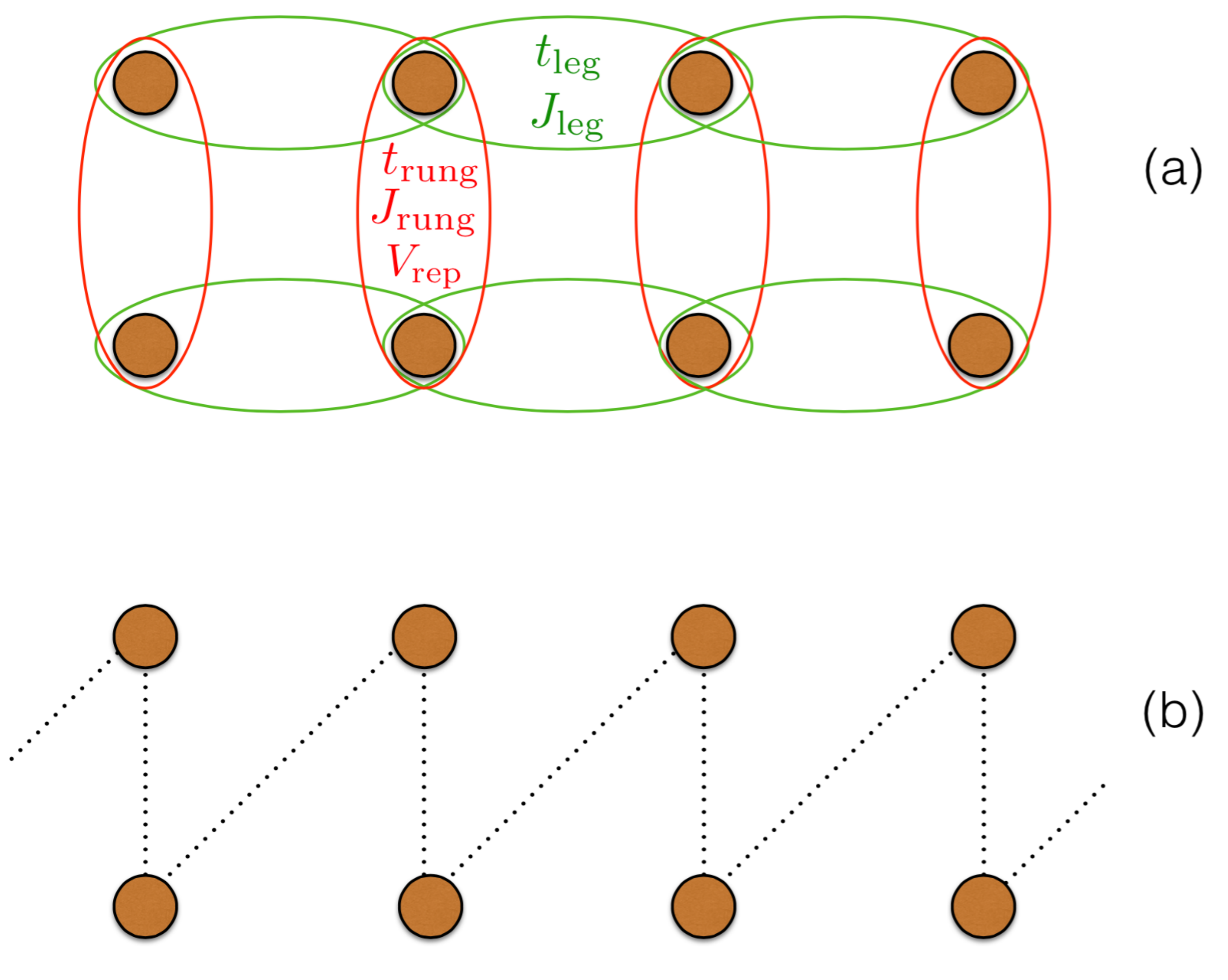}
\caption{Two-leg ladder: (a) Interactions along the leg and rung directions. (b) The zig-zag fusion path. The couplings have been indicated.}
\label{fig:ladder}
\end{center}
\end{figure}

\begin{figure*}[t]
\includegraphics[width=1.8\columnwidth]{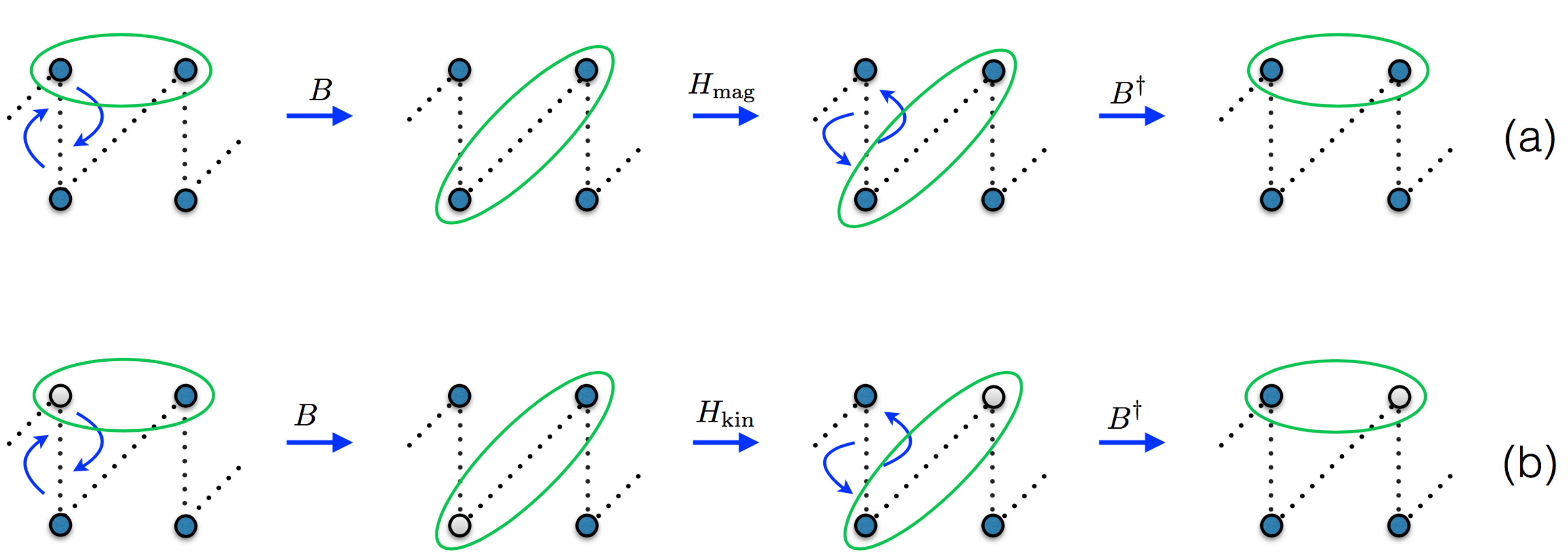}
\caption{Evaluation of Hamiltonian terms on a two-leg ladder along the leg direction that involve the braid operation. The blue circles represent \t particles white the white circles are for vacant sites. The green ellipses denote the particles that are interacting along the leg direction. (a) the magnetic term (b) the kinetic term.}
\label{fig:braid}
\end{figure*}

Our specific model of itinerant anyons is a generalization of the electronic $t$--$J$ model.\cite{zhang88}  Assuming a large on-site charging energy, we  eliminate the possibility of doubly occupied sites and allow anyons to only hop to empty sites. As in the case of electrons, the low-energy effective $t$--$J$ model allows hopping of the anyons to nearest neighbor vacant sites and an exchange interaction between nearest neighbor anyons analogous to the Heisenberg interactions explained above in Sec.~\ref{goldenchain}. The kinetic term can then be written as
$H_{\rm kin} =- t  h_{\rm kin}$, where $h_{\rm{kin}}$ is the (dimensionless) operator corresponding to the nearest neighbor hopping process shown in Fig.~\ref{fig:tJ}(b). A $t$--$J$ chain of itinerant anyons was studied for Ising and Fibonacci anyons,\cite{frac, itinerant_long} revealing a separation of excitations into charge and anyonic excitations, similar to spin-charge separation in its electronic counterpart.

\subsection{Ladders of Fibonacci anyons}\label{ladders}

\subsubsection{Undoped ladders}\label{undoped}

Ladders of Fibonacci anyons are formed by coupling  chains of localized anyons.\cite{ladder}  Anyons interact with their nearest neighbors along the leg and rung directions via \jl and $J_{\rm{rung}}$. These interactions are shown schematically in Fig.~\ref{fig:ladder}(a). As fusion path we choose the \emph{zig-zag path} shown in Fig.~\ref{fig:ladder}(b), since it minimizes the effective range of interactions on the fusion path. We choose periodic boundary conditions along the leg direction and open boundary conditions along the rungs. 

With this choice of fusion path, nearest neighbor interactions  on the rungs are also nearest neighbor along the fusion path, while those between anyons on the same leg are longer range along the fusion path.  Nearest neighbor rung interactions can be implemented in exactly the same way as for nearest neighbor on a chain (see Sec.~\ref{goldenchain}). In order to evaluate the interactions between \t particles on the  same leg, we have to implement a change of basis, this time by braiding them in a clockwise 
manner until they are neighbours along the fusion path.
This braiding is performed by the unitary braid matrix $B$ (see Eq.~(\ref{eq:braid})). 
Once the \t particles are nearest neighbours along the path they can interact with the same term as discussed above in Sec.~\ref{goldenchain}.
After carrying out the interaction, the anyons have to be braided back to their original positions. 

More specifically, for a two-leg ladder, adjacent anyons along the leg direction are next nearest neighbours along the fusion path (see Fig.~\ref{fig:ladder}(b)). Thus, one needs to implement one braid operation.
The Hamiltonian for the magnetic interactions between nearest neighbor rungs $r$ and $r+1$ on the upper leg is given by 
\begin{equation}
(H^1_{\rm mag})_{r}=\jl B^\dagger_{2r-1} (h_{\rm{mag}})_{2r} B_{2r-1} ,
\end{equation}
and on the lower leg as
\begin{equation}
(H^2_{\rm mag})_{r}=\jl B^\dagger_{2r+1} (h_{\rm{mag}})_{2r} B_{2r+1} ,
\end{equation}
 where $h_{\rm{mag}}$ has been defined in Eq.~(\ref{eq:hmag}) and $r$ labels the rungs
 (so that $i=2r$ labels the diagonal bonds along the path). 
In Fig.~\ref{fig:braid}(a), we summarize the magnetic interactions between nearest neighbor along the leg direction for a two-leg ladder. 

For a three-leg ladder, the leg interactions are  longer ranged interactions, since adjacent anyons on a leg are third neighbours along the fusion path. Thus, nearest neighbor leg interactions on a three-leg ladder ladder require two braids before particles are nearest neighbors  on the fusion path.\cite{ladder} One gets for interaction between rungs $r$ and $r+1$ on the upper leg,
\begin{equation}
(H^1_{\rm mag})_{r}=\jl B^\dagger_{3r-2} B^\dagger_{3r-1} (h_{\rm{mag}})_{3r} B_{3r-1} B_{3r-2} ,
\end{equation}
on the middle leg,
\begin{equation}
(H^2_{\rm mag})_{r}=\jl B^\dagger_{3r+1} B^\dagger_{3r-1} (h_{\rm{mag}})_{3r} B_{3r-1} B_{3r+1} ,
\end{equation}
and on the lower leg,
\begin{equation}
(H^3_{\rm mag})_{r}=\jl B^\dagger_{3r+2} B^\dagger_{3r+1} (h_{\rm{mag}})_{3r} B_{3r+1} B_{3r+2} ,
\end{equation}
where $i=3r$ labels the diagonal bonds along the path.

The full magnetic Hamiltonian on the legs is obtained by adding all contributions, $H^{\rm{leg}}_{\rm mag}=\sum_1^W H^l_{\rm mag}$, where $W$ is the number of legs. As an implementation detail we want to mention that the action of the  operator $H^l_{\rm mag}$ can generate up to $2^{2W-1}$ states for each bond interaction. The operator mixes spin labels, thus generates multiple images for each two-body interaction. This exponential increase in the number of resulting states leads to denser Hamiltonian matrices as one increases the width $W$ of the ladder and restricts us from exploring larger system sizes.

For ladders of Fibonacci anyons it was shown that similar odd-even effects as seen for $SU(2)$ spins continue to exist in the limit of strong AFM rungs.\cite{ladder} AFM coupled ladders of Fibonacci anyons with even number of legs are gapped while those with odd number of legs are critical and are described by the same CFT as the golden chain.
On the other hand, Fibonacci ladders with FM rung couplings are quite different from their $SU(2)$ counterparts.\cite{ladder} Fibonacci ladders with a width that is a multiple of three are gapped since the rungs form singlets $(j=0)$. Other widths are gapless since the isolated rungs effectively form $\tau$'s, thereby yielding a gapless chain as the effective low-energy model. This is different from $SU(2)$ spins, for which an even number form a singlet ground state, and where thus all even width ladders are gapped.

\subsubsection{Doped ladders}\label{doped}

In this paper we focus on  itinerant  ladders of Fibonacci anyons. As schematically represented in Fig.~\ref{fig:ladder}(a) we denote the interaction strengths along the leg direction by \jl and \tl for the magnetic and kinetic terms respectively. Along the perpendicular direction, the couplings \jr and \tr denote respectively the magnetic and kinetic terms. 

The magnetic interactions for the ladder have already been described in the section~\ref{undoped} and we thus only need to  discuss the kinetic terms. Once again, the two sites on a rung are adjacent on the fusion path and the hopping is thus  implemented as for the 1D $t$--$J$ chain (see section~\ref{tJ}). 
For a \t particle and hole lying on adjacent sites on the same leg of the ladder, we need to, once again, braid the sites to bring them to adjacent positions on the fusion tree. 
For a two-leg ladder, the kinetic term between rungs $r$ and $r+1$ on the upper leg is given by 
\begin{equation}
(H^1_{\rm kin})_{r}=\tl B^\dagger_{2r-1} (h_{\rm{kin}})_{2r} B_{2r-1} ,
\end{equation}
and on the lower leg as
\begin{equation}
(H^2_{\rm kin})_{r}=\tl B^\dagger_{2r+1} (h_{\rm{kin}})_{2r} B_{2r+1} ,
\end{equation}
 where $h_{\rm{kin}}$ has been defined above and $r$ labels the rungs. 
In Fig.~\ref{fig:braid}(b), we summarize the kinetic process between nearest neighbor sites along 
the leg direction for a two-leg ladder. 

For a three-leg ladder,  a \t particle and hole lying on adjacent sites on a leg are third neighbours along the fusion path. Thus, nearest neighbor leg interaction on a three-leg ladder requires two braids before the \t particle and the hole 
are nearest neighbors on the fusion path. One gets for kinetic terms between rungs $r$ and $r+1$ on the upper leg,
\begin{equation}
(H^1_{\rm kin})_{r}=\tl B^\dagger_{3r-2} B^\dagger_{3r-1} (h_{\rm{kin}})_{3r} B_{3r-1} B_{3r-2} ,
\end{equation}
on the middle leg,
\begin{equation}
(H^2_{\rm kin})_{r}=\tl B^\dagger_{3r+1} B^\dagger_{3r-1} (h_{\rm{kin}})_{3r} B_{3r-1} B_{3r+1} ,
\end{equation}
and on the lower leg,
\begin{equation}
(H^3_{\rm kin})_{r}=\tl B^\dagger_{3r+2} B^\dagger_{3r+1} (h_{\rm{kin}})_{3r} B_{3r+1} B_{3r+2} ,
\end{equation}
where $i=3r$ labels the diagonal bonds along the path.

Note that, in the above equations, some of the braids may be trivial, in contrast to the case of the magnetic interactions.
The full kinetic Hamiltonian on the legs is obtained by adding all contributions, $H^{\rm{leg}}_{\rm kin}=\sum_1^W H^l_{\rm kin}$. Here, the action of the  operator $H^l_{\rm kin}$ can generate 
up to $2^{2W-2}$ states for each bond interaction. 

\begin{figure}[t]
\includegraphics[width=0.9 \columnwidth]{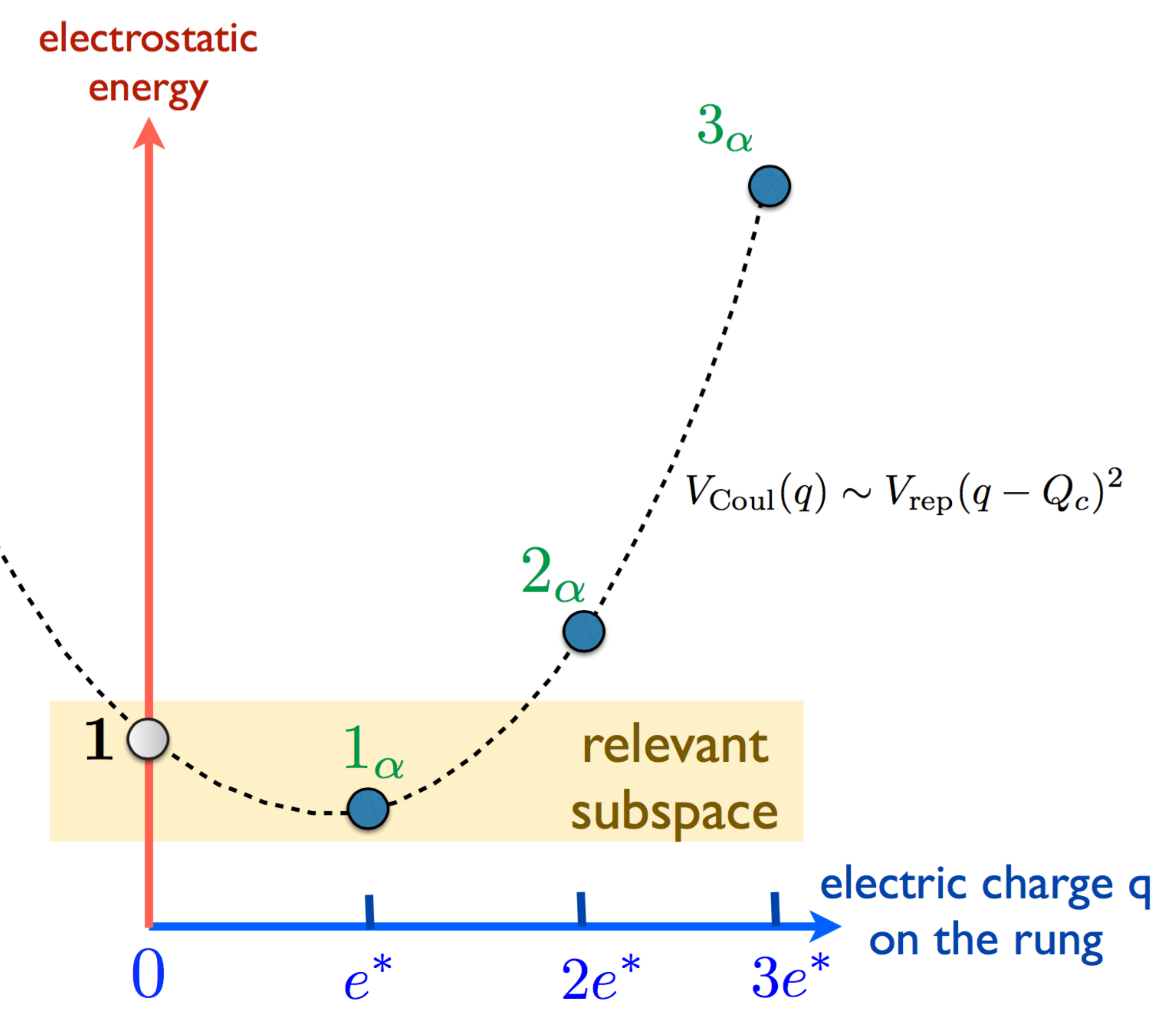}
\caption{Schematic energy spectra in the presence of a parabolic Coulombic charging energy. One can tune the chemical potential and the repulsion between the particles in order to gap out the higher energy sectors. This allows restricting the calculations to a low energy subspace.
}
\label{fig:parabola}
\end{figure}

We also consider models with an additional rung charging term
\begin{equation}
V_{\rm{Coul}} (q) \sim  \vr (q-Q_c)^2  ,
\end{equation}
where $q$ is the number of anyons on a rung and $Q_c$ is determined by the implicit chemical potential. Fig.~\ref{fig:parabola} shows the energy profile for a given rung composition on the ladder. For the three-leg ladder under consideration, this term acts pairwise between all the three possible pairs of particles that can exist on the three-site rung. Assuming a charging energy which is much larger compared to the exchange energy, we can consider the limit $\vr \rightarrow \infty$. In this limit we can restrict our calculations to just two values of the occupation on the rung, $n$ and $n+1$, with $n$ ranging from $0$ to $W-1$.  This reduces the Hilbert space and thus allows us to perform simulations of larger ladders.

\subsection{Simulation algorithm}

Our numerical simulations have been performed by exact diagonalization using the Lanczos algorithm.
We exploit periodic boundary conditions along the leg direction to implement translation symmetry. This allows us to block-diagonalize the Hamiltonian into $L$ blocks labeled by the total momentum $K=2\pi \frac{m}{L}$ ($m$ being an integer), which reduces the size of the Hilbert space, which is the major limiting factor in the simulations, by a factor  $L$. 

\section{Phase diagrams}
\begin{figure}[t]
\includegraphics[width=\columnwidth]{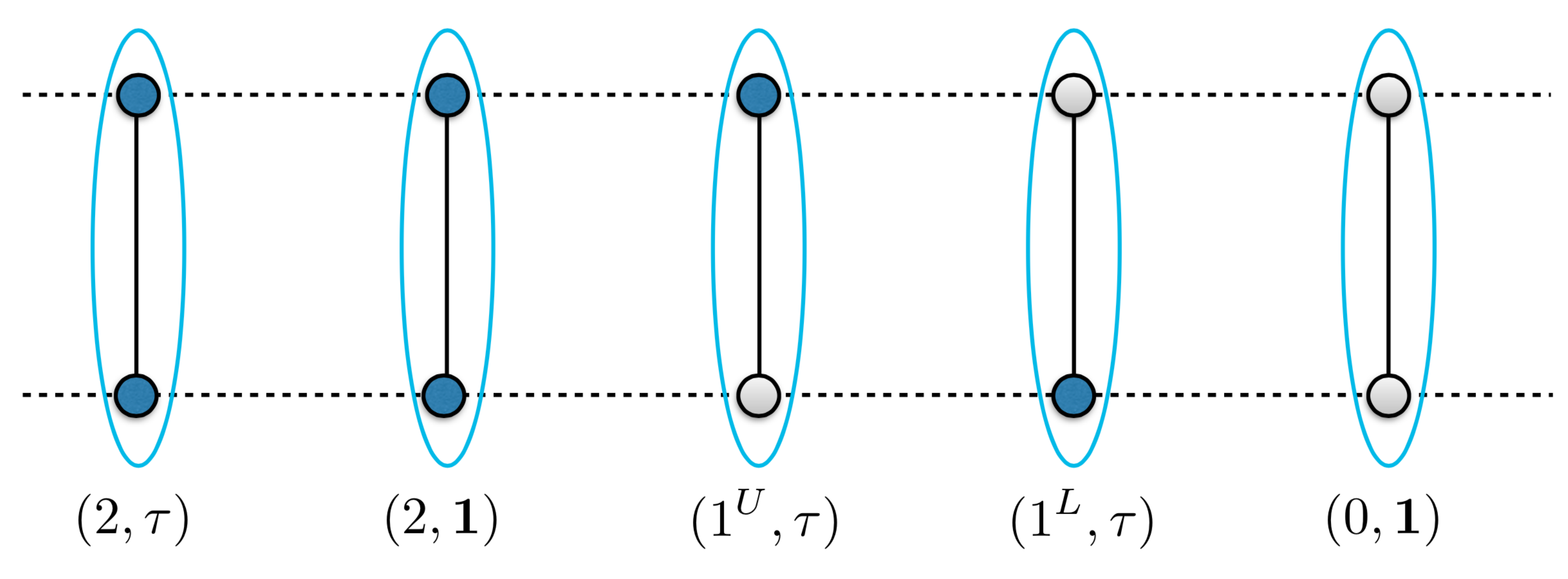}
\caption{All possible fusion outcomes for a rung of a two-leg ladder. The blue circles represent $\tau$'s while white circles represent vacant sites. The $U(1)$ and topological charges for the different configurations are denoted in the parenthesis. The superscript $U(L)$ refers to the \t lying on the upper (lower) leg. }
\label{fig:eff_particles_2}
\end{figure}

\subsection{Isolated rung limit}\label{isolated_rungs}

Analogous to standard electronic $t$--$J$ ladders the physics of anyonic $t$--$J$ ladders can be understood starting from the strong rung coupling limit.\cite{spinladders,Greven96} We thus begin by identifying the low-lying states of  isolated rungs.

In Fig.~\ref{fig:eff_particles_2}, we show the five rung configurations on isolated rungs for a two-leg ladder, and the total $U(1)$ and anyonic charges that are possible for these rung configurations.
When there are two holes on a rung, both of these charges are trivially zero. When the rung is occupied by two \t particles, the net $U(1)$ charge is 2, however the topological charge may be either $\mathbf{1}$ or $\t$, giving rise to two different quantum states $|2,\tau\rangle$ (named ``heavy $\tau$") and $|2,\mathbf{1}\rangle$ (named ``heavy hole" -- an empty rung being a ``light hole").
In the case when there is a single \t on the rung, it can be either on the upper or on the lower leg with charges denoted as $(1^U,\tau)$ and $(1^L,\tau)$  respectively. The corresponding quantum states are respectively $|1^U,\tau\rangle$ and $|1^L,\tau\rangle$.
The bonding and anti-bonding states $|1^\pm,\tau\rangle$ (named ``light $\tau$") are formed by linear superpositions of the configurations with charges $(1^U,\t)$ and $(1^L,\t)$ given by 
\begin{equation} 
|1^\pm,\tau\rangle = \frac{1}{\sqrt2}\big(|1^U,\tau\rangle \pm |1^L,\tau\rangle\big) \, .
\end{equation}

\begin{figure}[t]
\includegraphics[width=\columnwidth]{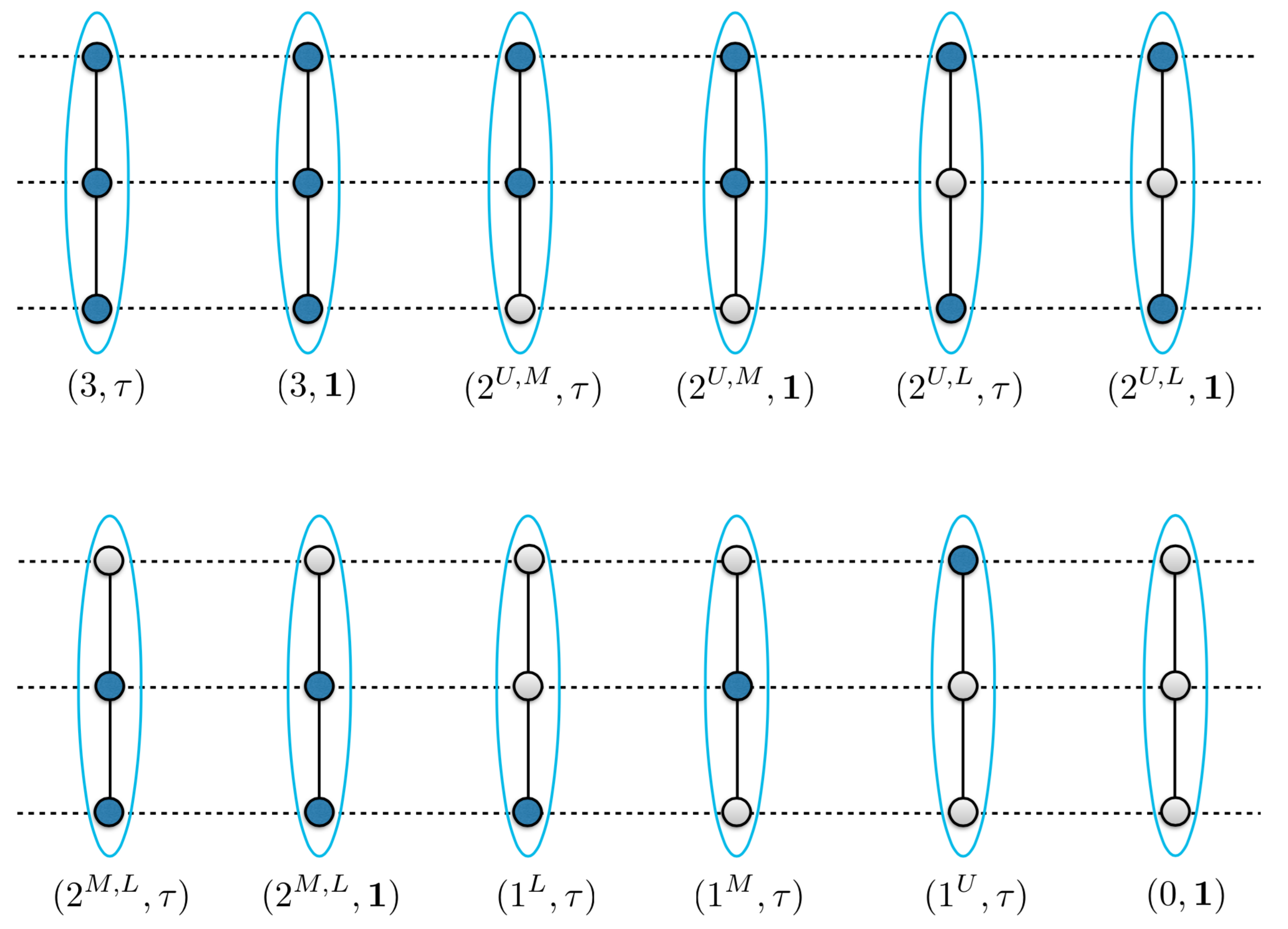}
\caption{The various fusion outcomes possible for a doped three-leg ladder. 
The blue circles represent $\tau$'s while white circles represent vacant sites. The first labels in the parenthesis signify the $U(1)$ charge, corresponding to the number of $\tau$'s present on each rung and the second refers to their fusion outcome. 
The superscripts $U,M,L$ refer to the positions of the \t on the different legs (upper, middle, lower) of the ladder.}
\label{fig:eff_particles_3}
\end{figure}

\begin{figure*}[t!]
\includegraphics[width=1.65 \columnwidth]{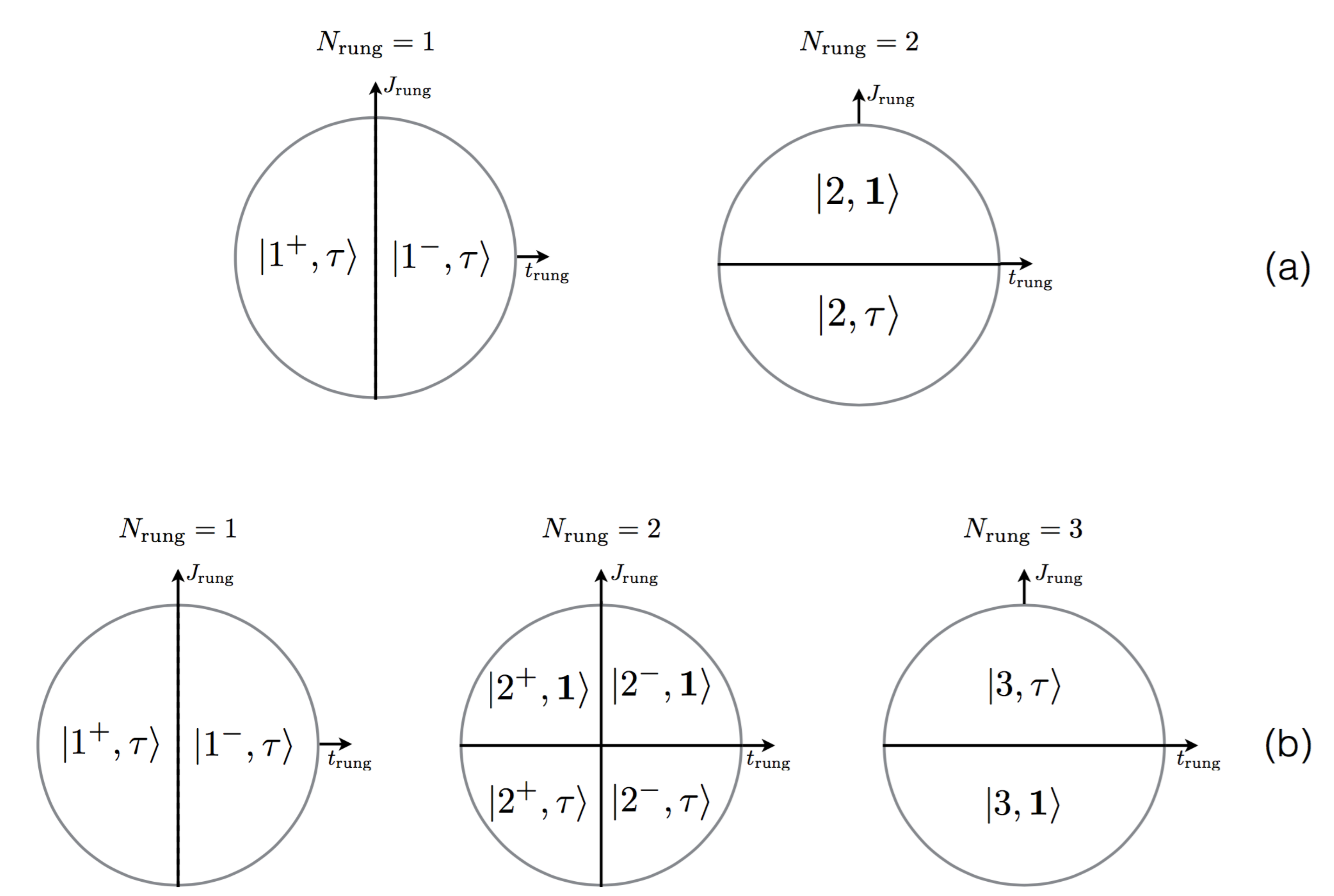}
\caption{Ground state phase diagram for (a) two-leg ladder and (b) three-leg ladder in the isolated rung limit (with or without charging energy). The lowest energy sectors for isolated rungs with different anyon numbers (labelled by $N_{\rm{rung}}$) on a ladders of itinerant Fibonacci anyons are indicated. The notations are the same as in Figs.~\ref{fig:eff_particles_2} and~\ref{fig:eff_particles_3}.   }
\label{fig:isolated_rung}
\end{figure*}

For a three-leg ladder, many more states are possible, as shown in Fig.~\ref{fig:eff_particles_3}. When there is a single \t on the rung $(N_{\rm{rung}}=1)$, the ``light $\tau$" quantum states formed by linear superpositions of the three different positions of the \t particle are given by 
\begin{eqnarray}
|1^\pm,\tau\rangle &=& \frac{1}{2}\big(|1^U,\tau\rangle + |1^L,\tau\rangle \pm \sqrt2 |1^M,\tau\rangle\big)\, ,\\
|1^0,\tau\rangle &=& \frac{1}{\sqrt2}\big(|1^U,\tau\rangle-|1^L,\tau\rangle \big)\, .
\end{eqnarray}
Depending on the sign of the hopping, one of the states $|1^\pm,\tau\rangle$ acquires the lowest energy.
Likewise, when there are two \t anyons on a rung $(N_{\rm{rung}}=2)$ we can form quantum states as linear superpositions of the states with the same $U(1)$ and topological charges.  
For $\jr >0$, one of the ``heavy hole" states 
\begin{eqnarray}
|2^\pm,\mathbf{1}\rangle = \frac{1}{\sqrt{2+\alpha^2}}\big(|2^{U,M},\mathbf{1}\rangle 
&+& |2^{M,L},\mathbf{1}\rangle \nonumber\\ 
&\pm& \alpha |2^{U,L},\mathbf{1}\rangle\big) 
\end{eqnarray} 
is the ground state, with 
\begin{equation}\label{eq:alpha}
\alpha =  \frac{\sqrt{\jr^2+8\tr^2}-\jr}{2\tr}.
\end{equation}
If for simplicity, we consider $\jr = t_{\rm rung}$, then $\alpha =1$. For $\jr <0$ either of the ``heavy $\tau$" states 
\begin{equation}
|2^\pm,\tau\rangle = \frac{1}{2}\big(|2^{U,M},\tau\rangle + |2^{M,L},\tau\rangle \pm \sqrt2 |2^{U,L},\tau\rangle\big)
\end{equation}
has lowest energy depending on the sign of the hopping $t_{\rm{rung}}$. When the rung is occupied by three \t particles, the total $U(1)$ charge on the rung is 3, and the possible quantum states are $|3,\tau\rangle$ (named ``super-heavy" $\tau$) and  $|3,\mathbf{1}\rangle$ (named ``super-heavy hole") depending on the net fusion outcome. 

In the limit of independent rungs and by parametrising the rung couplings as $t_{\rm{rung}}=\cos\theta$ and $J_{\rm{rung}}=\sin\theta$, we have mapped out the parameter space on a unit circle. 
In Fig.~\ref{fig:isolated_rung} we show the ground state phase diagram for an isolated rung on a two-leg or a three-leg ladder for different number $N_{\rm{rung}}$ of \t anyons on the rung. Note that these phase diagrams do not depend on the charging energy which only gives a constant energy shift $V_{\rm rep}(N_{\rm rung}e^*-Q_c)^2$ (depending on the $N_{\rm rung}$ sector). 

\subsection{Phase diagrams of weakly coupled rungs}\label{coupled_rungs}

\begin{figure*}[t!]
a)\includegraphics[width= \columnwidth]{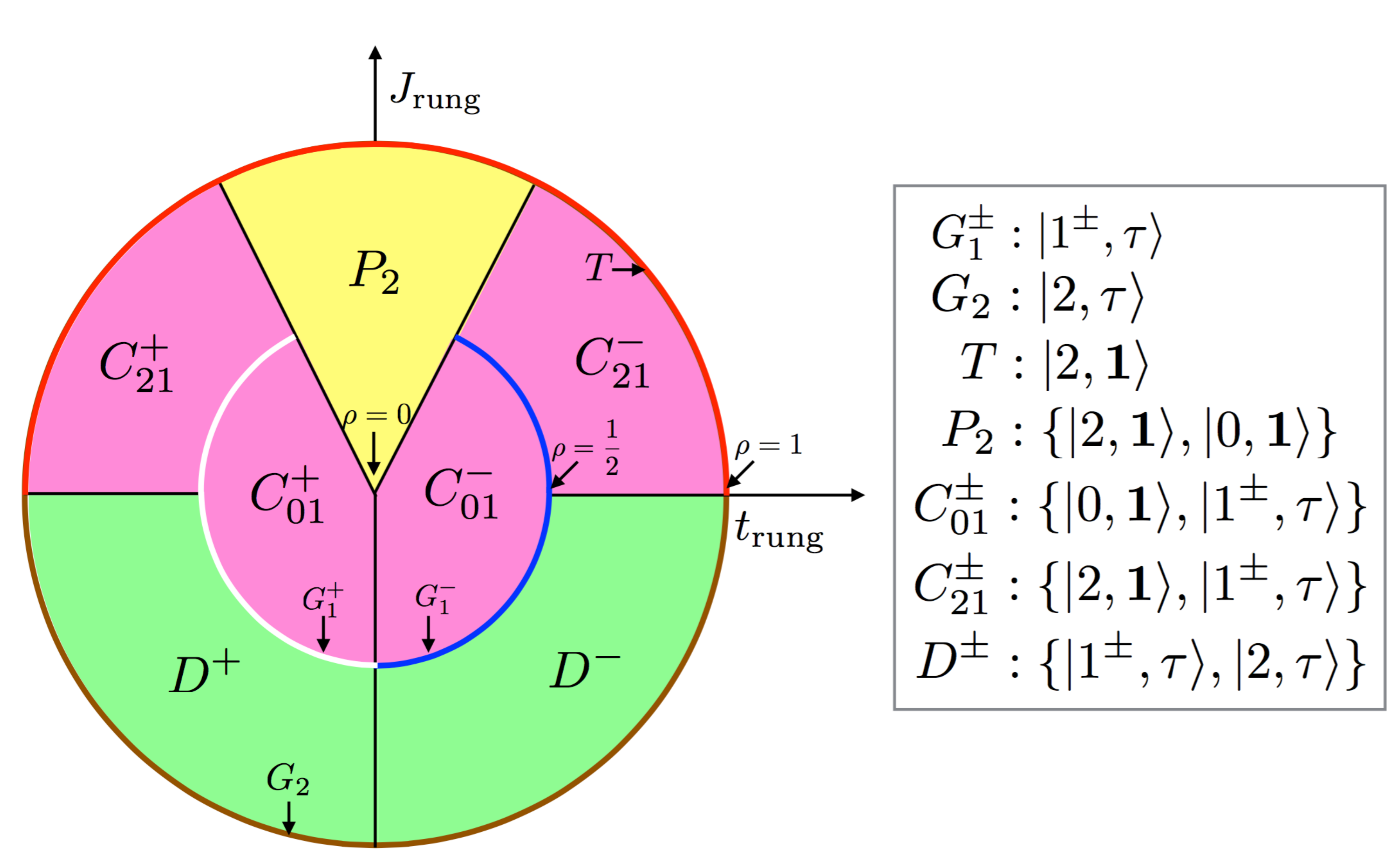}
b)\includegraphics[width= 0.62\columnwidth]{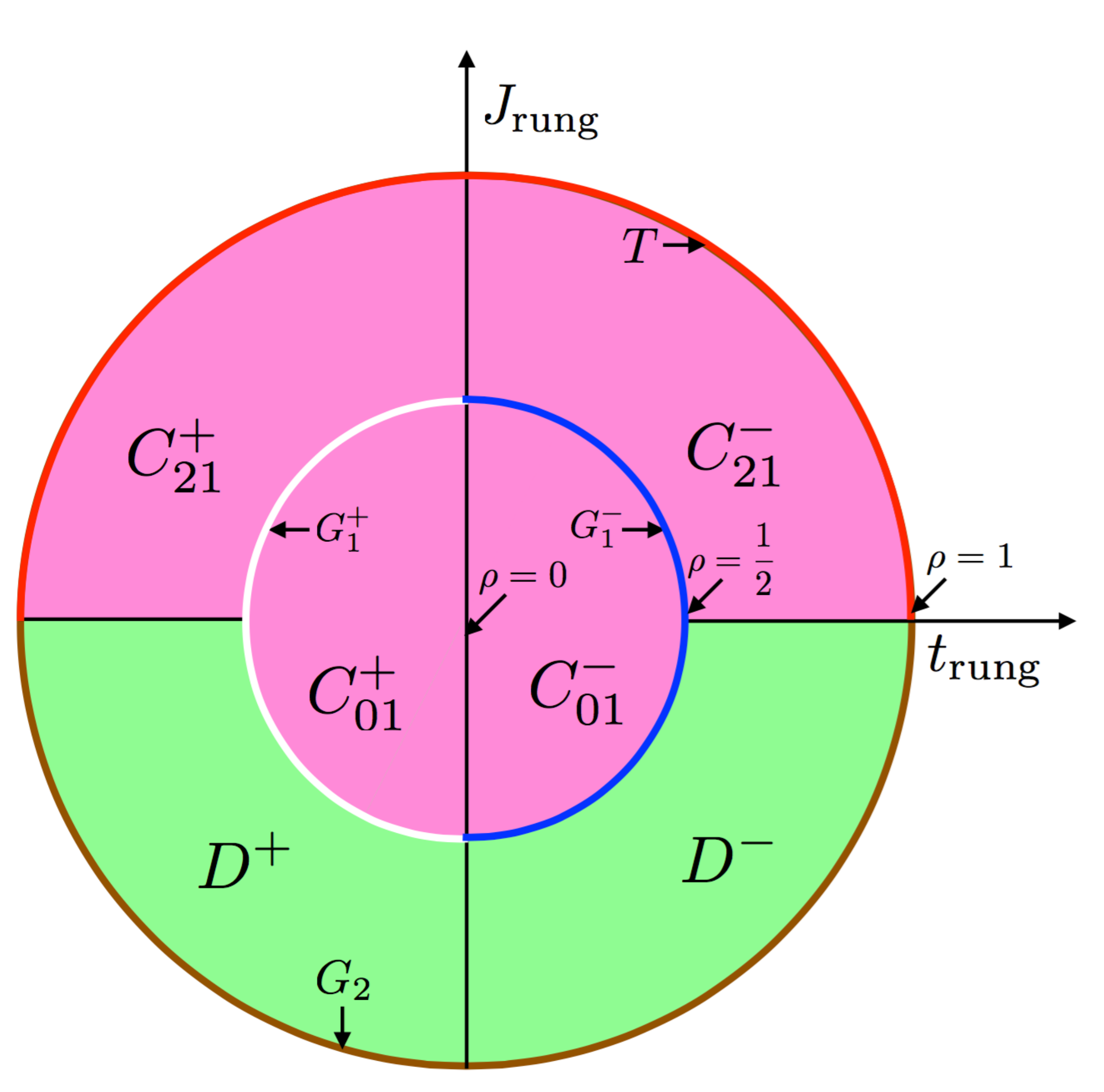}
\caption{Phase diagrams of the two-leg ladder in the strong rung coupling limit. a) is without a rung charging term and b) with a large rung charging term $V_{\rm rep}$. Here the radius denotes the density of anyons. Depending on filling and coupling several phases can be distinguished: a totally gaped phase (T), effective golden chain models (G), effective $t$--$J$ chains (C), paired phases (P), and a phase with two different types of $\t$ anyons (D). The legend indicates which rung states are relevant in the various phases. See the text for details.}
\label{fig:phase_diag2}
\end{figure*}

\begin{figure*}[t!]
a)\includegraphics[width= \columnwidth]{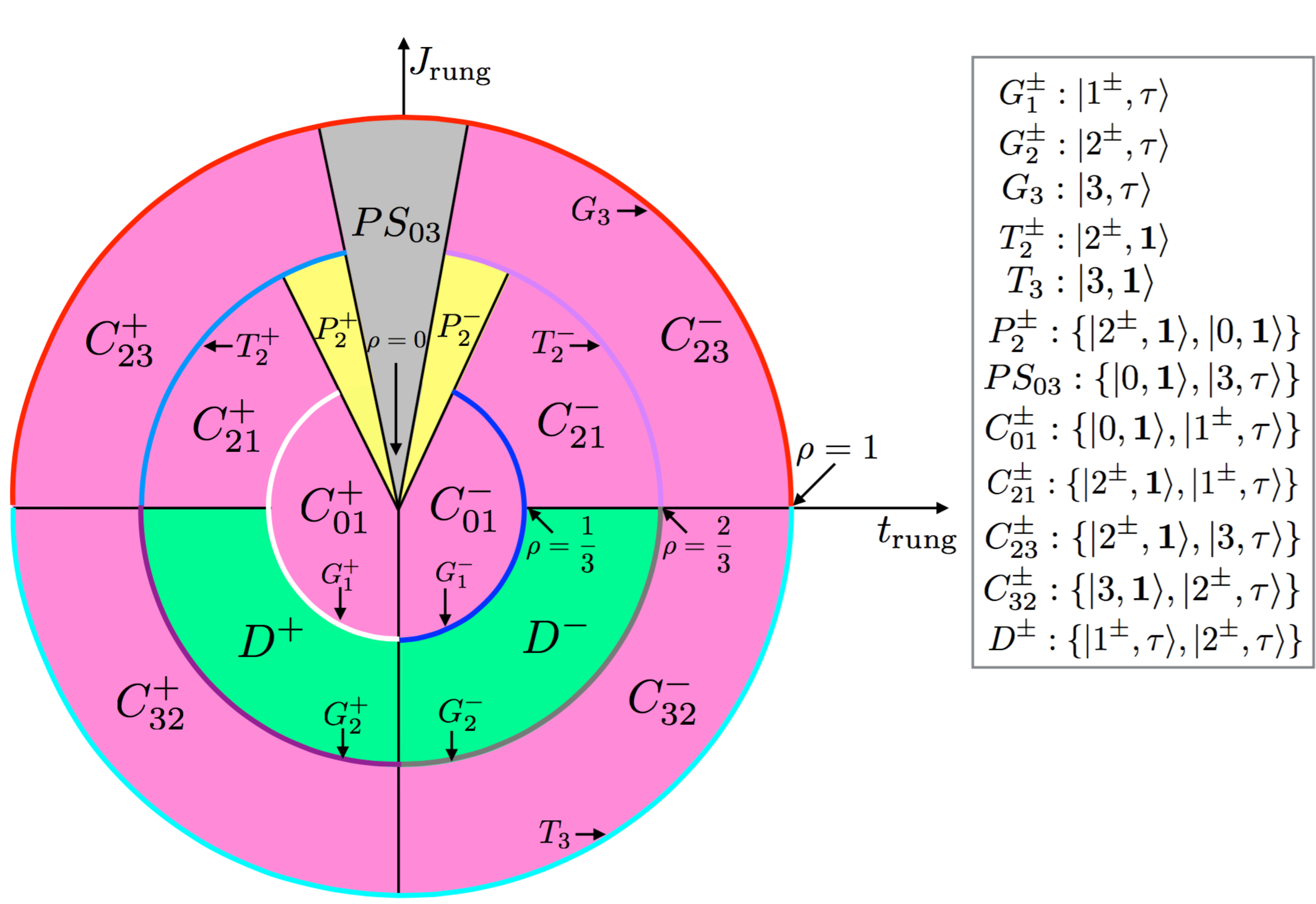}
b)\includegraphics[width= 0.72\columnwidth]{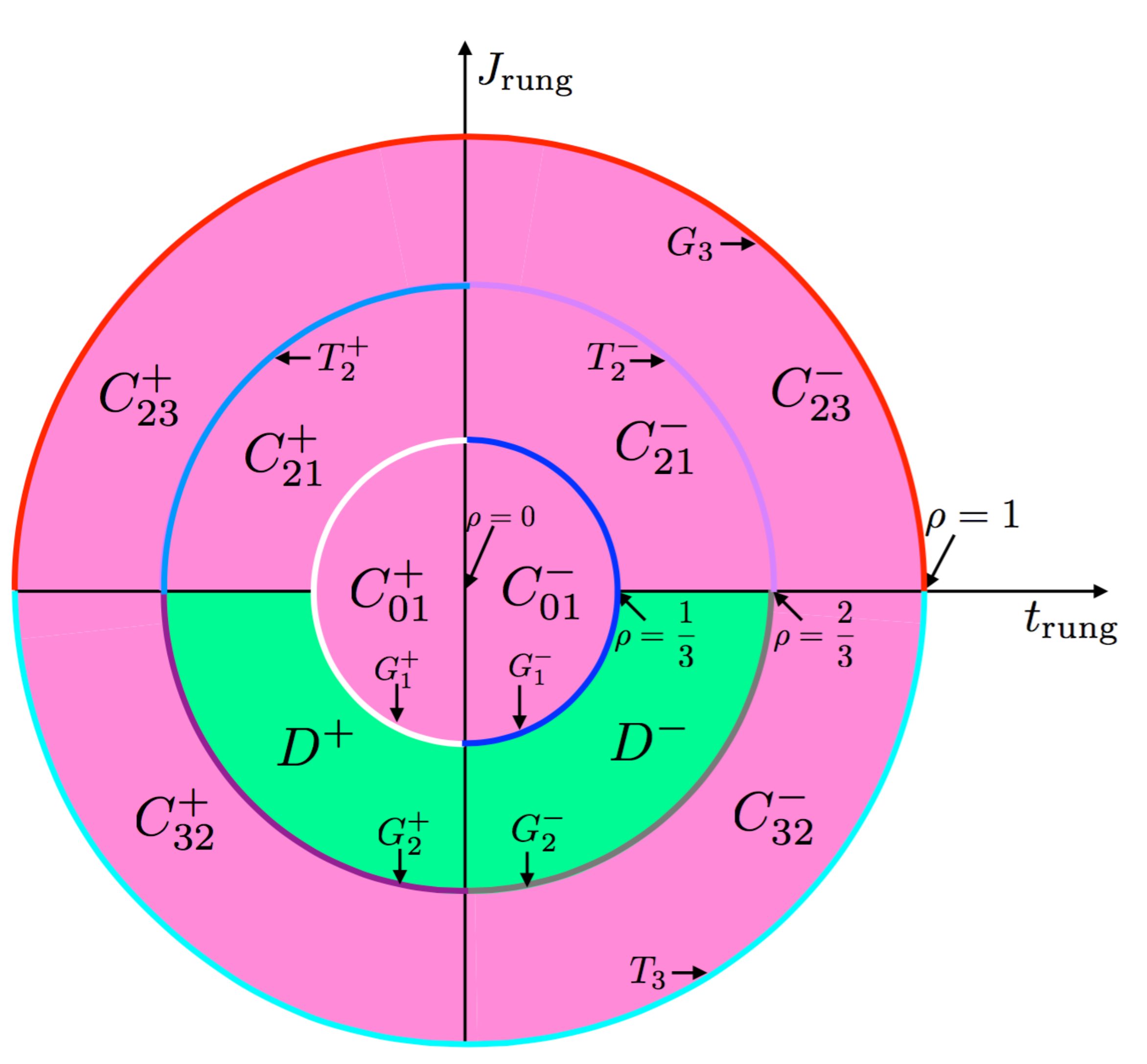}
\caption{Phase diagrams of the three-leg ladder in the strong rung coupling limit. a) is without a rung charging term and b) with a large rung charging term $V_{\rm rep}$. Here the radius denotes the density of anyons. Depending on filling and coupling several phases can be distinguished: a totally gaped phase (T), effective golden chain models (G), effective $t$--$J$ chains (C), paired phases (P), phase separated phases (PS) and a phase with two different types of $\t$ anyons (D). The legend indicates which rung states are relevant in the various phases.  See the text for details.}
\label{fig:phase_diag3}
\end{figure*}

Our goal is now to understand the phase diagram of anyonic ladders by starting from the isolated rung coupling limits. We do this by turning on 
small couplings \tl and \jl between the strongly coupled rungs such that $|\tl|, |\jl| << |\tr|,|\jr|$, in order to ensure that there is no transition to  excited state of the isolated rungs. Figures \ref{fig:phase_diag2} and \ref{fig:phase_diag3} summarize the phase diagrams for two and three-leg ladders.

Depending on the low-energy states on each rung we find six different types of phases:
\begin{itemize}
\item Totally gapped phases ($T$) appear when there are exactly two (for $\jr>0$) or three (for $\jr<0$) anyons per rung that fuse into the trivial channel. These phases will not be discussed further.
\item Effective golden chains ($G^\pm$) when there are exactly $n$ \t anyons on every rung that fuse into a total $\t$. An optional $\pm$ superscript indicates whether the particles are in a bonding (+) or antibonding($-$) state on a rung. These phases will be discussed in Sec. \ref{sec:effgolden}.
\item Paired phases ($P$) where two anyons on a rung fuse in the trivial channel, forming hard-core bosons. These phases will be discussed in Sec. \ref{sec:effpaired}
\item Effective $t$--$J$ chains ($C_{nm}^\pm$) consisting of an effective hole that arises from $n$ anyons on a rung fusing in the trivial channel  and  an effective \t anyon arising from $m$ anyons fusing in the \t channel. The $\pm$ superscript indicates whether the particles on a rung are in a  bonding or antibonding state. These phases will be discussed in Sec. \ref{sec:efftj}
\item Effective models consisting of two flavors of \t anyons ($D_{mn}^\pm$) that are formed by fusing $m$ and $n$ anyons on a rung respectively. Again the $\pm$ superscript indicates whether the particles on a rung are in a  bonding or antibonding state. 
\item A phase separated region $PS_{03}$ originated from an effective $t$--$J$ chain with dominant attraction between the effective super-heavy \t anyons.
\end{itemize}

The effect of a large rung charging energy $\vr$ is to suppress pairing and phase separation in two-leg and three-leg ladders. The other phases are unchanged when adding this term. We will thus use $\vr=\infty$ to reduce the Hilbert space dimension when numerically investigating the latter phases.

\section{Pairing and effective hard-core boson models}
\label{sec:effpaired}
We start our detailed discussion with paired phases that arise when two \t anyons on a rung fuse into an effective trivial particle and are then described by mobile hard core bosons. In the case of the two-leg ladder this phase appears when $\jr > 2|t_{\rm rung}|$. 
Identifying $|0,{\bf 1}\rangle$ with an empty site and  $|2,{\bf 1}\rangle$ with a hard-core boson we end up with an effective hard-core boson (HCB) model, 
\begin{equation}\label{eq:hcb_mat}
 H_{\rm HCB}=t \sum_i (b_i^\dagger b_{i+1} + h.c.) + V \sum_i n_i (1-n_{i+1}) , 
\end{equation}
where $b_i^\dagger$ creates a boson at site $i$ and $n_i=b_i^\dagger b_i$ is the boson density.
The effective hopping matrix element of this hard-core boson model is
obtained from second-order perturbation theory in $\tl$ to be
\begin{equation}
t=-\frac{2\tl^2}{\jr - 2|t_{\rm rung}|}.
\end{equation}
An effective nearest neighbor attraction between different types of holes comes also in second-order and is given by
\begin{equation}
V = -\frac{2\tl^2}{\jr - 2|t_{\rm rung}|}.
\end{equation}
This is similar to fermionic $t$--$J$ ladders mapping to a Luther-Emery liquid of Cooper pairs.

In the three leg ladder, similar paired phases described by the same hard-core boson model are found when $\rho<2/3$ and the rung couplings satisfy
\begin{equation}
3\phi |t_{\rm rung}|>\jr>\frac{3}{\sqrt2}|t_{\rm rung}|\, .
\end{equation}
The effective hopping matrix element of this hard-core boson model is
\begin{equation}
t=-\frac{2\tl^2}{E_D},
\end{equation}
where $E_D = - 2\sqrt2 |t_{\rm rung}|+\frac{\jr}{2}+\frac{\sqrt{\jr^2+8\tr^2}}{2}$ and a proof is provided in Eq. (\ref{eq:t2order}).
An additional effective attraction between different types of holes is given by
\begin{equation}
V = -\frac{2\tl^2}{E_D},
\end{equation}
as shown in Eq. (\ref{eq:v2order}).

\section{Effective golden chains}
\label{sec:effgolden}

If all rungs are at the same integer filling $n$, the effective model is either in a trivial gapped phase if the $n$ anyons fuse into the trivial channel, or an effective golden chain model if they form a total $\t$. We label the latter phases as $G_n$ or $G_n^\pm$, where the optional $\pm$ index indicates whether the anyons are in a bonding state ($+$) or antibonding state ($-$) on the rung.

 The phase $G_1^\pm$ appears in the two-leg ladder at unit filling (outside of the paired phase). 
With two \t particles per rung and $\jr<0$ we obtain the phase $G_2$.

Analogously, the three leg ladder has effective golden chain phases for specific densities $\rho=1/3,2/3,1$ on the ladder. In the case when all the rungs have exactly one \t particle, the phase $G_1^{\pm}$ is obtained.
When there are two \t anyons per rung and $\jr<0$, the phase $G_2^{\pm}$ appears.
Finally, for three \t particles per rung and $\jr>0$, we obtain the $G_3$ phase. 

All coupling constants $J$ of the effective Golden chains are listed in Table \ref{tab:goldenchain}. 

\begin{table}[t]
\caption{Effective couplings for the various golden chain phases as a function of the ladder width $W$ and density $\rho$.} 
\begin{tabular}{|c|c|c|c|c|}
\hline\hline
$W$ & $\rho$ & Phase & $J/\jl$ & Derivation Eq. \\ [1ex]
\hline
2 & $\frac{1}{2}$ & $G_1^{\pm}$ & $\frac{1}{2}$ & (\ref{eq:g1})\\[1ex]
2 & 1 & $G_2$ & $\frac{2}{\phi^2}$ & (\ref{eq:g2})\\[1ex] 
\hline
3 & $\frac{1}{3}$ & $G_1^{\pm}$ & $\frac{3}{8}$ & (\ref{eq:g3})\\[1ex]
3 & $\frac{2}{3}$ & $G_2^{\pm}$ & $\frac{11}{8\phi^2}$ & (\ref{eq:g4})\\[1ex]
3 & 1 & $G_3$ & 1& (\ref{eq:g5b})\\[1ex]
\hline
\end{tabular}
\label{tab:goldenchain}
\end{table}

\section{Effective $t$--$J$ models}
\label{sec:efftj}

\subsection{Two-leg ladder}\label{two_leg}

\subsubsection{Effective t--J models}

Doping the $\rho=1/2$ $G_1^\pm$ golden chain with (light) holes one obtains an effective anyonic $t$--$J$ chain ($C_{01}^\pm$).
The magnetic coupling $J$ is the same as for the $G_1^\pm$ golden chain.

Increasing the U(1) charge density ($\rho>1/2$), by effectively doping the $G_1^\pm$ golden chain with heavy holes, one obtains a similar $C_{21}^\pm$ $t-J$ chain for AFM rung couplings $\jr>0$ with a sign change of the hopping term.

Coupling constants are summarized in Table \ref{tab:t-J_2leg}.

\begin{table}[b]
\caption{Effective couplings for the various $t$--$J$ phases of the two-leg ladder as a function of the density $\rho$. } 
\begin{tabular}{|c|c|c|c|c|}
\hline\hline
Filling & Phase & ${J}/{\jl}$ & ${t}/{\tl}$ & Derivation Eq. \\ [1ex]
\hline
$\rho<\frac{1}{2}$ & $C_{01}^{\pm}$ & $\frac{1}{2}$ & $\frac{1}{2}$ & (\ref{eq:t1})\\[1ex]
$\rho>\frac{1}{2}$ & $C_{21}^\pm$ & $\frac{1}{2}$ & $-\frac{1}{2}$ & (\ref{eq:t2})\\[1ex] 
\hline
\end{tabular}
\label{tab:t-J_2leg}
\end{table}

\subsubsection{Effective model for charge degrees of freedom}\label{charge}

From the mapping of a doped ladder to an effective 1D $t$--$J$ chain we expect its spectrum to fractionalize into charge and anyon (called also ``spin'') degrees of freedom. To investigate spin-charge separation in the ladder, we first examine the pure $U(1)$ charge spectrum when $\jl=0$.
 
In the $J=0$ limit  of an anyoninc $t$--$J$ chain the itinerant anyons  behave like HCBs which can be mapped onto a system of spinless fermions. 
Adding an external flux 
in the ring, the HCB spectrum is therefore given by charge excitation parabolas,  
\begin{equation}
E_{\rm{HCB}} (p,\phi_{\rm ext}) = -2t \sum_{j(p)} \cos \bigg[\frac{2\pi}{L}\Big(j+\frac{1}{2}\Big) + \frac{\phi_{\rm{ext}}}{L} \bigg]  ,
\label{eq:hcb1}
\end{equation}
where $\{j(p)\}$ is a set of integers (labelled by the branch index $p$) which determine the continuous momenta, given by 
$K = \frac{2\pi}{L}\sum_{j(p)}(j+\frac{1}{2})  + \tilde\rho\phi_{\rm{ext}}$, $\tilde\rho$ being the density of particles in the system.
In the $J=0$ limit we must be careful since the fusion tree labels make the anyons distinguishable. Thus, in the absence of magnetic interactions the energy levels show a high degree of degeneracy that arises due to the built in non-Abelian nature of the Fibonacci anyons.  Moving an anyon across the boundary cyclically translates the fusion tree labels. All $N$ particles must be translated over the boundary to be able to have the original labeling. This brings about a phase shift of $\phi_n=2 \pi n/N$, $n$ being an integer. The charge spectrum of the anyonic chain can then be described as a union of all HCB spectra for all discrete values of $\phi_n$, with no external flux: 
\begin{equation}
E^{p,n}_{\rm{charge}}=E_{\rm{HCB}}(p,\phi_n)  .
\label{eq:hcb2}
\end{equation}
The states are labelled by their  total momentum $K_{p,n} = K_p + 2\pi \frac{n}{N}$.

\begin{figure}[b]
\includegraphics[width= \columnwidth]{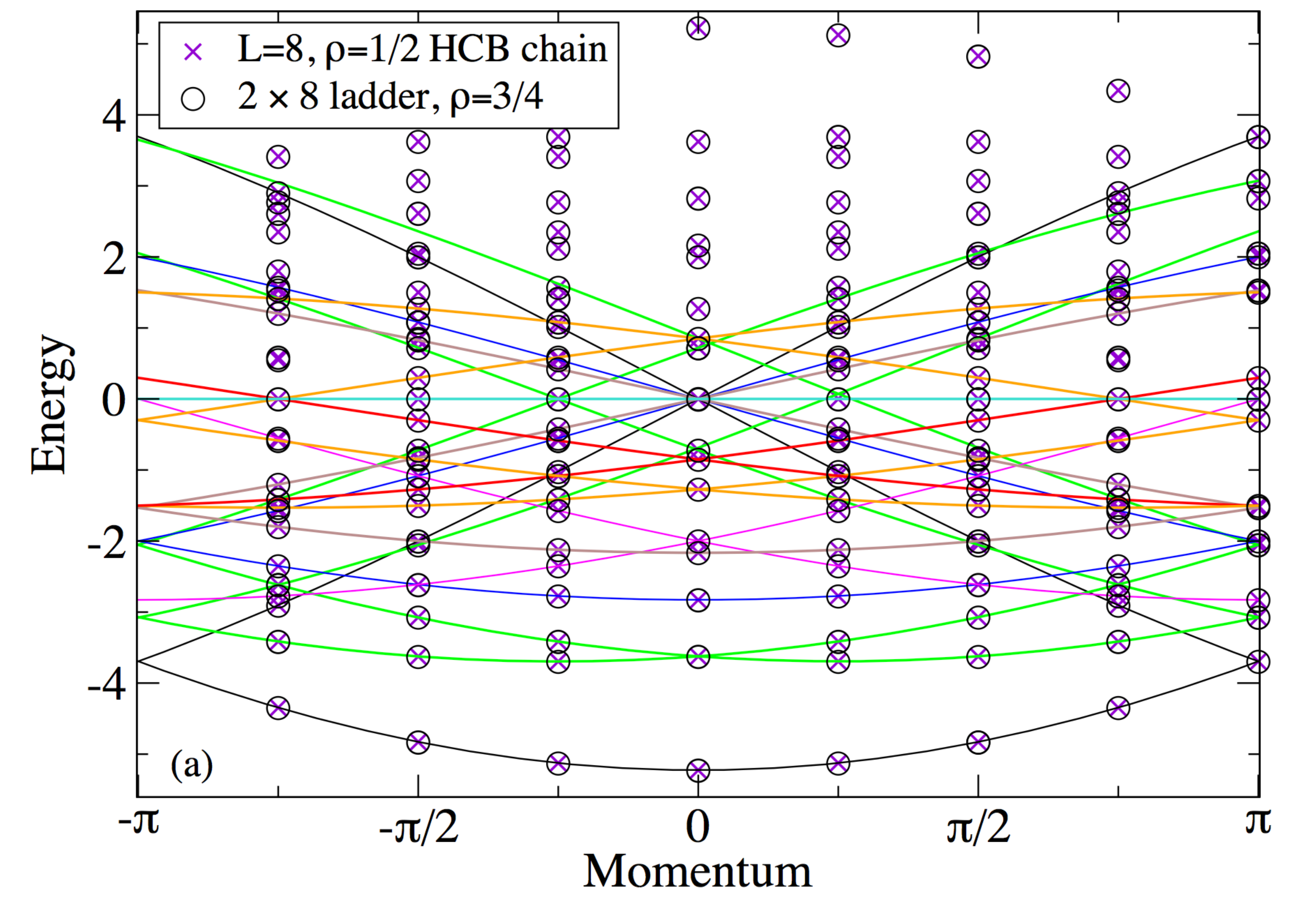}
\vskip 0.5truecm
\caption{Charge spectrum at $J_{\rm{rung}}=t_{\rm{rung}}=1000, t_{\rm{leg}} =1, J_{\rm{leg}}=0, \vr=\infty$. The solid lines denote the HCB spectrum (with an external flux) given by Eq.~(\ref{eq:hcb1}), different colors corresponding to the different charge branches
(labelled by $p$ in Eq.~(\ref{eq:hcb1}). The black circles denote the spectrum of a $2\times8$ ladder with $\rho=3/4$. 
The blue crosses correspond to the effective chain spectrum for $L=8$, $\tilde\rho=1/2$ (see  Eq.~(\ref{eq:hcb2})).}
\label{fig:charge28}
\end{figure}

\begin{figure}[t]
\includegraphics[width= \columnwidth]{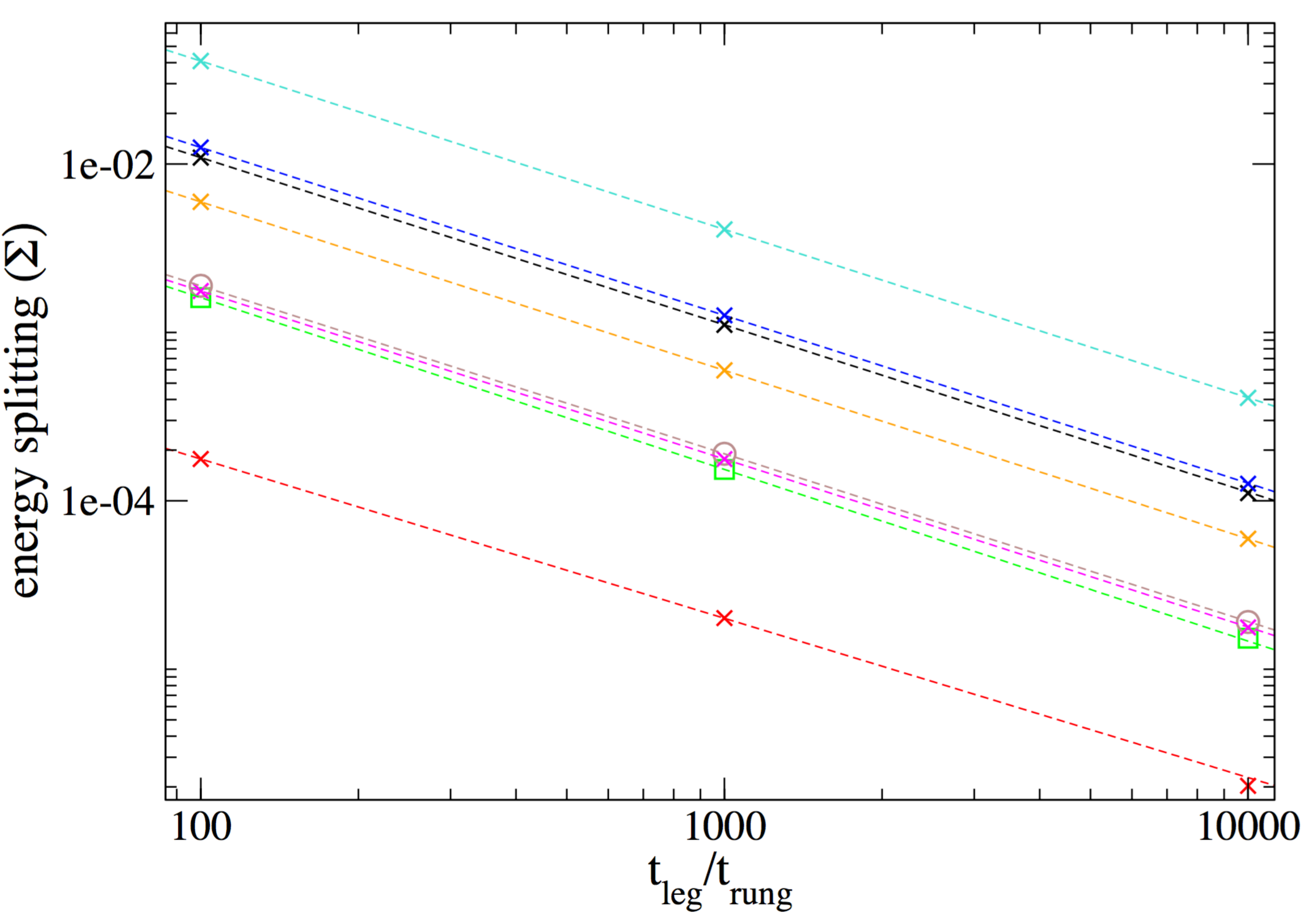}
\caption{
Energy splittings of each of the eight $E\le 0$ energy levels 
at $K=0$ of the $2\times 8$ ladder with $\rho=3/4$ (symbol/line colors here match the colors of the parabolic branches of
 Fig.~\protect\ref{fig:charge28}) as a function of $t_{\rm{leg}}/t_{\rm{rung}}$. 
The values of the leg couplings are $t_{\rm{leg}} =1, J_{\rm{leg}}=0, \vr=\infty$ and $t_{\rm{rung}}(= J_{\rm{rung}})$ takes different values from 100 to 10000. The fits correspond to 
the expected $t_{\rm{rung}}^{-1}$ behavior. }
\label{fig:split_2leg}
\end{figure}

\begin{figure*}[t]
\includegraphics[width=\columnwidth]{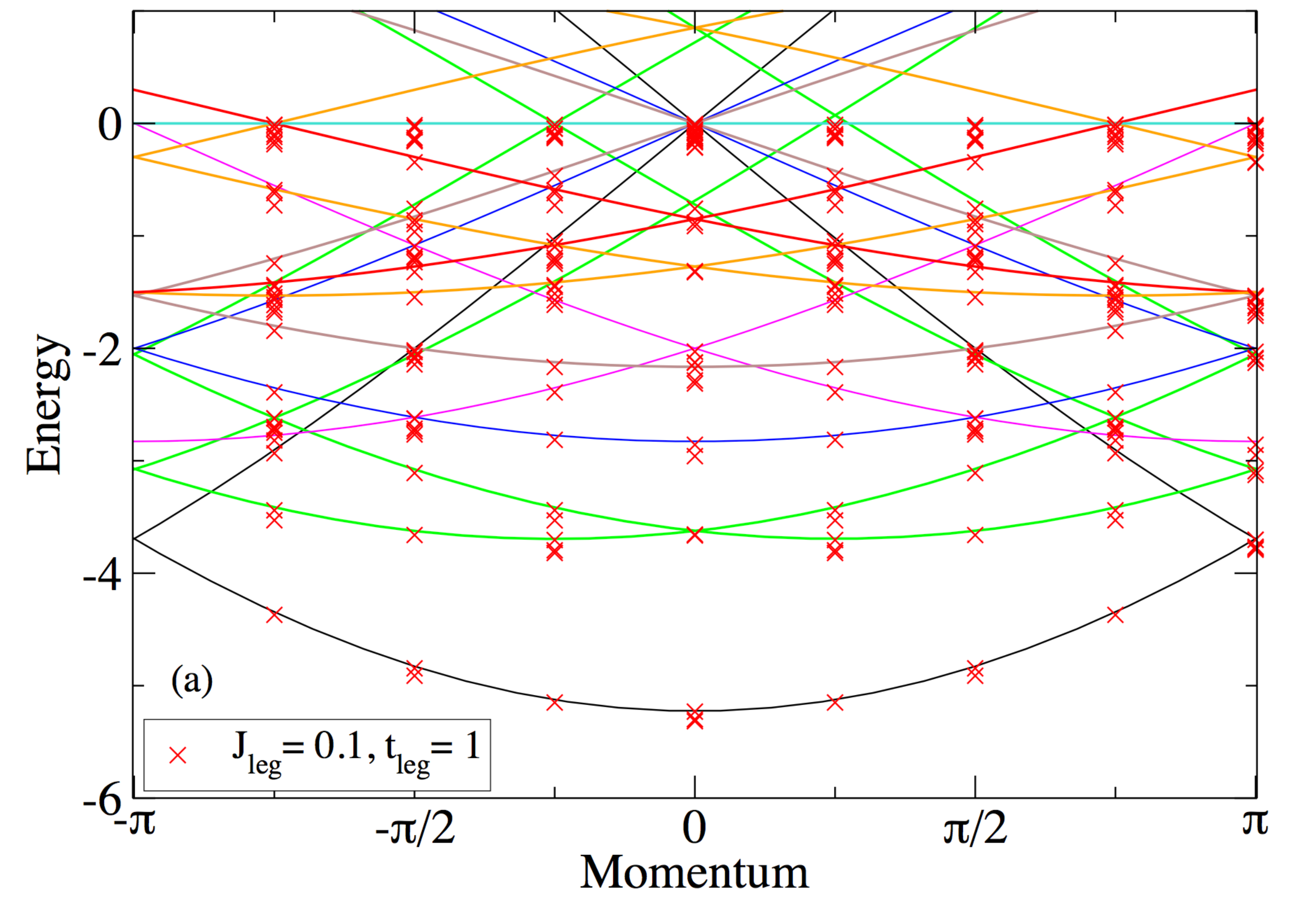}
\includegraphics[width=\columnwidth]{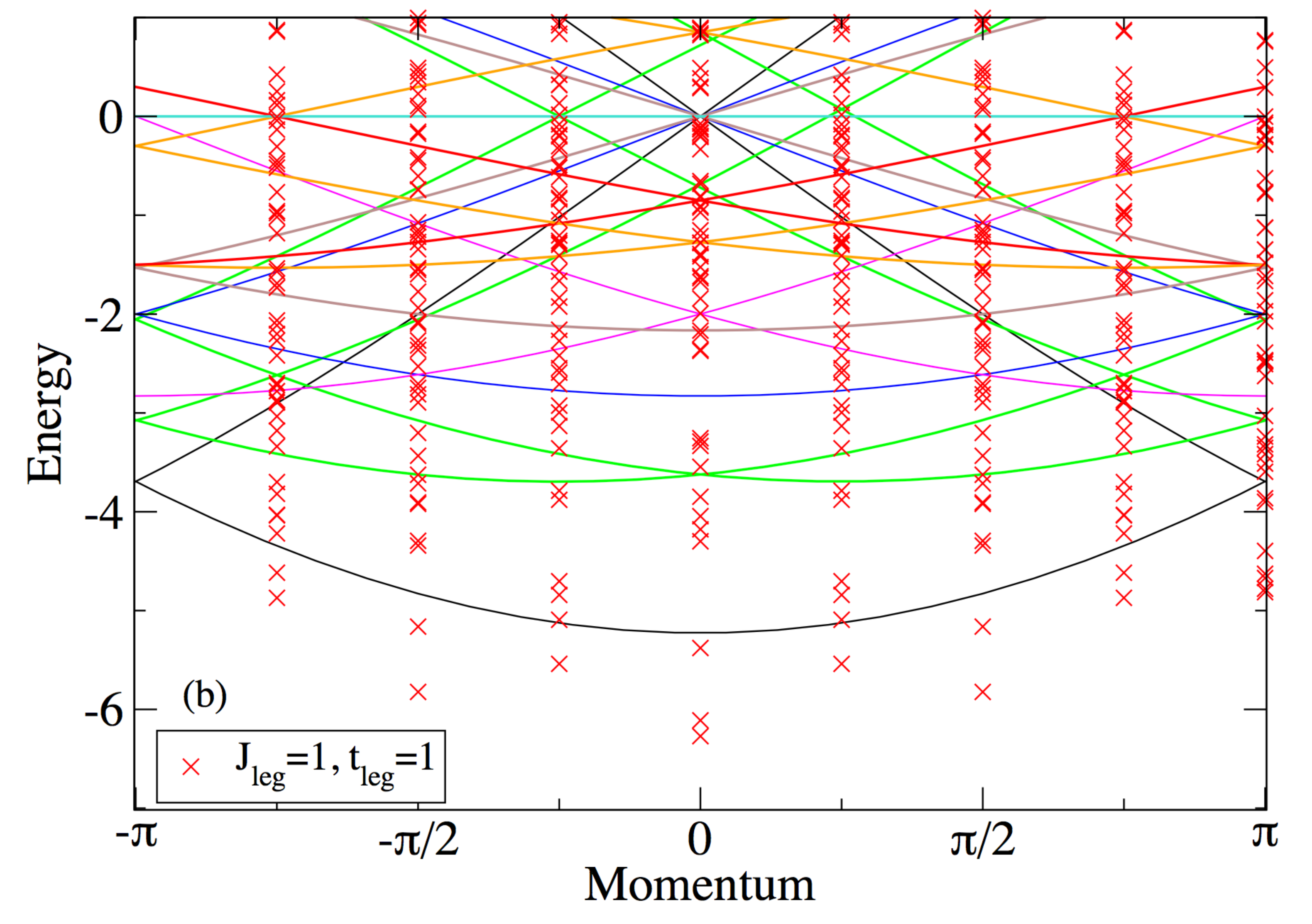}
\caption{Splitting of the degenerate levels on switching on different values of $J_{\rm{leg}}$ on a $2\times 8$ ladder with $\rho=3/4$ and $J_{\rm{rung}}=t_{\rm{rung}}=1000, t_{\rm{leg}} =1, \vr=\infty$. The parabolas show the continuous HCB spectrum relevant for $J_{\rm{leg}}=0$ (see Fig.~\ref{fig:charge28}) and the red crosses represent the ladder spectrum at (a) $J_{\rm{leg}}=0.1$ (b) $J_{\rm{leg}}=1$. 
}
\label{fig:splitting_smallJ}
\end{figure*}

Our numerical results show that, as expected, the charge spectrum of the ladder 
corresponds exactly to that of the effective chain.
As an example, Fig.~\ref{fig:charge28} shows the $\jl=0$ spectrum of a $2 \times 8$ ladder with $\rho=3/4$, which is in perfect agreement with the 
spectrum of an effective $J=0$ chain with $L=8$ and $\tilde\rho=1/2$ and with the HCB spectrum given by the cosine branches according to Eqs.~(\ref{eq:hcb1}) and (\ref{eq:hcb2}).  
Note that there is a global shift in energy between the ladder and the chain spectra given by \begin{equation} E_{\rm{shift}}=-\tr N_s -\jr N_d  , \end{equation} where $N_s$ ($N_d$) are the number of rungs carrying a single (two) \t(s).

Lastly, we would like to mention that the mapping is exact only in the limit when the rung couplings tend to infinity. For large, yet finite, rung couplings the energy levels of the ladder model in each parabola are split into an exponential number of energy levels over a finite energy range $\Sigma$. This is due to second order processes to higher energy states that give rise to
a broadening of order $\Sigma \sim t_{\rm{leg}}^2/t_{\rm{rung}}$ and $t_{\rm{leg}}^2/J_{\rm{rung}}$ of the energy levels, as shown in
Fig.~\ref{fig:split_2leg}.

\subsubsection{Numerical comparison between microscopic model and 1D t-J model}\label{anyon}

\begin{figure}[t]
\includegraphics[width=\columnwidth]{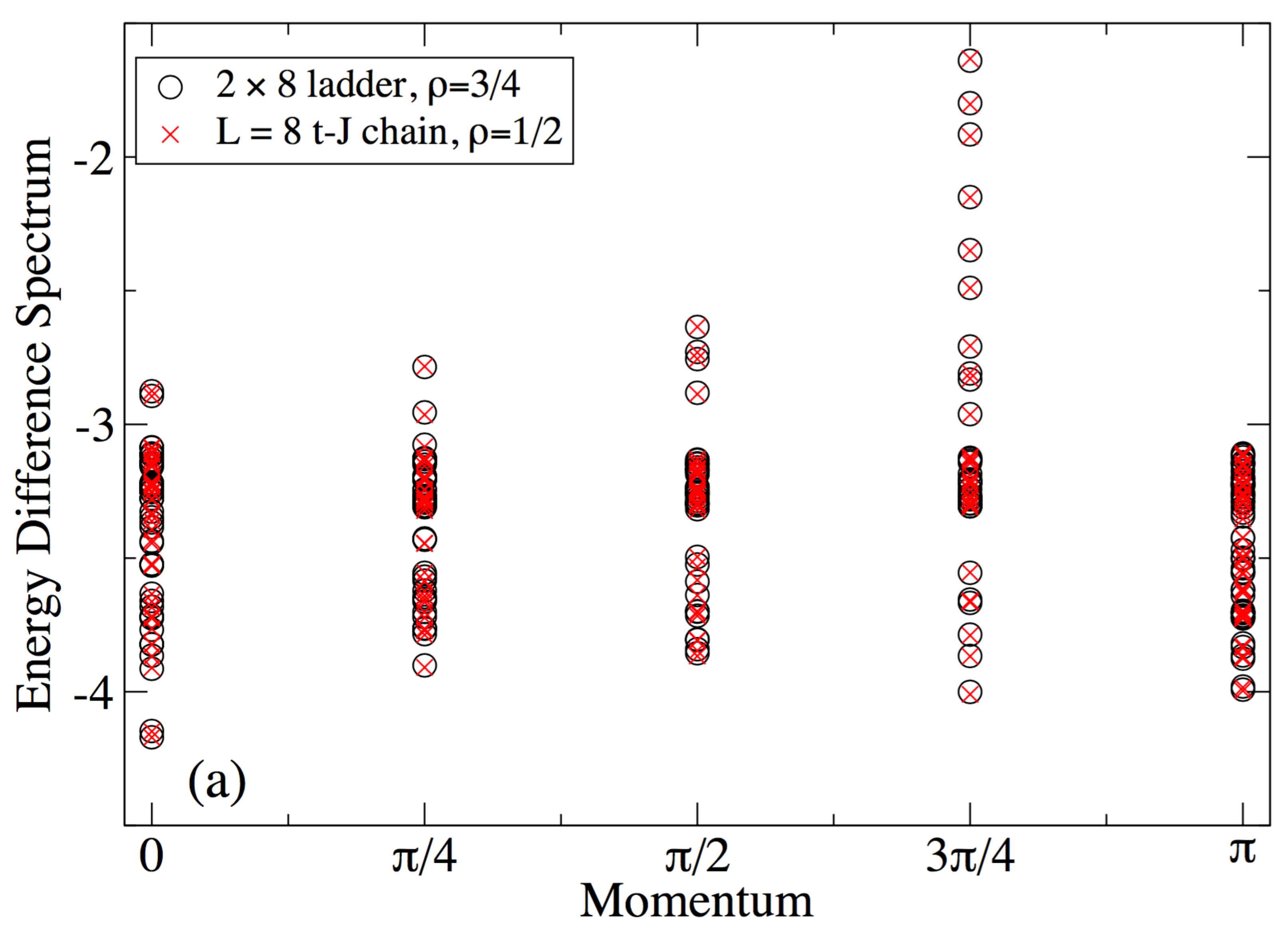}
\vskip 0.2truecm
\includegraphics[width=\columnwidth]{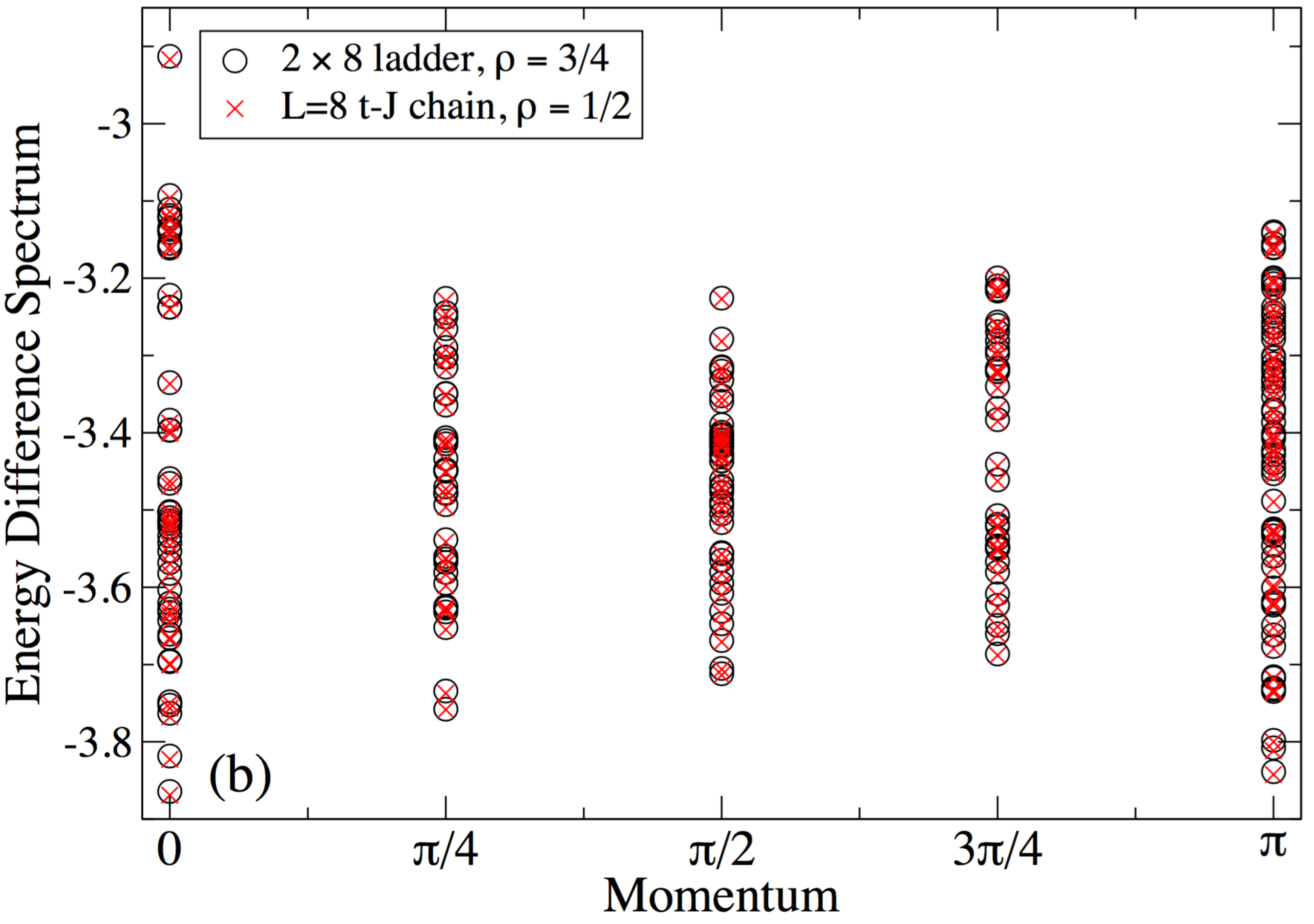}
\vskip 0.2truecm
\caption{
Comparison of the energy difference spectra, after subtracting the charge contribution to the energy of each state, of a $2\times 8$ ladder, $\rho=3/4$ with that of the effective $t$--$J$ chain $L=8, \tilde\rho=1/2$. The couplings on the ladder are $J_{\rm{rung}}=t_{\rm{rung}}=1000, t_{\rm{leg}} =1, \vr=\infty$ and (a) $J_{\rm{leg}}=0.1$ and (b) $J_{\rm{leg}}=1$.  }
\label{fig:anyon_smallJ}
\end{figure}

\begin{figure}[tb]
\includegraphics[width= \columnwidth]{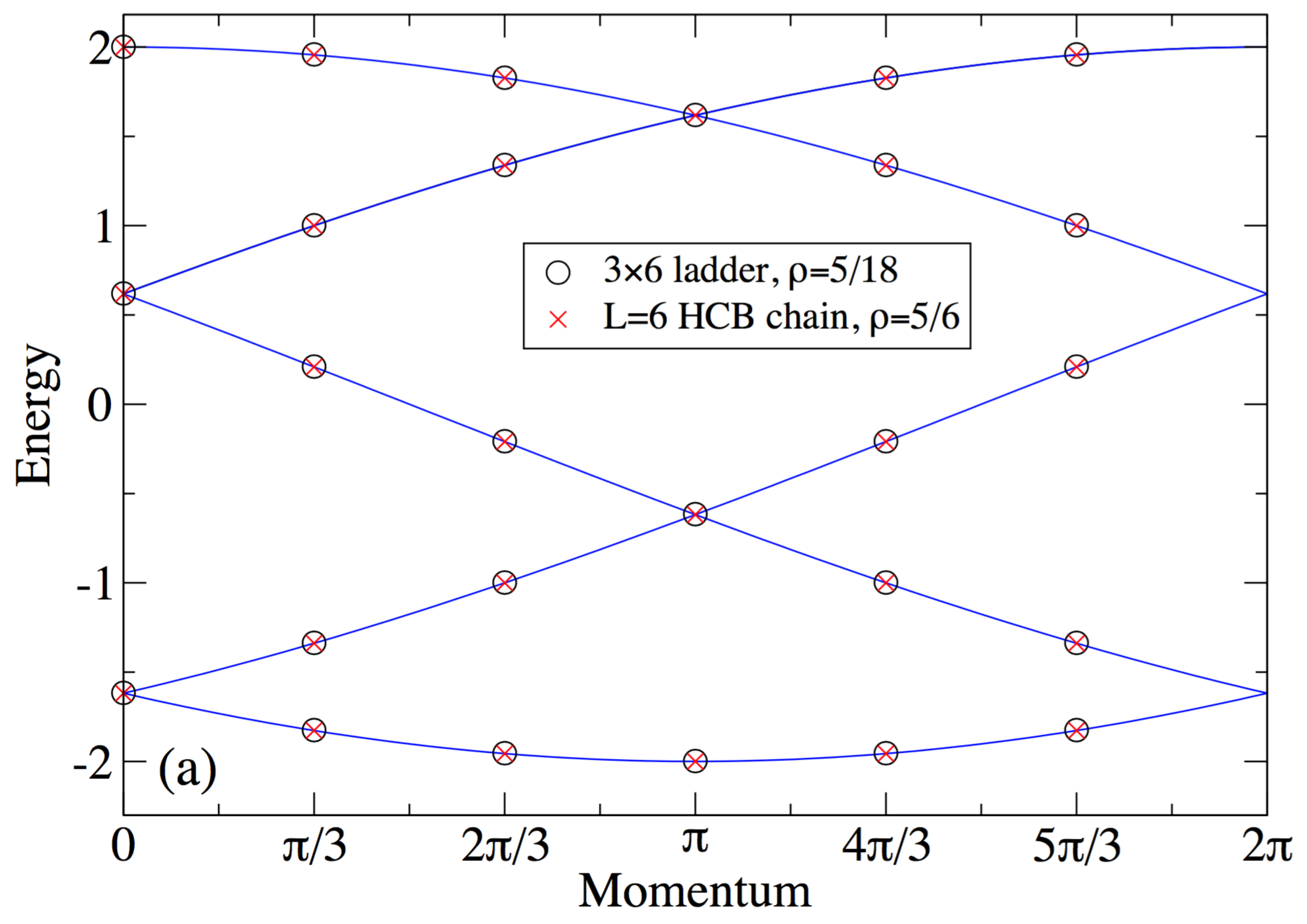}
\vskip 0.3truecm
\includegraphics[width= \columnwidth]{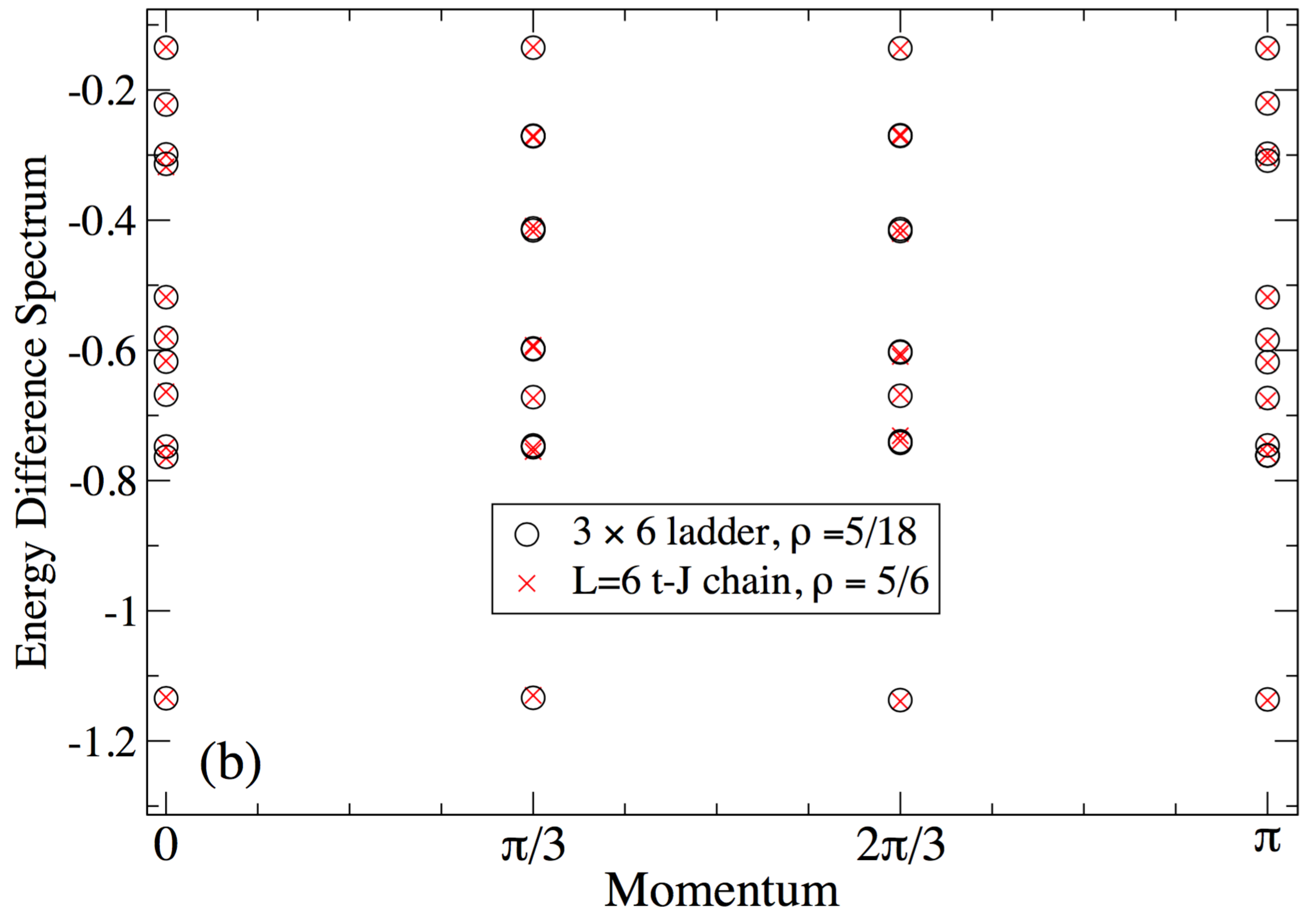}
\caption{
Three-leg ladder: The black circles denote the spectrum of a $3\times6$ ladder with $\rho=5/18$ while the red crosses are for the effective chain with $L=6$ and $\tilde\rho=5/6$. The couplings on the ladder are $J_{\rm{rung}}=t_{\rm{rung}}=1000, \tl=1, \vr=\infty$. (a) Charge spectrum at $J_{\rm{leg}}=0$. The solid lines denote the HCB spectrum. (b) energy difference spectrum for \jl = 0.1.}
\label{fig:charge36}
\end{figure}

We next turn on a small $J_{\rm{leg}}$ and adiabatically follow the splitting of the charge parabolas. Fig.~\ref{fig:splitting_smallJ}(a) zooms into the low energy spectrum to show how the highly degenerate energy levels are split by a small $J_{\rm{leg}}=0.1$. Fig.~\ref{fig:splitting_smallJ}(b) shows results for a larger coupling $J_{\rm{leg}}=1$.
We see that magnetic interactions lift the degeneracy of the states with an energy spread proportional to $LJ_{\rm leg}$. 
This is consistent with the behavior of the effective $t$--$J$ chain exhibiting ``spin-charge separation":
in Refs.~\onlinecite{frac,itinerant_long} we showed that the full excitation spectrum of an itinerant anyon chain is made up of two independent contributions originating from the charge degrees of freedom (described in section~\ref{charge}) and the anyon degrees of freedom which are given by a \textit{squeezed} (undoped) anyon chain of length $L_{\rm{a}} = \tilde\rho L$ where $\tilde\rho$ is the anyon density on a $L$ site $t$--$J$ chain of anyons.

We now perform a quantitative comparison of the spectra of the anyonic ladder and  its corresponding 
effective anyonic chain. 
As the charge spectra match, we focus on the energy difference spectrum (EDS) obtained by subtracting the (supposed) charge excitation component to each state. By construction, the EDS then carries  the information about the anyon degrees of freedom. 
Note that this procedure is only possible at low energy and for small enough $J_{\rm leg}$, {\it i.e.} when a well defined parabolic charge branch can be assigned unambiguously to the levels we consider. 
However, even when a large splitting of the energy levels is seen as
in Fig.~\ref{fig:splitting_smallJ}(b), we have been able to identify exactly the charge excitations corresponding to the various levels in the low energy spectrum up to an excitation energy of order $5t_{\rm leg}$ and hence obtain the corresponding EDS. The numerical results for the EDS on a $2\times 8$  ladder with anyon density $\rho=3/4$ for small \jl $= 0.1 t_{\rm{leg}}$ and intermediate \jl = \tl are shown in  Figs.~\ref{fig:anyon_smallJ}(a) and (b) respectively. We find the EDS of the ladder and of the effective chain to be in perfect agreement. The perfect mapping of the two-leg ladder physics to the physics of the chain hence implies straightforwardly that 
the concept of spin-charge fractionalization is not strictly 1D but also applies to the two-leg anyonic ladder,
in contrast to the electronic ladder analog.

Note that the EDS subtracted spectrum must not be confused with the actual energy spectrum of the corresponding squeezed golden chain. In our prescription, subtracting the charge excitations from the full spectrum, we get the spectrum corresponding to the anyon degrees of freedom as a function of the total momentum of the ladder/$t$--$J$ chain, rather than that of the squeezed golden chain. Thus the spectra shown in Figs.~\ref{fig:anyon_smallJ}(a),(b) are qualitatively different from the golden chain spectra.

\subsection{Three-leg ladder}\label{three_leg}

All results about the $t$--$J$ phases are summarized in Table \ref{tab:t-J_3leg} and we provide a short description below. 

\subsubsection{Density $\rho<1/3$}

For $\rho<1/3$, the effective model upon doping the effective golden chain $G_1^\pm$ is again a $t$--$J$ chain ($C_{01}^\pm$) independent of the sign of the couplings. Our numerical spectra agree well with the effective model, as shown in Fig.~\ref{fig:charge36} for a  3$\times$6 ladder with $\rho=5/18$. 

\begin{widetext}
\begin{table*}[t]
\begin{tabular}{|c|c|c|c|c|c|c|c|}
\hline\hline
Filling & \jr & Phase & ${J}/{\jl}$ & ${t}/{\tl}$ & Derivation Eq. & ${V}/{\jl}$ & Derivation Eq.\\ [1ex]
\hline
$\rho<\frac{1}{3}$ & any & $C_{01}^{\pm}$ & $\frac{3}{8}$ & $1$ & (\ref{eq:t4})& --- & ---\\[1ex]
$\frac{1}{3}<\rho<\frac{2}{3}$ & $>0$ & $C_{21}^\pm$ & $\frac{3}{8}$ & $\frac{1}{(2+\alpha^2)2\phi} \Big[(3+\alpha^2+2\sqrt2\alpha)\bigg(\cos\frac{4\pi}{5}\bigg)  + 1\Big]$ & (\ref{eq:t5})& --- & ---\\[1ex] 
$\rho>\frac{2}{3}$ & $>0$ & $C_{23}^\pm$ & $1$ & $\frac{1}{2+\alpha^2}\bigg[\frac{1}{\phi^{2}}+\frac{\alpha^2}{2\phi}+\frac{\alpha^2}{\phi^{3}}+1\bigg]$ & (\ref{eq:t6})& $-\frac{2}{\phi^{2}}$ & (\ref{eq:g5a}) \\[1ex] 
$\rho>\frac{2}{3}$ & $<0$ & $C_{32}^\pm$ & $\frac{11}{8\phi^2}$ & $\frac{1}{2\phi}-\frac{1}{4}$ & (\ref{eq:t7})& $\frac{11}{8\phi^3}$ & (\ref{eq:p3}) \\[1ex] 
any & $> 3\phi |t_{\rm rung}|$ & $PS_{03}$ & 1 & $\frac{\tl^2}{E_D^2}(3\phi^{-2}+2\phi^{-3}+\phi^{-2}e^{8\pi i/5})$ & (\ref{eq:t3order})& $-\frac{2}{\phi^{2}}$ & (\ref{eq:g5a}) \\[1ex]
\hline
\end{tabular}
 \caption{Effective couplings for the $t$--$J$ phases of the three-leg ladder. Here  $E_D = (-\phi+\frac{1}{2})\jr + \sqrt{2}\tr+\frac{\sqrt{\jr^2+8\tr^2}}{2}$.} 
\label{tab:t-J_3leg}
\end{table*}
\end{widetext}

\begin{figure}[tb]
\includegraphics[width=  \columnwidth]{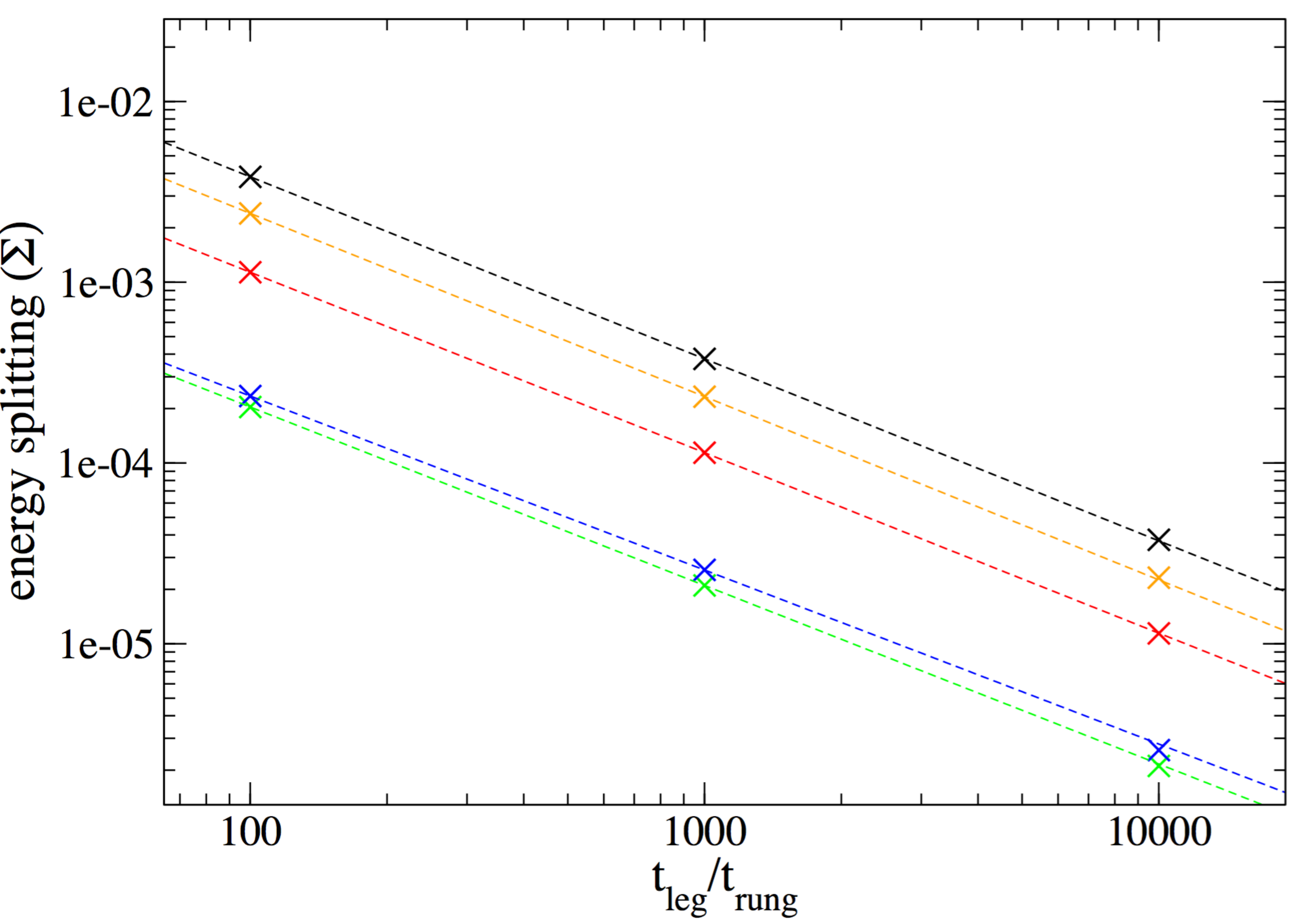}
\caption{
Energy splittings of all $E\le 0$ energy levels 
at $K=0$ of the $3\times 6$ ladder with $\rho=5/18$ as a function of $t_{\rm{leg}}/t_{\rm{rung}}$. 
The values of the couplings are $t_{\rm{leg}} =1, J_{\rm{leg}}=0, \vr=\infty$ and $t_{\rm{rung}}(= J_{\rm{rung}})$ takes different values from 100 to 10000. }
\label{fig:split_3leg}
\end{figure}

Note that, analogously to the case of the two-leg ladder, the mapping to the effective model is exact only in the limit
of infinite rung couplings. In Fig.~\ref{fig:split_3leg}, we show the log-log plot for the broadenings $\Sigma$ for each $E\le 0$ energy levels in the $K=0$ sector as a function of the inverse rung couplings for a three-leg ladder. The slope of $-1$ shows again that, for large but finite rung couplings,  there are second order effective processes involving higher energy states of the rungs.

\subsubsection{Density $1/3 < \rho < 2/3$ and $\jr>0$}

In the density regime $1/3<\rho<2/3$, we find an effective $t$--$J$ model ($C_{21}^\pm$) upon doping the $\rho=1/3$ $G_1^\pm$ golden chain with heavy holes (increasing the U(1) charge density)
or, equivalently, doping the totally gapped $\rho=2/3$ $T_2^\pm$ phase with effective light $\tau$ particles (reducing the U(1) charge density). Numerical simulations of three-leg ladders in this density regime gives an effective hopping 
that agrees very well with the analytical estimate.   

\subsubsection{Density $\rho>2/3$ and $\jr>0$}

In the strong antiferromagnetic  $\jr>0$ rung coupling limit, increasing the U(1) charge density starting from the $T_2^\pm$ gapped chain of heavy holes,
introduces super-heavy $\t$ anyons ({\it i.e.} $|3,\tau\rangle$ states).
The system is described by a $C_{23}^\pm$ (modified) $t$--$J$ chain.
Unlike in the previous simple $t$--$J$ chains,
the super-heavy $\tau$'s experience an effective nearest neighbor {\it attractive} potential. 

\subsubsection{Density $\rho>2/3$ and $\jr<0$}

For ferromagnetic $\jr<0$ at a density $\rho>2/3$, we map to an effective (modified) $t$--$J$ chain $C^\pm_{32}$ phase of heavy $\tau$'s and super-heavy holes. The heavy $\tau$'s experience a nearest neighbor 
{\it repulsive} potential. Our numerical simulations match very well the analytical estimates.

\subsubsection{Phase separation at large $\jr>0$}
In the absence of a rung charging energy $\vr$ an additional phase $PS_{03}$, that exhibits phase separation, appears for large $\jr>3\phi|t_{\rm rung}|$.
The physics of the interacting light holes and super-heavy $\tau$ particles is described by an extended 1D $t$--$J$--$V$ model that
contains the usual couplings of the regular $t$--$J$ model and, in addition, the 
nearest neighbor attractive potential $V$ between the super-heavy $\tau$ particles.
The dominant attraction $V$ leads  to phase separation between an empty and a completely filled ladder.

\section{Effective model of heavy and light Fibonacci anyons}\label{fm_rungs}

\subsection{The model}\label{fm_model}

We now discuss a new model which appears for strong FM rung couplings on a two-leg ladder with $\rho>1/2$
and a large $V_{\rm rep}$. A similar effective model also describes the three-leg ladder with FM rung couplings and $1/3<\rho<2/3$.
When the rung couplings are FM, the fusion of two $\t$'s results in a $\t$ charge.  One thus obtains an effective model with two $\it{different}$ Fibonacci particles, the $\it{heavy}$ and $\it{light}$ $\tau$'s distinguished by their $U(1)$ charge.  
For two-leg and three leg ladders, the magnetic interactions (similar to the Golden chain) and the potentials between the different flavors of \t particles are listed in Table \ref{tab:heavy_light}.

\begin{table}[b]
\begin{tabular}{|c|c|c|c|c|c|c|}
\hline\hline
Width & $\tau_1$ & $\tau_2$ & ${J}/{\jl}$ & Eq. & ${V}/{\jl}$ & Eq.\\ [1ex]
\hline
2 & heavy & heavy & $\frac{2}{\phi^{2}}$ & (\ref{eq:g2}) & $\frac{2}{\phi^{3}}$& (\ref{eq:p1}) \\[1ex]
2 & heavy & light & $-\frac{1}{\phi}$ & (\ref{eq:hl21}) &$\frac{1}{\phi}$ & (\ref{eq:p2}) \\[1ex] 
2 & light & light & $\frac{1}{2}$ & (\ref{eq:g1}) & ---& --- \\[1ex] 
\hline
3 & heavy & heavy & $\frac{11}{8\phi^2}$ & (\ref{eq:g4}) & $\frac{11}{8\phi^3}$& (\ref{eq:p3}) \\[1ex]
3 & heavy & light & $-\frac{5}{8\phi}$ & (\ref{eq:g6}) & $\frac{5}{8\phi}$ & (\ref{eq:p4}) \\[1ex] 
3 & light & light & $\frac{3}{8}$ & (\ref{eq:g3}) & ---& ---  \\[1ex] 
\hline
\end{tabular}
 \caption{Interactions between heavy and light $\tau$'s in two-leg and three-leg ladders. The labels $\tau_1$ and $\tau_2$ indicate the type (heavy or light) of the two interacting particles.} 
\label{tab:heavy_light}
\end{table}

\begin{figure}[t]
\includegraphics[width=\columnwidth]{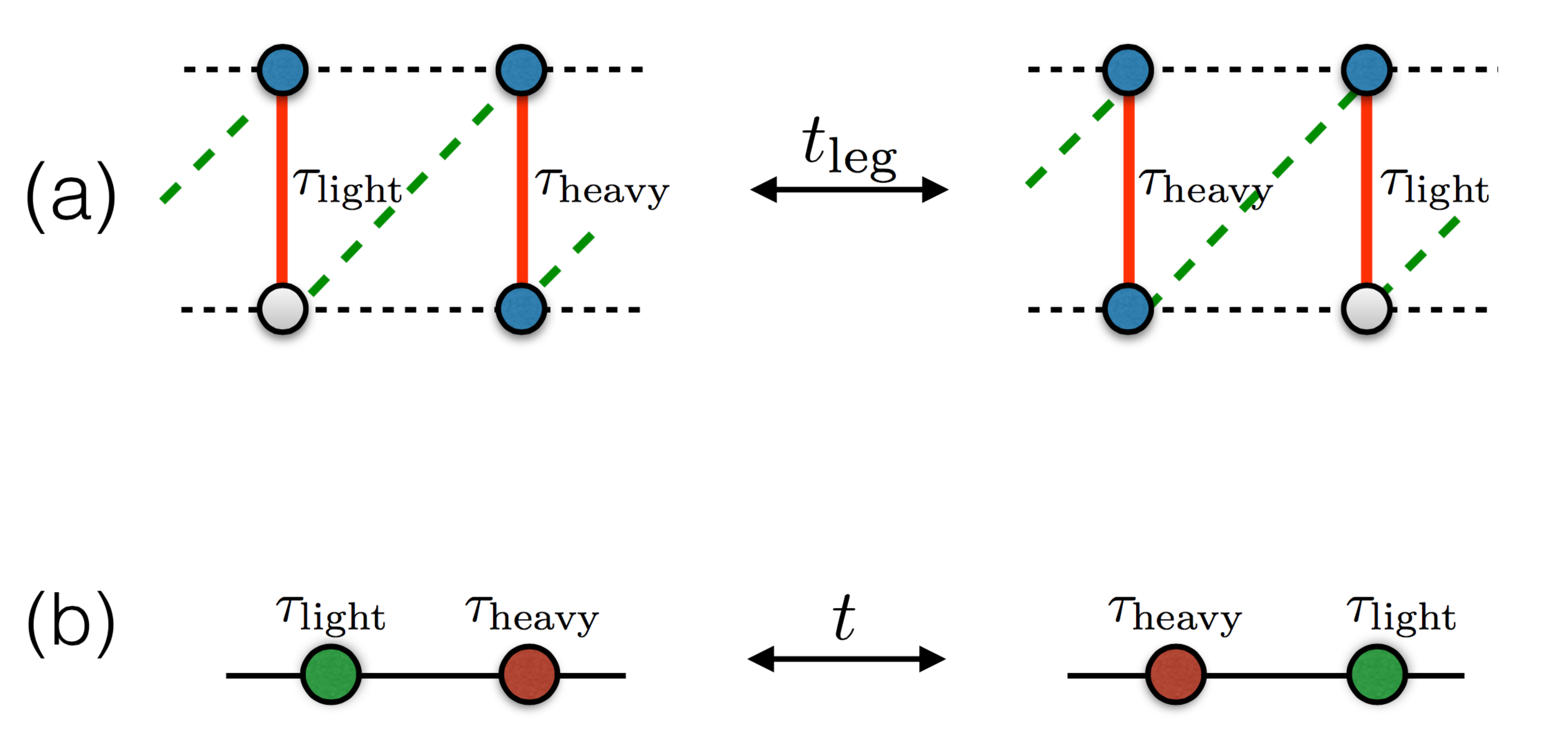}
\caption{Hopping of heavy (or equivalently light) $\tau$'s.  (a) Microscopic ladder model: the blue circles denote \t particles and the white circles are vacant sites on the ladder. (b) The effective chain.}
\label{fig:hl_hop}
\end{figure}

In addition to the magnetic and potential terms one also gets a kinetic process exchanging heavy and light $\tau$'s on nearest neighbor rungs. This process is shown schematically in Fig.~\ref{fig:hl_hop}(a),(b) for the microscopic ladder model and the effective chain respectively. 
Note that, in the $t$--$J$ chain, with holes and $\tau$'s, the hopping process shown in Fig.~\ref{fig:3figs}(c) moves the entire particle along with its charges and spin labels. 
Whereas now, the scenario is very different, the spin labels mix with each other as the heavy $\tau$'s hop over to exchange positions with the light 
$\tau$'s. 
The effective 1D model (in the basis of Eq.~(\ref{eq:basis})) for the hopping of heavy $\tau$'s is described by the Hamiltonian $H_{\rm HL}$ given below
\begin{equation}\label{eq:ham_hl}
 H_{\rm{HL}}=t\begin{bmatrix}
1 &  &  &  & \\
    & 0 &  &  & \\
    &  & 0 && \\
    &&&\phi^{-2} & \phi^{-3/2}\\
    &&&\phi^{-3/2}&\phi^{-1}
\end{bmatrix}  ,
\end{equation}
where  $t$ is the rescaled hopping amplitude. 
We have found that, for a two-leg ladder, the effective hopping amplitude is 
\begin{equation}
t=\cos(3\pi/5) t_{\rm{leg}} .
\end{equation}
A proof of this is provided in Eq. (\ref{eq:t3}).

For a three-leg ladder, the same model applies in the density regime $1/3<\rho<2/3$ (see Fig.~\ref{fig:phase_diag3}(b)
showing the two new low energy states on the rungs) and the hopping amplitude is 
\begin{equation}
t=\frac{1}{8}\big[1+9\cos(3\pi/5)\big] t_{\rm{leg}},
\end{equation}
as shown in Eq. (\ref{eq:t8}).
Note that there exists a symmetry between the two kinds of $\tau$'s {\it i.e.} the number of heavy and light $\tau$'s in the system can be swapped, leaving the physics unchanged. 
We provide a detailed analytical derivation of this Hamiltonian in Appendix~\ref{app:eff}.

\begin{figure}[t]
\includegraphics[width=\columnwidth]{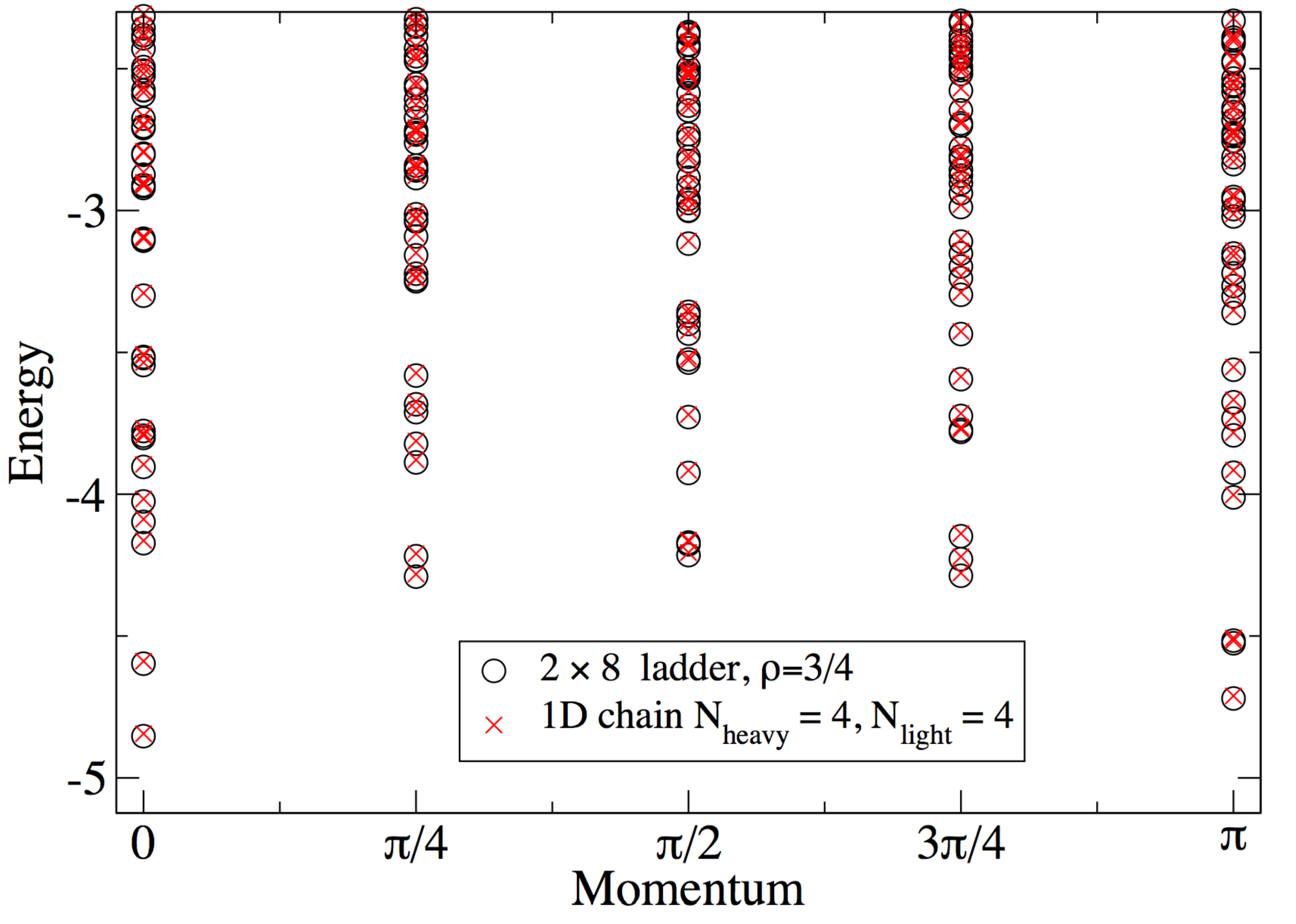}
\caption{Low energy spectrum of the two-leg ladder with strong FM rung couplings compared to the 1D heavy and light $\tau$ model. The black circles are for a 2$\times$8 ladder and anyon density $\rho=3/4$, while the red crosses are for the effective chain that it maps to ($L=8, \tilde\rho=1/2$). The values of the couplings are $t_{\rm{leg}} =1, J_{\rm{leg}}=0, \vr=\infty$ and $J_{\rm{rung}} =-2t_{\rm{rung}} = -2000$.}
\label{fig:hl}
\end{figure}

These analytical considerations are found to be in very good agreement with our numerical results for $\jl=0$.
Fig.~\ref{fig:hl} shows the comparison between the low energy spectra of the effective single chain with heavy and light taus and the ladder model. 
Note that for $\jl=0$ there is an exact E $\rightarrow$ -E symmetry in the spectrum (not shown here) that emanates from a  symmetry of exchanging  heavy and light $\tau's$. When there is an odd number of both  particle types, the momenta are shifted by $\pi$ under this exchange. 

The 1D model allows us to numerically solve larger systems with smaller finite size corrections.
We, however, restrict ourselves to the case $J_{\rm{leg}}=0$ when there are no magnetic interactions between \t particles along the leg direction but only a small hopping \tl operates between the rungs since already this simple model raises several open questions. 

\subsection{Single particle dispersion}

\begin{figure}[t]
\includegraphics[width=   \columnwidth]{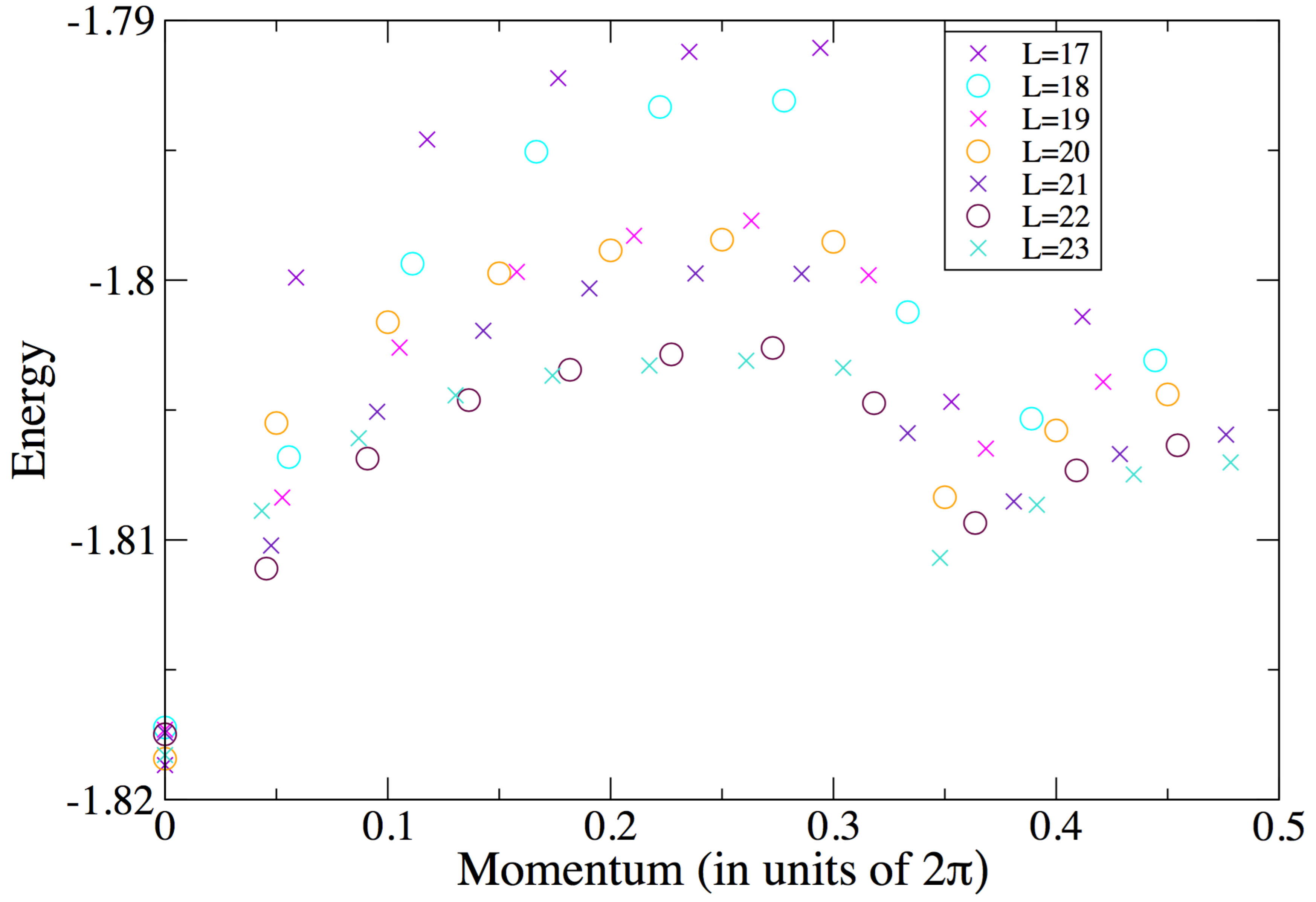}
\caption{ Energy dispersions of a single heavy (or light) \t  amidst $L-1$ light (or heavy) $\tau$'s for even (circles) and odd (crosses) length chains, with couplings $t=-1$
and $J=0$.}
\label{fig:nh1}
\end{figure}

We start with the simplest problem of a single light (heavy) $\t$ moving in a background of $L-1$ $\tau$'s of heavy (light) $\t$. We choose $t=-1$ in order to avoid even-odd chain length effects (although 
for $L$ even the sign of $t$ is irrelevant). 
In the spectra shown in Fig.~\ref{fig:nh1} for several chain lengths we observe that the dispersion minimum is always at $K=0$ and a local minimum 
appears around an incommensurate momentum. The  bandwidth is about $0.015 |t|$, independent of whether we consider a single heavy or light $\t$.

\subsection{Critical phase at generic fillings}\label{critical}

\begin{figure}[top]
\includegraphics[width=\columnwidth]{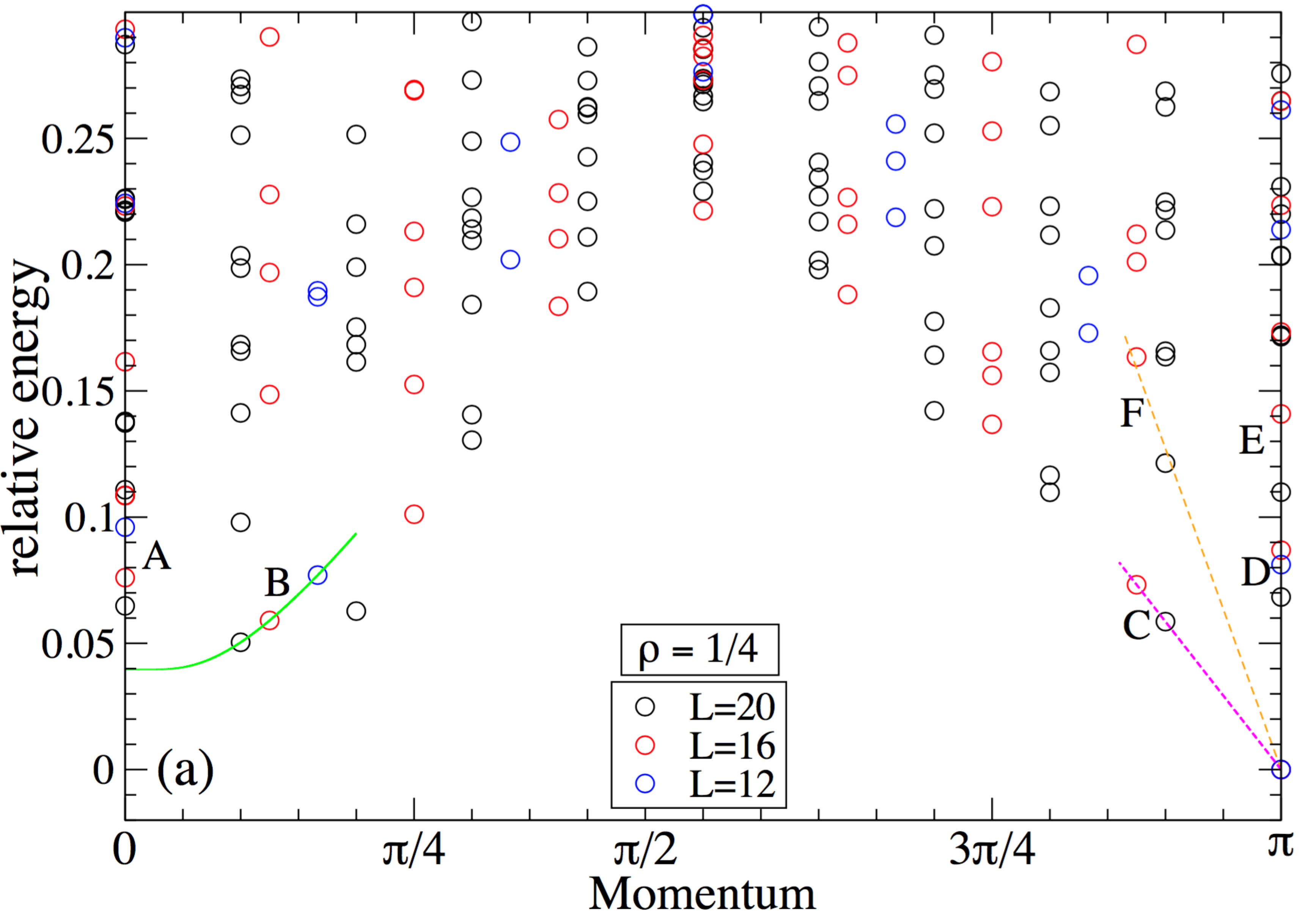}
\vskip 0.5truecm
\includegraphics[width=\columnwidth]{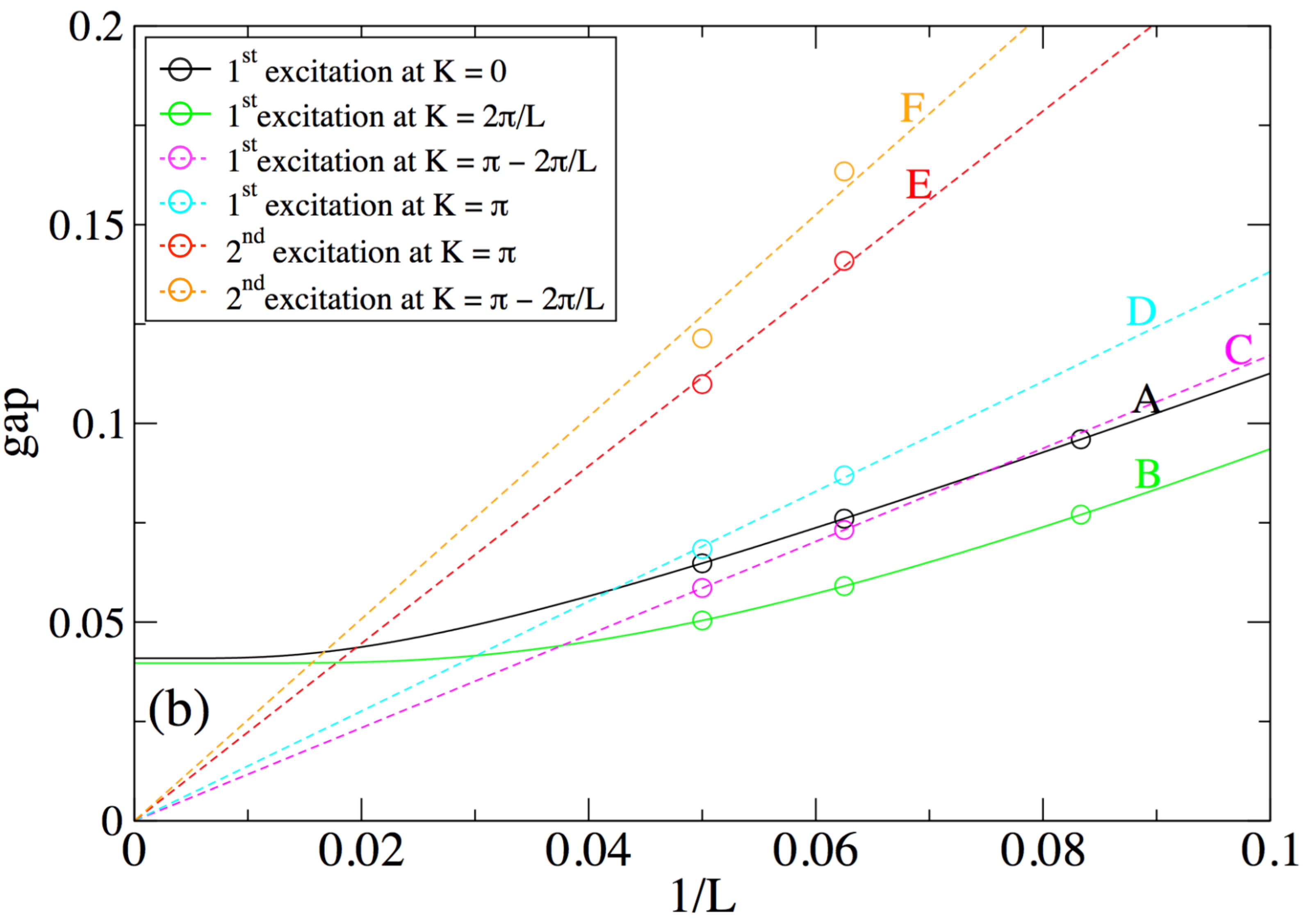}
\caption{(a) Spectra for the effective chains for heavy and light $\tau's$ at $\tilde{\rho}=1/4$ with $t=1,J=0$. Note the momenta have 
been shifted by $\pi$ for $L$ odd to make all spectra similar and the ground state energy has been subtracted out. 
The lowest energy levels at momenta $0$, $2\pi/L$, $\pi-2\pi/L$ and $\pi$ have been tagged as A, B, C and D, respectively.
The second excitations at momenta $\pi$ and $\pi-2\pi/L$ are labeled by E and F, respectively.
(b) Finite-size scaling analysis of the A, B, C, D, E and F energy excitations. Linear (dashed lines) and exponential (full lines) 
fits are shown around $K=\pi$ and $K=0$, respectively (see text). The scalings of the B, C and F gaps are also reported in (a). }
\label{fig:rho_fourth}
\end{figure}

We next consider a finite density $\tilde{\rho}$ of heavy  $\tau's$ and a corresponding filling of  $1-\tilde{\rho}$  light $\tau's$. Note that due to the symmetry between the heavy and light $\tau$'s, densities $\tilde{\rho}$ and $1-\tilde{\rho}$ are equivalent. We expect the same behavior for all densities except for the half-filled case $\tilde{\rho}=1/2$, which we shall consider separately in the next section. For simplicity we thus choose $\tilde{\rho}=1/4$ since it allows us to perform a finite size analysis using three different chain lengths  $L = 12$, 16, 20. The corresponding spectra are shown in Fig.~\ref{fig:rho_fourth}(a).

One-dimensional gapless systems are often described by a CFT and their lowest energy levels are then given by
\begin{equation}\label{eq:cft}
E(L)=e_TL + \frac{2\pi v}{L}(-\frac{c}{12}+h_L+h_R)  .
\end{equation}
where $c$ is the central charge and $h_L$, $h_R$ are the scaling dimensions of the `primary fields' of the CFT. 
The (thermodynamic) ground state energy per site $e_T$ and 
the velocity $v$ are non-universal constants. The finite size ground state energy $E_0(L)$ corresponds to $h_L=h_R=0$. 

To test the CFT prediction, we performed a finite-size scaling analysis of the first few energy gaps vs $1/L$. As shown in Fig.~\ref{fig:rho_fourth}(b), we observe
that the gaps around $K=\pi$ show a linear scaling with $1/L$,  suggesting gapless modes. 
This behavior is, in principle, consistent with the CFT scaling of Eq.~(\ref{eq:cft}). However, at this point, we could not identify the CFT that describes our model, being limited in Lanczos exact diagonalizations to system sizes of less then twenty sites.  Using the density matrix renormalization group 
(DMRG) might help to obtain the central charge, but is left for future studies.

The energy spectrum around $K=0$ shows a different behavior:  as shown in Fig.~\ref{fig:rho_fourth}(b), 
the finite size gaps of the first excited states at momentum $K=0$ and $K=2\pi/L$ 
could be fitted as $\Delta(L)=\Delta(\infty) + C/L \exp (-L/\xi)$, where $\Delta(\infty)\simeq 0.04$ is a finite energy gap and $\xi>10$ is a correlation length. This  suggests that, at density $\tilde{\rho}=1/4$, the energy spectrum shows both a gapless mode with linear disperson, described by a CFT, and additional gapped modes.

%%%%%%%%%%%%%%%%%%%%%%%%%%%%%%%%%%%%%%%%%%%%%%%%%%%%%%%%%%%%%%%%

\subsection{Possible topological gapped phase at $\tilde{\rho}=1/2$}\label{topo}

\begin{figure}[t]
\includegraphics[width= 0.9\columnwidth]{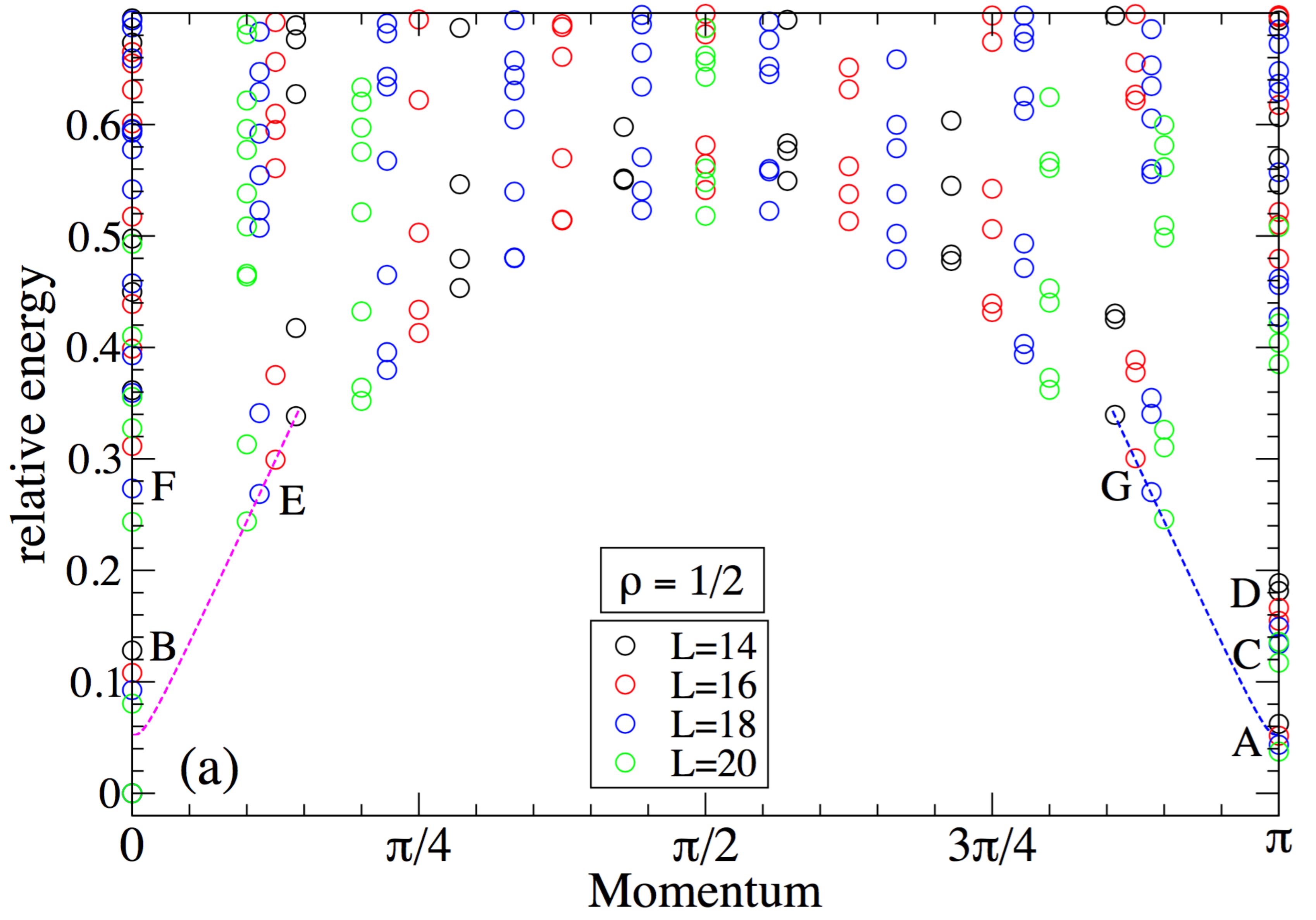}
\includegraphics[width=0.9 \columnwidth]{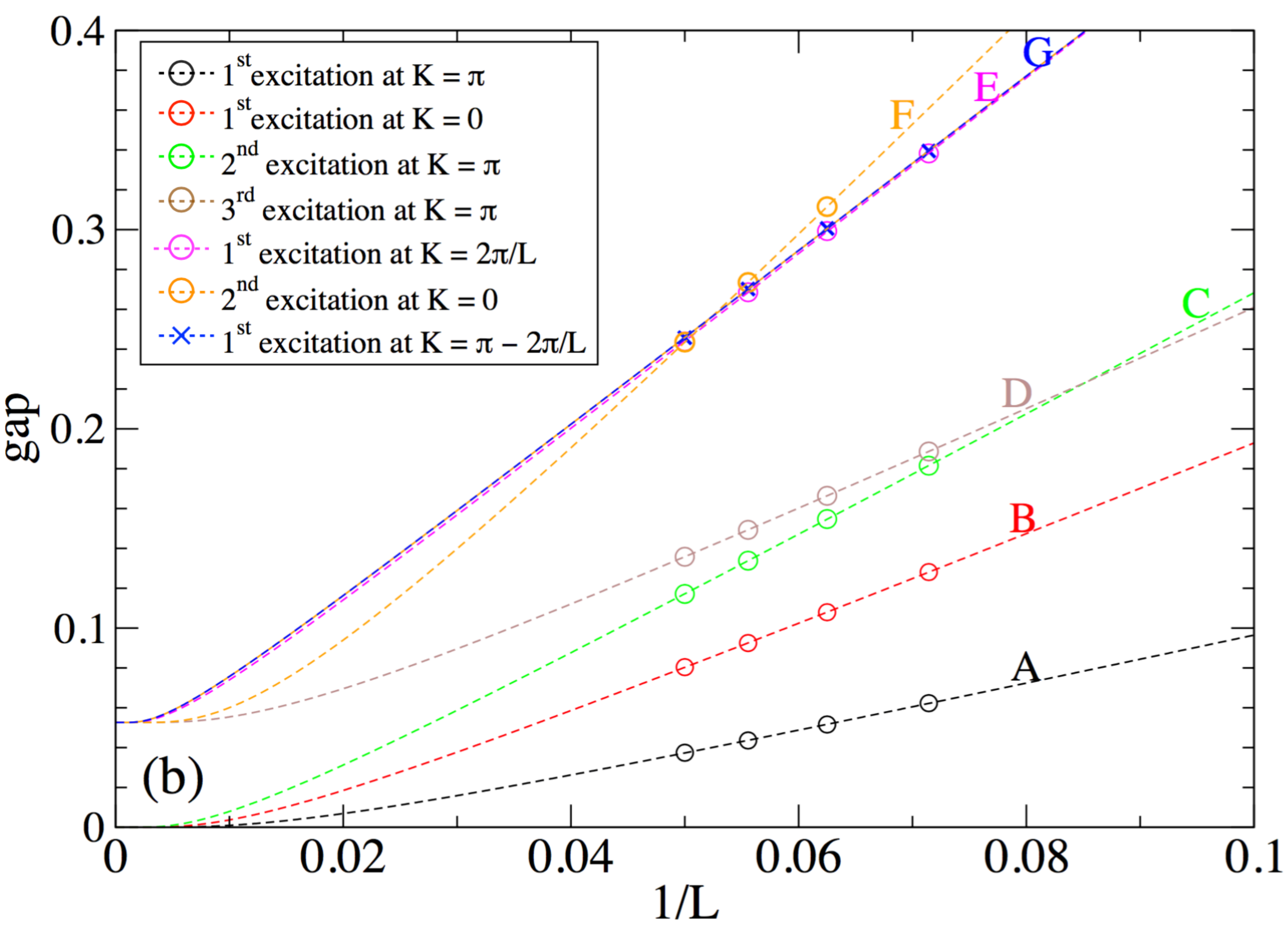}
\caption{
(a) Spectra of the chains of heavy and light $\tau$'s of different sizes at $\tilde{\rho}=1/2$ and $t=1,J=0$. The ground state energy has been subtracted out and the momenta are shifted by $\pi$ for odd number of particles of each type so as to get the same zero ground state momentum
in all cases. 
The lowest energy levels at momenta $0$, $2\pi/L$, $\pi-2\pi/L$ and $\pi$ have been tagged as B, E, G, A, respectively.
The second excitations at momenta $0$ and $\pi$ are labeled by F and C, respectively. The third excitation at momentum $\pi$ 
is labeled by D. 
(b) Finite-size scaling analysis of the energy gaps of (a) vs $1/L$ (see text).  The scalings of the E and G gaps are also reported in (a).  }
\label{fig:rho_half}
\end{figure}

 Next we consider the density $\tilde{\rho}=1/2$ where there is an equal number of heavy and light  $\tau's$. We simulated chains with lengths $L = 14$, 16, 18, and 20 and show these spectra in Fig.~\ref{fig:rho_half}(a), revealing low energy excitations at momenta $K=0$ and $K=\pi$.
Performing a finite size scaling analysis on the low lying states using 
system sizes $L$ ranging from 14 to 20 sites, as shown in Fig.~\ref{fig:rho_half}(b) we find that an exponential form like $\Delta(L)=\Delta(\infty) + C/L \exp (-L/\xi)$ provides reasonably good fits of the data. 
These fits suggest that three of the gaps extrapolate to zero and the next higher energy excitations extrapolate to 
a finite value $\Delta(\infty)\sim 0.05$. Note however, that the correlation lengths extracted from the fits are of the order of the system size so that our extrapolations have to be taken with caution. However, if correct, our findings would indicate a topological gapped phase with a four-fold degenerate ground state, although dimerization is not excluded (since ground state momenta are both $0$ and $\pi$). 
In any case, we believe half-filling is a special case and very different from the other density regimes we considered. This behavior is also notably different from the golden chains which are known to be gapless for both FM and AFM leg couplings.

\section{Conclusions and outlook}\label{conclude}
We have  studied  ladders of itinerant Fibonacci anyons, determined their phase diagrams and constructed effective low energy models. In the limit of strong rung couplings we find  several different phases:  totally gaped phases, paired phases described by hard core bosons, golden chain phases, $t-J$ phases that carry \t anyons and trivial particles and lastly the heavy and light \t phase that carries two flavors of Fibonacci anyons. We have derived effective low-energy 1D chain models for all of these phases.

The golden chain phases ($G$) appear for special commensurate fillings, where each rung of the ladder carries a topological charge $\tau$ in the same quantum state. The ladder can hence be mapped onto a golden chain with rescaled couplings.  Depending on the sign of the rung couplings, we also obtain totally gapped phases ($T$), where all the rungs carry a trivial topological charge.

We obtain additional paired phases ($P$) described by an effective hard-core boson model, similar to fermionic $t$--$J$ ladders mapping to a Luther-Emery liquid of Cooper pairs. 
This phase is obtained in a two-leg ladder when ever anyons on a rung always combine to a totally trivial charge.

The most common phase are described by effective $t-J$ models ($C_{nm}$).
In the case of a three-leg ladder with $\jr>3\phi |t_{\rm rung}|$ a dominant attraction leads to phase separation. All other phases show itinerant behavior. The mapping onto a $t$-$J$ chain shows that spin-charge fractionalization, which was initially found for 1D  chains of non-Abelian anyons, also occurs in ladders.

For FM rung couplings, a new phase occurs in which the low-energy model contains Fibonacci anyons carrying different electric charges but the same topological charge. The corresponding effective model is hence very different. This phase ($D_{mn}$) is seen when $\jr <0$ and $\rho>1/2$ for a two-leg ladder while $1/3<\rho<2/3$ for a three-leg ladder. 
This model allows magnetic interactions between the different flavors of \t anyons in addition to an exchange of the two particle species, where a heavy and light \t  swap positions. We have analyzed this model in the limit when there are no magnetic interactions between the different flavors of \t particles, since it is very rich in itself leading to several open questions, including a potential topological gapped phase at half-filling.
For other densities, we find evidence for gapless modes described by a CFT. At this stage, it is unclear what is the central charge and the CFT that describes our model. These issues deserve further studies.

\section*{Acknowledgments}
M.S. and D.P. acknowledge the support from NQPTP Grant No. ANR-0406-01 from the French Research Council (ANR) and the HPC resources of CALMIP (Toulouse) for CPU time under project P1231.  M.T. acknowledges hospitality of the Aspen Center for Physics, supported by NSF grant 1066293.

\appendix\label{appendix}
\section{Effective models}\label{app:eff}

In this Appendix we  explain the derivation of the effective models. We perturbatively derive the nearest neighbor couplings of the effective 1D models assuming  large rung couplings and small leg couplings: $|J_{\rm{rung}}|,|t_{\rm{rung}}| \gg |J_{\rm{leg}}|,|t_{\rm{leg}}|$.

\subsection{Magnetic and Potential Interactions}\label{app:mag}

Magnetic interactions will only be present in effective models if two neighboring sites each have a total anyonic $\t$ charge.  A rung state with a total anyonic $\t$ charge can have one (light $\tau$), two (heavy $\tau$) or three (super heavy $\tau$) $\tau$ particles on the rung, as shown in Table~\ref{table}. The possible magnetic interactions are marked by pink arrows. Below we investigate all possible combinations 
for two neighboring rungs in two-leg and three-leg ladders, respectively and compute the effective interactions from
first-order perturbation in $J_{\rm leg}$.

\begin{table}[t!]
\includegraphics[width= 0.7\columnwidth]{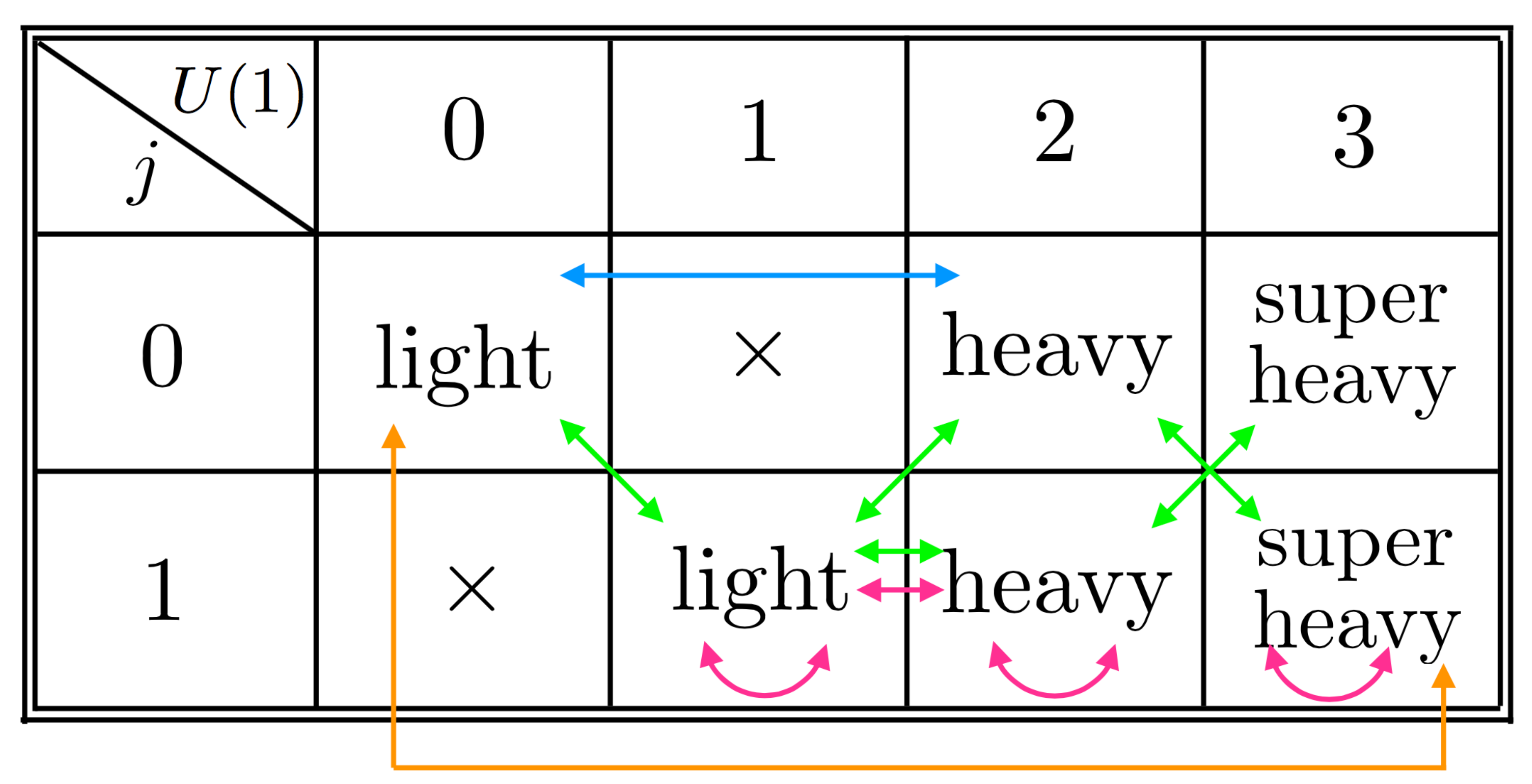}
\caption{The various rung states classified according to their U(1) charge (columns) and topological charge (lines). 
$j=0$ ($j=1$) corresponds to holes ($\tau$ particles). 
The possible magnetic interactions (in the present models) are indicated by pink arrows and the green arrows represent the possible kinetic terms arising in first-order perturbation in $t_{\rm leg}$.  
The leading kinetic and potential interactions can also arise only in second or third-order in
$t_{\rm leg}$, marked by blue and orange arrows respectively.}
\label{table}
\end{table}

\subsubsection{Two-leg ladder}\label{2leg:all_cases}

For the two-leg ladder the rungs have to be in the light $\tau$ state 
\begin{equation}
|1^\pm, \t\rangle = \frac{1}{\sqrt2}\big(|1^U,\tau\rangle \pm |1^L,\tau\rangle\big) \equiv \frac{|U\rangle \pm  |L\rangle}{\sqrt2},
\label{eq:s}
\end{equation}
or in the heavy $\tau$ state $|2, \t\rangle$,
where we introduced short hand notations $|U\rangle$ and $|L\rangle$ for an anyon on the upper or lower site of a rung, respectively. We review three possible cases below.

{\bf Light $\tau$ -- light $\tau$~:} We first evaluate the matrix element $ \langle\psi |H^\text{leg}_{\rm magn}|\psi\rangle$ of the state
\begin{eqnarray}
|\psi\rangle &=& |1^\pm, \t\rangle  \otimes|1^\pm, \t\rangle  \nonumber \\
 &=& \frac{1}{2}(|UU\rangle \pm |LU\rangle \pm |UL\rangle + |LL\rangle)  ,
\end{eqnarray}
where $H^\text{leg}_{\rm magn}$ is the (part of the) Hamiltonian that describes the magnetic interaction between two anyons once they are nearest neighbors along a leg of the ladder (see main text).

The only non-vanishing matrix elements for the magnetic interaction in the effective model arise when the two anyons are on the same leg of the ladder. Thus, we have
\begin{eqnarray}
 \langle\psi |H^\text{leg}_{\rm magn}|\psi\rangle &=& \frac{1}{4}(\langle UU|H^\text{leg}_{\rm magn}|UU \rangle + \langle LL|H^\text{leg}_{\rm magn}|LL \rangle) \nonumber\\
 &=& \frac{1}{2}\langle UU|H^\text{leg}_{\rm magn}|UU \rangle  .
\end{eqnarray}
The second step follows since the contributions from magnetic interactions on the upper and lower legs of the ladder are equal in magnitude. 
It follows immediately that the effective magnetic interaction is half the bare interaction:
\begin{equation}\label{eq:g1}
J = \frac{1}{2}J_{\rm{leg}} .
\end{equation} 

{\bf Heavy $\tau$ -- heavy $\tau$~:} Let us now consider the case where both rungs are in the heavy $\tau$ state.  
The state $|\psi\rangle$ is now defined by 
\begin{equation}
|\psi\rangle = |2,\tau\rangle \otimes |2,\tau\rangle \, .
\end{equation}
We calculate the matrix element $ \langle\psi |H^\text{leg}_{\rm magn}|\psi\rangle$ for this state to obtain 
\begin{eqnarray}\label{eq:fm1}
 \langle\psi |H^\text{leg}_{\rm magn}|\psi\rangle &=& (\langle UU|H^\text{leg}_{\rm magn}|UU \rangle + \langle LL|H^\text{leg}_{\rm magn}|LL \rangle)\nonumber \\ 
 &=& 2 \langle UU|H^\text{leg}_{\rm magn}|UU \rangle\, ,
\end{eqnarray}
where the second steps follows since both the terms have the same contribution.

To evaluate the contribution of such a term, we need to calculate explicitly the matrix elements for all bond labels. The bond labels belong to the set ${\cal S}$ defined in Eq.~(\ref{eq:basis}), which we rewrite here for convenience
\begin{equation}\label{eq:basis2}
{\cal S}=\{|{\bf 1}\tau{\bf 1}\rangle, |{\bf 1}\tau\tau\rangle, |\tau\tau{\bf 1}\rangle, |\tau{\bf 1}\tau\rangle, |\tau\tau\tau\rangle\}  . \end{equation}
We denote the initial quantum state (including site and bond labels) as:
\begin{equation}\label{eq:in}
|\Psi_{\alpha}\rangle=\big |\psi;\xi_\alpha\rangle ,
\end{equation}
where $|\xi_\alpha\rangle\in {\cal S}$. The final states after the magnetic process are given by :
\begin{equation}\label{eq:fin}
|\Psi_{\beta}\rangle=\big |\psi;\xi_\beta\rangle  ,
\end{equation}
where $|\xi_\beta\rangle\in{\cal S}$.

Then, the matrix elements for this process in the basis $\mathcal{S}$ are given by:
\begin{equation}
   H^{\rm leg}_{\rm magn}=-\jl\begin{bmatrix}
   \phi^{-1} &  &  &  & \\
    & \phi^{-3} &  &  & \\
    &  & \phi^{-3} && \\
    &&&\phi^{-2} & \phi^{-7/2}\\
    &&&\phi^{-7/2}&\phi^{-2}+\phi^{-5} 
\end{bmatrix}  .
\end{equation}

This effective Hamiltonian is proportional to the golden chain Hamiltonian (upto an overall shift) and the contribution to the effective coupling on each leg is  $\phi^{-2} J_{\rm leg}$.
Thus, the effective potential between the two heavy $\tau$'s is given by
\begin{equation}\label{eq:p1}
V = 2 \phi^{-3} J_{\rm leg},
\end{equation}
and the 
effective magnetic interaction for the entire process obtained by combining contributions from both the legs (see Eq.~(\ref{eq:fm1})), is thus given by
\begin{equation}\label{eq:g2}
J = 2 \phi^{-2} J_{\rm leg} \, .
\end{equation}

{\bf Light $\tau$ -- heavy $\tau$~:} Finally, we consider the case where we have a heavy \t and light $\tau$ on neighboring rungs.
The state $|\psi\rangle$ is now defined as
\begin{eqnarray}
|\psi\rangle &=& |1^\pm, \t\rangle  \otimes|2, \t\rangle \, . 
\end{eqnarray}
We calculate the matrix element $ \langle\psi |H^\text{leg}_{\rm magn}|\psi\rangle$ for this state to obtain 
\begin{eqnarray}\label{eq:fm}
 \langle\psi |H^\text{leg}_{\rm magn}|\psi\rangle &=& \frac{1}{2}(\langle UU|H^\text{leg}_{\rm magn}|UU \rangle + \langle LL|H^\text{leg}_{\rm magn}|LL \rangle)\nonumber \\ 
 &=&  \langle UU|H^\text{leg}_{\rm magn}|UU \rangle\, ,
\end{eqnarray}
where the second steps follows since both the terms have the same contribution.
As before we denote the initial and final states as defined in Eq.~(\ref{eq:in}) and Eq.~(\ref{eq:fin}) respectively.
Then, the matrix elements for this process in the basis $\mathcal{S}$ are given by:
\begin{equation}
   H^{\rm leg}_{\rm magn}=-\jl\begin{bmatrix}
  0 &  &  &  & \\
    & \phi^{-1} &  &  & \\
    &  & \phi^{-1} && \\
    &&&\phi^{-2} & -\phi^{-5/2}\\
    &&&-\phi^{-5/2}&\phi^{-3} 
\end{bmatrix}  .
\end{equation}

This Hamiltonian matrix is equivalent to the golden chain Hamiltonian with the effective coupling
\begin{equation}\label{eq:hl21}
J = - \phi^{-1} J_{\rm leg} \,
\end{equation}
and an effective potential
\begin{equation}\label{eq:p2}
V = \phi^{-1} J_{\rm leg} \,
\end{equation}

\subsubsection{Three-leg ladder}\label{3leg:all_cases}

For three legs, a rung has a total anyonic $\tau$ charge if it is in one of the $|1^\pm,\tau\rangle$ (light), 
$|2^\pm,\tau\rangle$ (heavy) or $|3,\tau\rangle$ (super heavy) state. Below we consider all possible cases for
 neighboring rungs both with a total anyonic $\tau$ charge.

{\bf Light $\tau$ -- light $\tau$~:} We first assume that the two rungs are both in the state 
\begin{eqnarray}
|1^\pm,\tau\rangle &=& \frac{1}{2}\big(|1^U,\tau\rangle + |1^L,\tau\rangle \pm \sqrt2 |1^M,\tau\rangle\big)  \nonumber \\
&=&\frac{1}{2}{|L\rangle} \pm \frac{1}{\sqrt2}{|M \rangle}+\frac{1}{2} |U \rangle,
\label{eq:s3leg}
\end{eqnarray} 
where  the states $|L\rangle, |M\rangle, |U\rangle$ denote the position of the anyon on the lower, middle or upper leg respectively, and $\pm$ depends on the sign of $t_{\rm rung}$. 
As above we calculate the matrix element for the state
\begin{equation}
|\psi\rangle = |1^\pm,\tau\rangle \otimes |1^\pm,\tau\rangle
\end{equation}
and obtain
\begin{eqnarray}
 \langle\psi |H^{\rm leg}_{\rm{magn}}|\psi\rangle &=& \frac{1}{16}\big(\langle UU|H^{\rm leg}_{\rm{magn}}|UU \rangle \nonumber \\
 &+& \langle LL|H^{\rm leg}_{\rm{magn}}|LL \rangle \nonumber \\
 &+&\frac{1}{4}\langle MM|H^{\rm leg}_{\rm{magn}}|MM \rangle\big)\nonumber\\
 &=&\frac{3}{8}\langle UU|H^{\rm leg}_{\rm{magn}}|UU\rangle   ,
 \label{eq:effec}
\end{eqnarray}
resulting in an effective coupling
\begin{equation}\label{eq:g3}
J = \frac{3}{8}  {J_{\rm{leg}}} .
\end{equation}

{\bf Heavy $\tau$ -- heavy $\tau$~:} Let us now assume the effective $\tau$'s on the rungs are in the heavy \t state 
\begin{eqnarray}\label{eq:2pm}
|2^\pm,\tau\rangle &=& \frac{1}{2}\big(|1^{U,M},\tau\rangle + |1^{M,L},\tau\rangle \pm \sqrt2 |1^{U,L},\tau\rangle\big) \nonumber \\
&\equiv&\frac{1}{2}\big(|UM\rangle + |ML\rangle \pm \sqrt2 |UL\rangle\big),
\end{eqnarray}
where we have used the short hand notation $|AB\rangle$ denoting that the \t lies on the leg $A$ and $B$ and the net fusion channel outcome being a \t is implicit.

As above we calculate the matrix element for the state
\begin{eqnarray}
|\psi\rangle &=& |2^\pm,\tau\rangle \otimes |2^\pm,\tau\rangle \\ \nonumber
&=& \frac{1}{4}\big(|UM,UM\big>+|UM,ML\big>+\sqrt2|UM,UL\big>  \\ \nonumber
&+&|ML,UM\big>+|ML,ML\big>+\sqrt2|ML,UL\big> \\ \nonumber
&+&\sqrt2|UL,UM\big>+\sqrt2|UL,ML\big>+2|UL,UL\big> \big)
\end{eqnarray}
and obtain
\begin{eqnarray}
 \langle\psi |H^{\rm leg}_{\rm{magn}}|\psi\rangle &=& \frac{1}{16}\big(\langle UM,UM|H^{\rm leg}_{\rm{magn}}|UM,UM \rangle \nonumber \\
  &+& \langle ML,ML|H^{\rm leg}_{\rm{magn}}|ML,ML \rangle \nonumber \\
  &+&2 \langle UL,UL|H^{\rm leg}_{\rm{magn}}|UL,UL \rangle \nonumber \\
 &+& \langle UM,ML|H^{\rm leg}_{\rm{magn}}|UM,ML \rangle \nonumber \\
  &+& \sqrt2 \langle UM,UL|H^{\rm leg}_{\rm{magn}}|UM,UL \rangle \nonumber \\
   &+&  \langle ML,UM|H^{\rm leg}_{\rm{magn}}|ML,UM \rangle \nonumber \\
    &+& \sqrt2 \langle ML,UL|H^{\rm leg}_{\rm{magn}}|ML,UL \rangle \nonumber \\
     &+& \sqrt2 \langle UL,UM|H^{\rm leg}_{\rm{magn}}|UL,UM \rangle \nonumber \\
      &+& \sqrt2 \langle UL,ML|H^{\rm leg}_{\rm{magn}}|UL,ML \rangle \big) \, .\nonumber \\
      &&
  \label{eq:effec23}
\end{eqnarray}

The first three terms have two magnetic interactions each along the leg direction while all the others have only one. Moreover all these magnetic interactions are equal in magnitude, the contribution of a single term giving the effective coupling $J = \phi^{-2}J_{\rm leg}$ (see Eq.~(\ref{eq:g2})). Taking into account the contributions from all the terms, it follows immediately that the effectively magnetic coupling is
\begin{align}\label{eq:g4}
J &= \frac{11}{8}  \phi^{-2} {J_{\rm{leg}}} 
\end{align}
and the effective potential is given by
\begin{align}\label{eq:p3}
V&= \frac{11}{8}  \phi^{-3} {J_{\rm{leg}}}\, .
\end{align}

{\bf Super-heavy $\tau$ -- super-heavy $\tau$~:} Let us now assume the states on the rungs with total anyonic charge \t are both given by 
$|3,\tau\rangle$ (super-heavy $\tau$'s). Then, we calculate the matrix element for the state
\begin{equation}
|\psi\rangle = |3,\tau\rangle \otimes |3,\tau\rangle
\end{equation}
and obtain
\begin{eqnarray}
 \langle\psi |H^{\rm leg}_{\rm{magn}}|\psi\rangle &=& \big(\langle UU|H^{\rm leg}_{\rm{magn}}|UU \rangle \nonumber \\
 &+& \langle LL|H^{\rm leg}_{\rm{magn}}|LL \rangle \nonumber \\
 &+&\langle MM|H^{\rm leg}_{\rm{magn}}|MM \rangle\big)\, .
 \label{eq:effec3}
\end{eqnarray}
Magnetic interactions on the upper and middle leg contribute to a potential in the effective Hamiltonian given by
\begin{equation}\label{eq:g5a}
V = -2\phi^{-2} J_{\rm leg} \, ,
\end{equation} 
while the magnetic process on the lower leg results 
in an effective magnetic coupling given by 
\begin{equation}\label{eq:g5b}
J = J_{\rm leg} \, .
\end{equation}

{\bf Light $\tau$ -- heavy $\tau$~:} Finally, we consider the case when the two $\tau$'s on the neighboring rungs 
are in the $|1^\pm,\tau\rangle$ and the $|2^\pm,\tau\rangle$ states. The state $|\psi\rangle$ is now given by
\begin{eqnarray}
|\psi\rangle &=& |2^\pm,\tau\rangle \otimes |1^\pm,\tau\rangle \\ \nonumber
&=& \frac{1}{2}\big(|UM\big>+|ML\big>+\sqrt2|UL\big>\big)  \\ \nonumber
&\otimes & \frac{1}{2}\big(|U\big>+|L\big>+\sqrt2|M\big>\big) \\ \nonumber
&=&|UM,U\big>+|UM,L\big>+\sqrt2|UM,M\big> \\ \nonumber
&+&|ML,U\big>+|ML,M\big>+\sqrt2|ML,M\big> \\ \nonumber
&+&\sqrt2|UL,U\big>+\sqrt2|UL,L\big>+2|UL,M\big>\, .
\end{eqnarray}

As before we calculate the matrix element 
\begin{eqnarray}
 \langle\psi |H^{\rm leg}_{\rm{magn}}|\psi\rangle &=& \frac{1}{16}\big(\langle UM,U|H^{\rm leg}_{\rm{magn}}|UM,U \rangle \nonumber \\
  &+& 2 \langle UM,M|H^{\rm leg}_{\rm{magn}}|UM,M \rangle \nonumber \\
       &+&  \langle ML,L|H^{\rm leg}_{\rm{magn}}|ML,L \rangle \nonumber \\
       &+& 2 \langle ML,M|H^{\rm leg}_{\rm{magn}}|ML,M \rangle \nonumber \\
       &+&2  \langle UL,U|H^{\rm leg}_{\rm{magn}}|UL,U \rangle \nonumber \\
       &+&2  \langle UL,L|H^{\rm leg}_{\rm{magn}}|UL,L \rangle \big) \, .\nonumber \\
      &&
  \label{eq:effechl3}
\end{eqnarray}
 The contributions of each of these terms to the effective coupling is $(-\phi^{-1})\jl$ (see Eq.~(\ref{eq:hl21})), thus the net effective coupling for this magnetic process is given by 
 \begin{equation}\label{eq:g6}
J = -\frac{5}{8  \phi} {J_{\rm{leg}}} 
\end{equation}
and the effective potential is 
 \begin{equation}\label{eq:p4}
V = \frac{5}{8  \phi} {J_{\rm{leg}}} 
\end{equation}

\subsection{Kinetic terms}\label{app:kin}

Whenever the total U(1) charges of the two neighboring rung states differ by $\pm 1$,
a hopping process occurs in first order in $t_{\rm leg}$.
It is the case for a charge-1 (light) $\tau$ and a charge-0 (light) or charge-2 (heavy) hole, a charge-1 (light) $\tau$ and a 
charge-2 (heavy) $\tau$, a charge-2 (heavy) hole and a charge-3 (super heavy) $\tau$, a charge-2 (heavy) $\tau$ and a 
charge-3 (super heavy) hole. All these are marked schematically by green arrows in Table.~\ref{table}. 
Below we investigate all these possibilities in two leg and three leg ladders, respectively.

\subsubsection{Two-leg ladder}\label{2leg:kin}
Let us first evaluate the effective hopping in a two-leg ladder arising when there is a \t particle in the state $ |1^\pm, \t\rangle$  and an effective hole in either the empty $|e\rangle=|0,{\bf 1}\rangle$ (case 1) or the
fully occupied $|f\rangle=|2,{\bf 1}\rangle$ (case 2) rung state on adjacent rungs.

{\bf Light hole -- light $\tau$~:} First, we need to evaluate the matrix element
 $\big<\Psi_1| H^{\rm leg}_{\rm kin}|\Psi_2\rangle$ between the two states 
 \begin{equation}
 |\Psi_1\rangle=| 1^\pm, \t\rangle \otimes|e\rangle  \equiv \frac{|Ue\rangle \pm  |Le\rangle}{\sqrt2}
 \end{equation}
 and 
  \begin{equation}
  |\Psi_2\rangle=|e\rangle \otimes|1^\pm, \t\rangle  \equiv \frac{|eU\rangle \pm  |eL\rangle}{\sqrt2},
   \end{equation} where $H^{\rm leg}_{\rm kin}$ is the kinetic part of the ladder Hamiltonian 
 living on the legs (see main text). Similar to the case of magnetic interactions one gets a factor $1/2$ and obtains for the effective hopping 
\begin{equation}\label{eq:t1} t = \frac{1}{2} t_{\rm{leg}} .\end{equation}

{\bf Heavy hole -- light $\tau$~:} Next, one evaluates the matrix element between the states
 \begin{equation}
 |\Psi_1\rangle=| 1^\pm, \t\rangle \otimes|f\rangle  \equiv \frac{|Uf\rangle \pm  |Lf\rangle}{\sqrt2}
 \end{equation}
 and 
  \begin{equation}
  |\Psi_2\rangle=|f\rangle \otimes|1^\pm, \t\rangle  \equiv \frac{|fU\rangle \pm  |fL\rangle}{\sqrt2} \, .
 \end{equation}
 The matrix elements are written explicity as
   \begin{equation}
\langle\Psi_2| H^{\rm leg}_{\rm kin}|\Psi_1\rangle =\frac{1}{2} \langle fU|H^{\rm leg}_{\rm kin}|Uf\rangle +\frac{1}{2} \langle fL|H^{\rm leg}_{\rm kin}|Lf\rangle \, .
  \end{equation}
The derivation is identical but involves two non-trivial braids. The contributions from the hopping on the two legs are thus given by
  \begin{eqnarray}
  \langle f{U}|H^{\rm leg}_{\rm kin}| U{f} \rangle &=& e^{4\pi i/5}\phi^{-1} t_{\rm{leg}}  \big<\mathbf1\t |h_{\rm kin}|\t \mathbf1\rangle,\nonumber\\
    \langle f{L}|H^{\rm leg}_{\rm kin}| L{f} \rangle &=&e^{-4\pi i/5} \phi^{-1} t_{\rm{leg}}  \big<\mathbf1\t |h_{\rm kin}|\t \mathbf1\rangle.\nonumber\\
  \end{eqnarray}

Adding both the terms we get an overall $-1$ factor, compared to the previous case:
\begin{equation}\label{eq:t2} t = -\frac{1}{2} t_{\rm{leg}} .\end{equation}

{\bf Light $\tau$ -- heavy $\tau$~:} Finally, let us consider two nearest neighbor rungs, one carrying a heavy \t  \ and the other with a light $\tau$. 
Here, the bond labels belong to the set ${\cal S}$ defined in Eq.~(\ref{eq:basis2}).
We denote the `initial' quantum state (including site and bond labels) with a light (heavy) \t on the first (second) rung as:
\begin{equation}
|\Psi_{i,\alpha}\rangle=\big |1_l2_h;\xi_\alpha\rangle ,
\end{equation}
where $i$ stands for initial and $|\xi_\alpha\rangle\in {\cal S}$. The `final' states after the hopping process are given by :
\begin{equation}
|\Psi_{f,\beta}\rangle=\big |1_h2_l;\xi_\beta\rangle  ,
\end{equation}
where $f$ is for final and $|\xi_\beta\rangle\in{\cal S}$.
One needs to compute all matrix elements $\big<\Psi_\beta|
H^{\rm leg}_{\rm kin}|\Psi_\alpha\rangle$.
The matrix elements for the hopping of a heavy and a light \t on the lower leg of the ladder are found to be: 
\begin{widetext}
\begin{eqnarray}
\big< 1_h2_l; \mathbf{1} \t \mathbf{1}|H^{\rm leg}_{\rm kin}|1_l2_h; \mathbf{1} \t \mathbf{1}\rangle&=& \tl  e^{-3\pi i/5} \langle\mathbf{1} \t|h_{\rm kin}|\t \mathbf{1}\rangle, \nonumber\\
\big< 1_h2_l; \mathbf{1}\t \t |H^{\rm leg}_{\rm kin}|1_l2_h; \mathbf{1}\t \t \rangle&=& -\tl  e^{-3\pi i/5} \phi^{-1} \langle\mathbf{1} \t|h_{\rm kin}|\t \mathbf{1}\rangle, \nonumber\\
\big< 1_h2_l; \t \t \mathbf{1}|H^{\rm leg}_{\rm kin}|1_l2_h; \t \t \mathbf{1}\rangle &=&  -\tl  e^{-3\pi i/5} \phi^{-1} \langle\mathbf{1} \t|h_{\rm kin}|\t \mathbf{1}\rangle, \nonumber\\
\big< 1_h2_l; \t  \mathbf{1} \t |H^{\rm leg}_{\rm kin}|1_l2_h; \t \t \t \rangle& =&  \big< 1_h2_l; \t  \t \t |H^{\rm leg}_{\rm kin}|1_l2_h; \t  \mathbf{1} \t \rangle = \tl  e^{-3\pi i/5} \phi^{-1/2} \langle\mathbf{1} \t|h_{\rm kin}|\t \mathbf{1}\rangle, \nonumber\\
\big< 1_h2_l; \t  \t \t |H^{\rm leg}_{\rm kin}|1_l2_h; \t \t \t \rangle& =&  \tl  e^{-3\pi i/5} \phi^{-2} \langle\mathbf{1} \t|h_{\rm kin}|\t \mathbf{1}\rangle   .
\end{eqnarray}
\end{widetext}
The matrix elements for hopping on the other leg remain the same except for the direction of the braid being reversed, {\it i.e.} all phase factors $e^{-3\pi i/5} \rightarrow e^{3\pi i/5}$.
As we sum up contributions from both the legs, the phase factors add up and the entire matrix (in the basis (\ref{eq:basis2})) is written as :
\begin{equation}
   H_{\rm{eff}}=t\begin{bmatrix}
   1 &  &  &  & \\
    & -\phi^{-1} &  &  & \\
    &  & -\phi^{-1} && \\
    &&&0 & \phi^{-1/2}\\
    &&&\phi^{-1/2}&\phi^{-2} 
\end{bmatrix}  ,
\end{equation}
where the effective hopping amplitude is\begin{equation}\label{eq:t3}t =  \big(\cos (3 \pi/5)\big) t_{\rm leg}  .\end{equation} 
A characteristic feature of this effective hopping Hamiltonian is that it mixes the spin labels.
This is remarkably different from the simple hopping process between a hole and a $\tau$.  In the text, we have quoted the effective Hamiltonian matrix for this process as $H_{\rm HL}$ described in Eq.~(\ref{eq:ham_hl}), which is related to $H_{\rm{eff}}$ by rescaling it such that the $2 \times 2$ block has eigenvalues 0 and 1. More precisely,  $H_{\rm HL} = a H_{\rm eff}+b \mathbb{I}$ with $a = \phi^{-1}$ and $b= \phi^{-2}$ (where $\mathbb{I}$ is the identity matrix).

\subsubsection{Three-leg ladder}\label{3leg:kin_all}
For a three-leg ladder, we can derive the relevant matrix elements in a similar way.

{\bf Light hole -- light $\tau$~:} Let us start with the simplest case and compute the matrix element $\big< \Psi_1| H^{\rm leg}_{\rm kin}|\Psi_2\rangle$
where $|\Psi_1\rangle=|s\rangle \otimes |e\rangle$,  $|\Psi_2\rangle=|e\rangle \otimes |s\rangle$,
$|s\rangle\equiv | 1^\pm, \t\rangle$ is defined in (\ref{eq:s3leg}) and $|e\rangle$ is the empty rung. Using obvious notations,
\begin{eqnarray}
\big< \Psi_1| H^{\rm leg}_{\rm kin}|\Psi_2\rangle&=&\frac{1}{4}\big( \big< eU| H^{\rm leg}_{\rm kin}|Ue\rangle\nonumber\\
&+& 2\big< eM| H^{\rm leg}_{\rm kin}|Me\rangle\nonumber\\
&+& \big< eL| H^{\rm leg}_{\rm kin}|Le\rangle\big) \nonumber\\
 &=&\big< eU| H^{\rm leg}_{\rm kin}|Ue\rangle  ,
\end{eqnarray}
Since all the $F$-moves are trivial because of the holes on the rungs and there would be no phase factors due to the braidings either, we get the effective hopping:
\begin{equation}\label{eq:t4} t = t_{\rm{leg}}  .\end{equation} 

{\bf Heavy hole -- light $\tau$~:} The calculation is slightly more involved when the effective 
effective hole state $|d\rangle$ involves two $\t$ anyons on the rung,
\begin{equation}\label{eq:d}
|d\rangle = \frac{1}{\sqrt{2+\alpha^2}}(|\overline{L}\rangle + |\overline{U}\rangle + \alpha|\overline{M}\rangle)  ,
\end{equation}
where $\overline{X}$ means a vacant site on the rung at position $X$.
The matrix element $\big< \Psi_1| H^{\rm leg}_{\rm kin}|\Psi_2\rangle$
now involves the initial and final states,
\begin{widetext}
\begin{eqnarray}
|\Psi_1\rangle &=& |s\rangle \otimes |d\rangle \nonumber\\
&=&\frac{1}{2\sqrt{2+\alpha^2}}\Big(|U\overline{U}\rangle + \sqrt2 |M\overline{U}\rangle + |L\overline{U}\rangle +\alpha |U\overline{M}\rangle +\sqrt2\alpha|M\overline{M}\rangle+\alpha|L\overline{M}\rangle +|U\overline{L}\rangle + \sqrt2 |M\overline{L}\rangle + |L\overline{L}\rangle \Big) \\
|\Psi_2\rangle &=& |d\rangle \otimes |s\rangle \nonumber\\
&=&\frac{1}{2\sqrt{2+\alpha^2}}\Big(|\overline{U}U\rangle + \sqrt2 |\overline{U}M\rangle + |\overline{U}L\rangle+
\alpha |\overline{M}U\rangle +\sqrt2\alpha|\overline{M}M\rangle+\alpha|\overline{M}L\rangle+
|\overline{L}U\rangle + \sqrt2 |\overline{L}M\rangle + |\overline{L}L\rangle \Big) \, .
\end{eqnarray}

After expanding both sides one gets,
\begin{eqnarray}
\langle\Psi_1|H^{\rm leg}_{\rm kin}|\Psi_2\rangle &=& \frac{1}{4(2+\alpha^2)}\Big(\langle \overline{L}L|H^{\rm leg}_{\rm kin}| U\overline{U} \rangle
%\nonumber\\&+&
+ \sqrt2\alpha \langle \overline{M} M|H^{\rm leg}_{\rm kin}| U\overline{U} \rangle 
%\nonumber\\&+& 
+ 2 \langle \overline{U} M|H^{\rm leg}_{\rm kin}| M\overline{U} \rangle \nonumber\\
&+&\langle \overline{U} L|H^{\rm leg}_{\rm kin}| L\overline{U} \rangle 
%\nonumber\\&+&
+\alpha^2\langle \overline{M} U|H^{\rm leg}_{\rm kin}| U\overline{M} \rangle 
%\nonumber\\&+&
+\sqrt2\alpha \langle \overline{L} L|H^{\rm leg}_{\rm kin}| M\overline{M} \rangle \nonumber\\
&+&\sqrt2 \alpha \langle \overline{U} U|H^{\rm leg}_{\rm kin}| M\overline{M} \rangle
 % \nonumber\\&+& 
+\alpha^2 \langle \overline{M} L|H^{\rm leg}_{\rm kin}| L\overline{M} \rangle
% \nonumber\\&+& 
 +\langle \overline{L} U|H^{\rm leg}_{\rm kin}| U\overline{L} \rangle \nonumber\\
&+&2  \langle \overline{L} M|H^{\rm leg}_{\rm kin}| M\overline{L} \rangle %\nonumber\\&+&
+\sqrt2 \alpha \langle \overline{M} M|H^{\rm leg}_{\rm kin}| L\overline{L} \rangle %\nonumber\\&+& 
+\langle \overline{U} U|H^{\rm leg}_{\rm kin}| L\overline{L} \rangle \Big) .
\label{eq:mat1}
\end{eqnarray}
%\end{widetext}

%\begin{widetext}
The contributions of the individual terms are as follows:
\begin{eqnarray}
\langle \overline{L}L|H^{\rm leg}_{\rm kin}| U\overline{U} \rangle &=&  \langle \overline{U} U|H^{\rm leg}_{\rm kin}| L\overline{L} \rangle=\phi^{-1} \tl \big< \mathbf{1}\t|h_{\rm kin}|\t \mathbf{1}\rangle,\nonumber \\
 \langle \overline{M} M|H^{\rm leg}_{\rm kin}| U\overline{U} \rangle &=& \langle \overline{U} M|H^{\rm leg}_{\rm kin}| M\overline{U} \rangle =  \langle \overline{L} U|H^{\rm leg}_{\rm kin}| U\overline{L} \rangle = e^{4\pi i/5}\phi^{-1} \tl \big< \mathbf{1}\t|h_{\rm kin}|\t \mathbf{1}\rangle,\nonumber\\
 \langle \overline{L} L|H^{\rm leg}_{\rm kin}| M\overline{M} \rangle &=& \langle \overline{U} L|H^{\rm leg}_{\rm kin}| L\overline{U} \rangle =  \langle \overline{M} M|H^{\rm leg}_{\rm kin}| L\overline{L} \rangle = e^{-4\pi i/5}\phi^{-1} \tl \big< \mathbf{1}\t|h_{\rm kin}|\t \mathbf{1}\rangle,\nonumber\\
 \langle \overline{M} U|H^{\rm leg}_{\rm kin}| U\overline{M} \rangle &=&  \langle \overline{U} U|H^{\rm leg}_{\rm kin}| M\overline{M} \rangle=e^{4\pi i/5}\phi^{-1} \tl \big< \mathbf{1}\t|h_{\rm kin}|\t \mathbf{1}\rangle,\nonumber\\
 \langle \overline{M} L|H^{\rm leg}_{\rm kin}| L\overline{M} \rangle &=&  \langle \overline{L} M|H^{\rm leg}_{\rm kin}| M\overline{L} \rangle=e^{-4\pi i/5}\phi^{-1} \tl \big< \mathbf{1}\t|h_{\rm kin}|\t \mathbf{1}\rangle .
\label{eq:mat2}
\end{eqnarray}
\end{widetext}
Replacing (\ref{eq:mat2}) into (\ref{eq:mat1}) one gets the effective hopping,
\begin{eqnarray}\label{eq:t5}
t&=&\frac{\langle\Psi_1|H^{\rm leg}_{\rm kin}|\Psi_2\rangle}{\big<\mathbf{1}\t |h_{\rm kin}|\t \mathbf{1}\rangle} \nonumber\\
&=& \frac{1}{(2+\alpha^2)2\phi} \Big[(3+\alpha^2+2\sqrt2\alpha)\bigg(\cos\frac{4\pi}{5}\bigg)  + 1\Big]  t_{\rm{leg}} \nonumber \\
\end{eqnarray}

{\bf Heavy hole -- super-heavy $\tau$~:} The third case corresponds to the effective hole state $|d\rangle$ defined in (\ref{eq:d}) and the 
effective \t particle state defined by the fully occupied rung $|f\rangle=|3,\t\rangle$. The initial 
and final states $|\Psi_1\rangle$ and $|\Psi_2\rangle$ are now given by :
\begin{eqnarray}
|\Psi_1\rangle &=& |d\rangle \otimes |f\rangle \nonumber\\
&=&\frac{1}{\sqrt{2+\alpha^2}}\Big(|\overline{U}f\rangle +\alpha|\overline{M}f\rangle + |\overline{L}f\rangle  \Big)  ,\\
|\Psi_2\rangle &=& |f\rangle \otimes |d\rangle \nonumber\\
&=&\frac{1}{\sqrt{2+\alpha^2}}\Big(|f\overline{U}\rangle + \alpha |f\overline{M}\rangle + |f\overline{L}\rangle \Big)  .
\end{eqnarray}

The matrix element for the kinetic Hamiltonian on the legs is now expressed as :
\begin{eqnarray}
\langle\Psi_2|H^{\rm leg}_{\rm kin}|\Psi_1\rangle &=& \frac{1}{2+\alpha^2}\Big(\langle f\overline{U}|H^{\rm leg}_{\rm kin}|\overline U{f} \rangle\nonumber\\
&+& \alpha^2 \langle f\overline{M} |H^{\rm leg}_{\rm kin}|\overline M{f} \rangle \nonumber\\
&+&\langle f\overline{L} |H^{\rm leg}_{\rm kin}|\overline L{f} \rangle \Big)  .
\end{eqnarray}

The individual contributions of these terms are:
\begin{eqnarray}
\langle f\overline{U}|H^{\rm leg}_{\rm kin}| U\overline{f} \rangle &=& \phi^{-2} t_{\rm{leg}}  \big<\mathbf1\t |h_{\rm kin}|\t \mathbf1\rangle,\nonumber\\
\langle f\overline{M} |H^{\rm leg}_{\rm kin}| M\overline{f} \rangle&=& \bigg(\frac{1}{2 \phi}+\phi^{-3}\bigg)  t_{\rm{leg}}\big<\mathbf1\t |h_{\rm kin}|\t \mathbf1\rangle,\nonumber\\
 \langle f\overline{L} |H^{\rm leg}_{\rm kin}| L\overline{f} \rangle &=& t_{\rm{leg}}\big<\mathbf1\t |h_{\rm kin}|\t \mathbf1\rangle  .
 \end{eqnarray}
Adding the contributions of the three terms we get:
\begin{eqnarray}\label{eq:t6}
t&=&\frac{\langle\Psi_2|H^{\rm leg}_{\rm kin}|\Psi_1\rangle}{\big<\mathbf1\t |h_{\rm kin}|\t \mathbf1\rangle} \nonumber\\
&=& \frac{1}{2+\alpha^2}\bigg[\frac{1}{\phi^{2}}+\frac{\alpha^2}{2\phi}+\frac{\alpha^2}{\phi^{3}}+1\bigg]  t_{\rm{leg}}  .
\end{eqnarray}

{\bf Super-heavy hole -- heavy $\tau$~:} The fourth case corresponds to a \t in the state  $|2^\pm, \t\rangle$ defined in Eq.~(\ref{eq:2pm}).
The effective hole on the ladder is defined by the fully occupied state $|f\rangle \equiv |3,{\mathbf 1}\rangle$. We calculate the matrix element of the states
\begin{eqnarray}
 |\Psi_1\rangle&=&| 2^\pm, \t\rangle \otimes|f\rangle \nonumber\\
  &\equiv& \frac{1}{2}\big({|UM,f\rangle +  |ML,f\rangle \pm \sqrt2|UL,f\rangle\big)}
 \end{eqnarray}
 and 
 \begin{eqnarray}
 |\Psi_2\rangle&=&|f\rangle\otimes| 2^\pm, \t\rangle  \nonumber\\
  &\equiv& \frac{1}{2}\big({|f,UM\rangle +  |f,ML\rangle \pm \sqrt2|f,UL\rangle\big)}.
 \end{eqnarray}
 The matrix elements are given by
   \begin{eqnarray}\label{eq:tFM}
\langle\Psi_2| H^{\rm leg}_{\rm kin}|\Psi_1\rangle &=&\frac{1}{4}\big( \langle f,UM|H^{\rm leg}_{\rm kin}|UM,f\rangle \nonumber \\
&+& \langle f,ML|H^{\rm leg}_{\rm kin}|ML,f\rangle \nonumber \\
&+&2 \langle f,UL|H^{\rm leg}_{\rm kin}|UL,f\rangle\big)\, .
  \end{eqnarray}
The contributions of the individual hopping terms are
   \begin{eqnarray}\label{eq:hopFM}
 \langle f,UM|H^{\rm leg}_{\rm kin}|UM,f\rangle &=& \phi^{-1}e^{6\pi i/5}{\big<\mathbf1\t |h_{\rm kin}|\t \mathbf1\rangle}, \nonumber \\
\langle f,UL|H^{\rm leg}_{\rm kin}|UL,f\rangle &=& \phi^{-1}e^{-6\pi i/5}{\big<\mathbf1\t |h_{\rm kin}|\t \mathbf1\rangle}, \nonumber \\
 \langle f,ML|H^{\rm leg}_{\rm kin}|ML,f\rangle &=&\phi^{-1} {\big<\mathbf1\t |h_{\rm kin}|\t \mathbf1\rangle}\, .
  \end{eqnarray}
Using Eqs.~\ref{eq:hopFM} in Eq.~(\ref{eq:tFM}), we get
\begin{eqnarray}\label{eq:t7}
t&=&\frac{\langle\Psi_2|H^{\rm leg}_{\rm kin}|\Psi_1\rangle}{\big<\mathbf1\t |h_{\rm kin}|\t \mathbf1\rangle} \nonumber\\
&=& \bigg(\frac{1}{2\phi}-\frac{1}{4}\bigg)  t_{\rm{leg}}  .
\end{eqnarray}

{\bf Light $\tau$ -- Heavy $\tau$~:} Finally, we consider two rungs with a light \t defined by the state $|s\rangle$ in (\ref{eq:s3leg}) and a heavy \t defined by the state $|h\rangle$ given as :
\begin{equation}\label {eq:h}
|h \rangle = \frac{1}{2}(|\overline{L}\rangle + \sqrt2 |\overline{M}\rangle+|\overline{U}\rangle)   .
\end{equation}

The initial and final states, formed by the tensor product of $|s\rangle$ and $|h\rangle$ are defined as:
\begin{eqnarray}
|\Psi_1\rangle &=& |s\rangle \otimes |h\rangle \nonumber\\
&=&\frac{1}{4}\Big(|U\overline{U}\rangle + \sqrt2 |M\overline{U}\rangle + |L\overline{U}\rangle \nonumber\\
&+&\sqrt2 |U\overline{M}\rangle +2|M\overline{M}\rangle+\sqrt2 |L\overline{M}\rangle \nonumber\\
&+&|U\overline{L}\rangle +\sqrt2 |M\overline{L}\rangle + |L\overline{L}\rangle \Big) \end{eqnarray}
and,
\begin{eqnarray}
|\Psi_2\rangle &=& |h\rangle \otimes |s\rangle \nonumber\\
&=&\frac{1}{4}\Big(|\overline{U}U\rangle + \sqrt2 |\overline{U}M\rangle + |\overline{U}L\rangle \nonumber\\
&+& \sqrt2 |\overline{M}U\rangle +\sqrt2|\overline{M}M\rangle+\sqrt2 |\overline{M}L\rangle \nonumber\\
&+&|\overline{L}U\rangle + \sqrt2 |\overline{L}M\rangle + |\overline{L}L\rangle \Big)  . 
\end{eqnarray}
The matrix elements corresponding to the hopping along the leg on the ladder can be expanded using the expression of the states to give :
\begin{widetext}
\begin{eqnarray}\label{eq:hl1}
\langle\Psi_1|H^{\rm leg}_{\rm kin}|\Psi_2\rangle &=& \frac{1}{16}\Big(\langle \overline{L}L|H^{\rm leg}_{\rm kin}| U\overline{U} \rangle
%\nonumber\\&+&
+ 2 \langle \overline{M} M|H^{\rm leg}_{\rm kin}| U\overline{U} \rangle %\nonumber\\&+& 
+2 \langle \overline{U} M|H^{\rm leg}_{\rm kin}| M\overline{U} \rangle %\nonumber\\&+&
+\langle \overline{U} L|H^{\rm leg}_{\rm kin}| L\overline{U} \rangle \nonumber\\
&+& 2\langle \overline{M} U|H^{\rm leg}_{\rm kin}| U\overline{M} \rangle %\nonumber\\&+&
+2 \langle \overline{L} L|H^{\rm leg}_{\rm kin}| M\overline{M} \rangle %\nonumber\\&+& 
+2 \langle \overline{U} U|H^{\rm leg}_{\rm kin}| M\overline{M} \rangle %\nonumber\\
+ 2\langle \overline{M} L|H^{\rm leg}_{\rm kin}| L\overline{M} \rangle \nonumber\\
&+& \langle \overline{L} U|H^{\rm leg}_{\rm kin}| U\overline{L} \rangle% \nonumber\\&+&
+2  \langle \overline{L} M|H^{\rm leg}_{\rm kin}| M\overline{L} \rangle %\nonumber\\&+& 
+2 \langle \overline{M} M|H^{\rm leg}_{\rm kin}| L\overline{L} \rangle %\nonumber\\&+& 
+\langle \overline{U} U|H^{\rm leg}_{\rm kin}| L\overline{L} \rangle \Big) .
\end{eqnarray}
%\end{widetext}
The contributions of the individual terms are as follows:
%\begin{widetext}
\begin{eqnarray}\label{eq:hl2}
\langle \overline{L}L|H^{\rm leg}_{\rm kin}| U\overline{U} \rangle &=&  \langle \overline{U} U|H^{\rm leg}_{\rm kin}| L\overline{L} \rangle =\tl \big< \mathbf{1}\t|h_{\rm kin}|\t \mathbf{1}\rangle, \nonumber\\
\langle \overline{M} M|H^{\rm leg}_{\rm kin}| U\overline{U} \rangle&=&  \langle \overline{U} M|H^{\rm leg}_{\rm kin}| M\overline{U} \rangle =\langle \overline{L} U|H^{\rm leg}_{\rm kin}| U\overline{L} \rangle=\tl e^{-3\pi i/5}\big< \mathbf{1}\t|h_{\rm kin}|\t \mathbf{1}\rangle,  \nonumber\\
\langle \overline{U} L|H^{\rm leg}_{\rm kin}| L\overline{U} \rangle& =& \langle \overline{L} L|H^{\rm leg}_{\rm kin}| M\overline{M} \rangle = \langle \overline{M} L|H^{\rm leg}_{\rm kin}| L\overline{M} \rangle = \tl e^{3\pi i/5}\big< \mathbf{1}\t|h_{\rm kin}|\t \mathbf{1}\rangle,  \nonumber\\
\langle \overline{M} U|H^{\rm leg}_{\rm kin}| U\overline{M} \rangle &=& \langle \overline{U} U|H^{\rm leg}_{\rm kin}| M\overline{M} \rangle =\tl  e^{-3\pi i/5}\big< \mathbf{1}\t|h_{\rm kin}|\t \mathbf{1}\rangle,  \nonumber\\
\langle \overline{M} M|H^{\rm leg}_{\rm kin}| L\overline{L} \rangle &=&  \langle \overline{L} M|H^{\rm leg}_{\rm kin}| M\overline{L} \rangle =\tl  e^{3\pi i/5}\big< \mathbf{1}\t|h_{\rm kin}|\t \mathbf{1}\rangle    .
\end{eqnarray}
Replacing (\ref{eq:hl2}) into (\ref{eq:hl1}) one gets the effective hopping,
\begin{eqnarray}\label{eq:t8}
t&=&\frac{\langle\Psi_1|H^{\rm leg}_{\rm kin}|\Psi_2\rangle}{\big<\mathbf1\t |h_{\rm kin}|\t \mathbf1\rangle} \nonumber\\
&=& \frac{1}{8}\big[1+9\cos(3\pi/5)\big] t_{\rm{leg}}  .
\end{eqnarray}

\end{widetext}

\subsection{Higher order terms}\label{app:higher_order}

In addition to the above cases, we can also have a kinetic and potential terms when the difference in the $U(1)$ 
charges on neighboring rungs is larger than 1. Such  process occurs e.g. (i) between a charge-0 (light) hole and a charge-2 (heavy) hole (marked by blue arrows in Table.~\ref{table}) in the $P_2$ paired phase
or (ii) between a charge-0 (light) hole and a charge-3 (super heavy) $\tau$ (marked by orange arrows in Table.~\ref{table}) in  the $PS_{03}$ phase.

In case (i), one needs to hop twice to be able to come back to a configuration $|{\Psi_J}\rangle$ that has the same $U(1)$ charges as the initial configuration $|{\Psi_I}\rangle$, so the effective Hamiltonian (leaving in the relevant subspace) is obtained in second-order perturbation,
in $t_{\rm leg}$\cite{iprm02,jpm14} 
\begin{equation}
\langle{\Psi_I}|H_{\rm eff}|{\Psi_J}\rangle =-\sum_{r}\frac{\langle{\Psi_I}|H_{\rm kin}^{\rm leg}|{\Psi_r}\rangle\langle{\Psi_r}|H_{\rm kin}^{\rm leg}|{\Psi_J}\rangle}{E_r-E_J},
\end{equation}
where the sum is on the intermediate states corresponding to two (light) $\tau$'s on neighboring sites. The energy denominator is given by 
$E_D = (-2\tr + J_{\rm rung})$
and matrix elements for all the intermediate states are ($-t_{\rm leg}$) 
so that one gets 
for $I=J$ a potential energy 
\begin{equation}\label{eq:v2order}
V=-\frac{2 \tl^2}{E_D}
\end{equation}
and for $I\ne J$ a hopping term 
\begin{equation}\label{eq:t2order}
t=-\frac{2\tl^2}{E_D} .
\end{equation}

In case (ii),  one needs to hop three times to be able to come back to a configuration that has the same $U(1)$ charges as the initial configuration, so the effective Hamiltonian is now obtained in third-order perturbation in $t_{\rm leg}$:\cite{iprm02,jpm14}
{\small{\begin{equation}
\langle{\Psi_I}|H_{\rm eff}|{\Psi_J}\rangle=
            \sum_{r,s}\frac{\langle{\Psi_I}|H_{\rm kin}^{\rm leg}|{\Psi_r}\rangle\langle{\Psi_r}|H_{\rm kin}^{\rm leg}|{\Psi_s}\rangle\langle{\Psi_s}|H_{\rm kin}^{\rm leg}|{\Psi_J}\rangle}{(E_s-E_J)(E_r-E_J)}, 
\end{equation}}}
where the intermediate states $|{\Psi_r}\rangle$ and $|{\Psi_s}\rangle$ carry a heavy hole and a light \t on neighboring rungs. The energy denominators given by the difference in energy between the initial (degenerate with the final) state and the intermediate states thus take the value 
\begin{equation}
E_D = (\phi-\frac{1}{2})\jr - \sqrt{2}\tr-\frac{\sqrt{\jr^2+8\tr^2}}{2} .
\end{equation} 
Starting from an initial state $|3,\tau\rangle \otimes |e\rangle$ a simple hopping yields the intermediate state $|2,\mathbf{1}\rangle \otimes |1,\tau\rangle$ with the matrix elements given by for hopping on the upper leg
\begin{equation}\label{eq:t1u}
t^1_U=\phi^{-1} t_{\rm leg},
\end{equation}
on the middle leg,
\begin{equation}\label{eq:t1m}
t^1_M= \phi^{-1} e^{-4i\pi/5} t_{\rm leg},
\end{equation}
and on the lower leg,
\begin{equation}\label{eq:t1l}
t^1_L= t_{\rm leg} \, .
\end{equation}
Subsequently, a second hopping yields the intermediate state $|1,\tau\rangle \otimes |2,\mathbf{1}\rangle$ with hopping matrix elements on the upper leg
\begin{equation}
t^2_U=\phi^{-1}e^{-4i\pi/5} t_{\rm leg},
\end{equation}
on the middle leg,
\begin{equation}
t^2_M= \phi^{-1} t_{\rm leg},
\end{equation}
and on the lower leg,
\begin{equation}
t^2_L=\phi^{-1}e^{4i\pi/5} t_{\rm leg} \, .
\end{equation}
Finally to come back to a state carrying the same $U(1)$ charges on the rungs, the hopping amplitudes for the three legs are just the complex conjugate of those described in Eqs.~(\ref{eq:t1u})-(\ref{eq:t1l}), thus giving for $I=J$ a potential energy 
\begin{equation}
V=\frac{(2\phi^{-2}+1) \tl^2}{E_D}
\end{equation} 
and for $I\ne J$ a hopping term 
\begin{equation}\label{eq:t3order}
t=\frac{\tl^3}{E_D^2} (3\phi^{-2}+2\phi^{-3}+\phi^{-2}e^{8\pi i/5}) .
\end{equation}

\end{document}